\documentclass[journal,onecolumn]{IEEEtran}

\usepackage[utf8]{inputenc}
\usepackage[T1]{fontenc}
\usepackage{textcomp}
\usepackage{gensymb}
\usepackage{blindtext}
\usepackage{enumitem}
\setlist{parsep=0pt,listparindent=\parindent}
\usepackage{amsmath}
\usepackage{float}
\usepackage{caption}
\usepackage{placeins}
\usepackage{lipsum}  
\usepackage{graphicx}
\usepackage{caption}
\usepackage{subcaption}
\extrafloats{100}
\usepackage{cite}
\usepackage{listings}
\usepackage{color}
\usepackage{listings}
\usepackage{amsfonts}

\newfloat{lstfloat}{htbp}{lop}
\floatname{lstfloat}{Listing}
 
\definecolor{codegreen}{rgb}{0,0.6,0}
\definecolor{codegray}{rgb}{0.5,0.5,0.5}
\definecolor{codepurple}{rgb}{0.58,0,0.82}
\definecolor{backcolour}{rgb}{0.95,0.95,0.92}
\lstdefinestyle{mystyle}{
    commentstyle=\color{codegreen},
    numberstyle=\tiny\color{codegray},
    stringstyle=\color{codepurple},
    basicstyle=\footnotesize,
    breakatwhitespace=false,         
    breaklines=true,                 
    captionpos=b,                    
    keepspaces=true,                 
    numbers=left,                    
    numbersep=5pt,                  
    showspaces=false,                
    showstringspaces=false,
    showtabs=false,                  
    tabsize=2
}
\begin{document}

\title{Flight Trajectory Planning for Fixed-Wing Aircraft \\ in Loss of Thrust Emergencies.}

\author{
\IEEEauthorblockN{
Saswata Paul\IEEEauthorrefmark{1},
Frederick Hole\IEEEauthorrefmark{2},
Alexandra Zytek\IEEEauthorrefmark{1} and
Carlos A. Varela\IEEEauthorrefmark{1}
}\\
\IEEEauthorblockA{
Department of Computer Science\IEEEauthorrefmark{1};
Department of Mechanical, Aerospace, and Nuclear Engineering\IEEEauthorrefmark{2}\\
Rensselaer Polytechnic Institute, Troy, New York, 12180\\
\{pauls4,
holef,
zyteka\}@rpi.edu,
cvarela@cs.rpi.edu
}
}

\maketitle

\begin{abstract}
Loss of thrust emergencies---e.g., induced by bird/drone strikes or fuel exhaustion---create the need for dynamic data-driven flight trajectory planning to advise pilots or control UAVs.  While total loss of thrust (gliding) trajectories to nearby airports can be pre-computed for all initial points in a 3D flight plan, dynamic aspects such as partial power, wind, and airplane surface damage must be considered for accuracy.
In this paper, we propose a new Dynamic Data-Driven Avionics Software (DDDAS) approach which during flight updates a damaged aircraft performance model, used in turn to generate plausible flight trajectories to a safe landing site.  Our damaged aircraft model is parameterized on a baseline glide ratio for a clean aircraft configuration assuming best gliding airspeed on straight flight.  The model predicts purely geometric criteria for flight trajectory generation, namely, glide ratio and turn radius for different bank angles and drag configurations.  Given actual aircraft flight performance data, we dynamically infer the baseline glide ratio to update the damaged aircraft model. Our new flight trajectory generation algorithm thus can significantly improve upon prior Dubins based trajectory generation work by considering these data-driven geometric criteria. We further introduce a trajectory utility function to rank trajectories for safety, in particular, to prevent steep turns close to the ground and to remain as close to the airport or landing zone as possible. 
As a use case, we consider the Hudson River ditching of US Airways 1549 in January 2009 using a flight simulator to evaluate our trajectories and to get sensor data (airspeed, GPS location, barometric altitude). In this example, a baseline glide ratio of 17.25:1 enabled us to generate trajectories up to 28 seconds after the birds strike, whereas, a 19:1 baseline glide ratio enabled us to generate trajectories up to 36 seconds after the birds strike. DDDAS can significantly improve the accuracy of generated flight trajectories thereby enabling better decision support systems for pilots in total and partial loss of thrust emergency conditions.
\end{abstract}

\section{Introduction}

Dynamic Data-Driven Applications and Systems (DDDAS) use data from sensor measurements to dynamically update a system's model, thereby, improving the model's accuracy and its effectiveness in prediction-based applications~\cite{darema}. Flight systems are quite complex, rendering pure model-based approaches insufficient to accurately predict an aircraft's performance upon system failures. In this paper\let\thefootnote\relax\footnote{This work was accepted as a full paper and presented in the Second International Conference on InfoSymbiotics / DDDAS (Dynamic Data Driven Applications Systems) held at MIT, Cambridge, Massachusetts in August, 2017.}, we investigate Dynamic Data-Driven Avionics Software (DDDAS), in decision support systems for unexpected loss of thrust (LOT) emergencies. LOT emergencies can result from fuel exhaustion as in Tuninter 1153's accident, but also from bird strikes as in the Hudson River ditching of US Airways 1549. Aerodynamic models can be used to plan flyable trajectories from the airplane's LOT initial location to an airport runway or to an off-airport landing site.  We propose a DDDAS approach to flight trajectory generation that distills a complex aerodynamics model into purely geometric constraints: glide ratio and radius of turns for different bank angles and drag configurations. Our damaged aircraft model can predict the different glide ratios and radii of turns using a single parameter: the aircraft's {\it baseline glide ratio, $g_0$}.  $g_0$ represents the glide ratio for a clean aircraft configuration assuming best gliding airspeed on straight flight. If we can infer the actual $g_0$ from sensor data on the fly, we can re-compute flight trajectories that will automatically consider dynamic factors, such as partial power, wind aloft, and airplane surface damages, since these factors will impact the inferred baseline glide ratio. By considering different bank angles and drag configurations in the generated plans, we can rank different plausible trajectories according to the risk they impose on flights. We develop safety metrics to rank flyable trajectories considering maximum distance from landing site, total length or trajectory, total time of trajectory, average altitude, length of extended runway segment and average bank angle over height. We use a flight simulator to evaluate generated trajectories, and to gather sensor data (e.g., airspeed, GPS location, and barometric altitude) to dynamically update the damaged aircraft performance model.

Our dynamic data driven trajectory generation approach generates a trajectory in the clean configuration with no thrust and gathers data from aircraft sensors for comparison. If there is an inconsistency between the two data, the observed data is sent to a model refinement component, which infers a new glide ratio and sends it to the damaged aircraft model. The damaged aircraft model uses this refined baseline glide ratio to create a new function to compute glide ratios for different bank angles and sends it to the trajectory generation algorithm to generate trajectories which are consistent with the actual capabilities of the aircraft. The damaged aircraft model takes as input the glide ratio for the clean configuration and a drag multiplier table for generating the glide ratios in dirty configurations and generates glide ratios and radii of turn for different values of bank angle, airspeed and drag configuration. Our trajectory planning algorithm generates trajectories from an initial point to a given runway or a set of runways in case of a LOT emergency situation. After the possible trajectories are generated, they are evaluated on the basis of several safety metrics and ranked according to their safety level. This is important because in case of a real emergency, the pilots have to choose a course of action in a matter of seconds and if trajectories are already ranked, it becomes easier for the pilots to make a fast and educated choice.    

\begin{table*}
\caption {Glide ratio and radius of turn for various bank angles at best glide speed(225 kts) for Airbus A320.} \label{tab:table1}
\centering
 \begin{tabular}{c c c c c c c } 
 Bank angle:& 0° & 10° & 20° & 30° & 45° & 60°\\ [1ex] 

 Glide ratio:& 17.25:1 & 16.98:1 & 16.21:1 & 14.92:1 & 12.19:1 & 8.62:1 \\ [1ex] 
 
 Radius of turn(feet):& \(\infty\) & 25430 & 12319 & 7766 & 4484 & 2588 \\ [1ex]
 \end{tabular}
 \caption {Glide ratio and radius of turn for various bank angles at best glide speed(65 kts) for Cessna 172.} \label{tab:table0}
\centering
 \begin{tabular}{c c c c c c c } 
 Bank angle:& 0° & 10° & 20° & 30° & 45° & 60°\\ [1ex] 

 Glide ratio:& 9:1 & 8.86:1 & 8.45:1 & 7.79:1 & 6.36:1 & 4.5:1 \\ [1ex] 
 
 Radius of turn(feet):& \(\infty\) & 2122 & 1028 & 648 & 374 & 216 \\ [1ex]
 \end{tabular}
\
\end{table*}
\begin{figure}[!htb]
  \centering
  \begin{subfigure}{.5\textwidth}
  \centering
  \includegraphics[width=.8\linewidth]{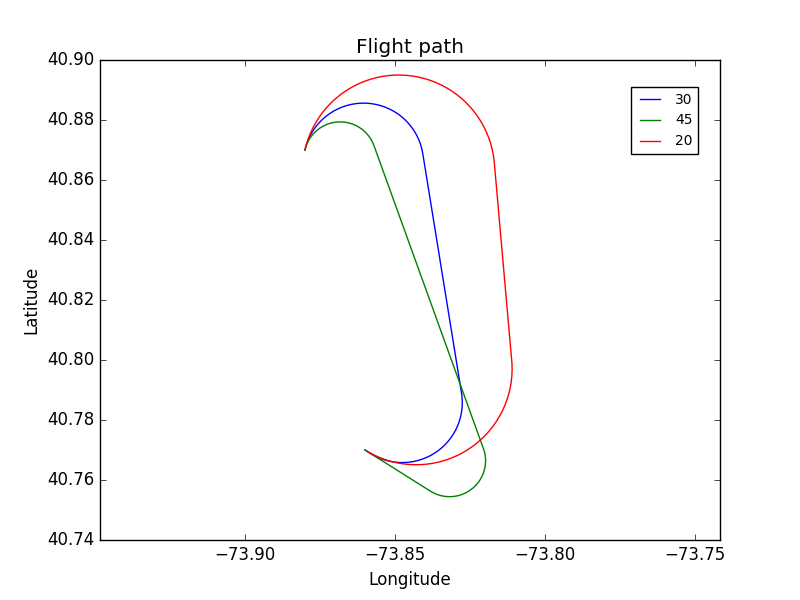}
  \caption{2D view.}
  \label{}
\end{subfigure}%
\begin{subfigure}{.5\textwidth}
  \centering
  \includegraphics[width=.8\linewidth]{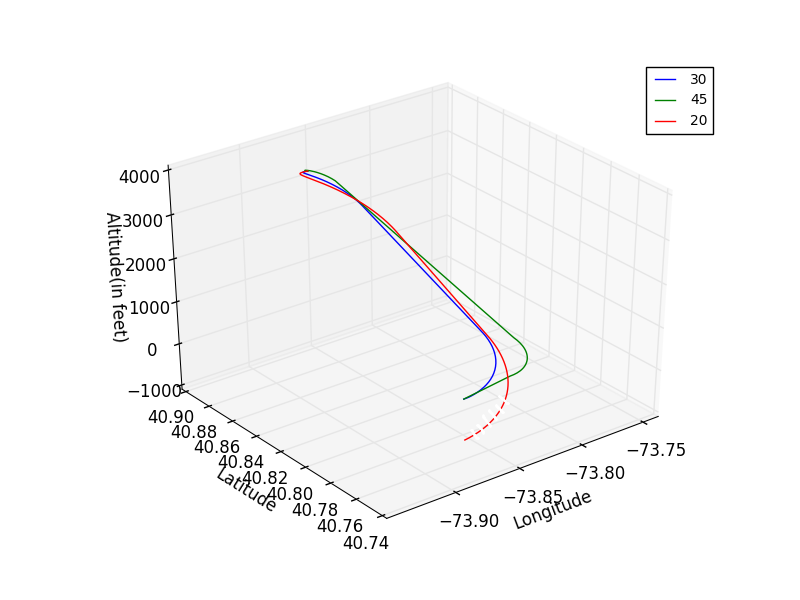}
  \caption{3D view}
  \label{}
\end{subfigure}
\caption{Effect of bank angle on trajectories.}
\label{fig:bank2d}
\end{figure}
Bank angle of turns has a major impact on the glide ratio and radius of turn of an aircraft (Table~\ref{tab:table1}, Table~\ref{tab:table0}). This difference in glide ratio and radius of turn in turn, has a major impact on the type of trajectories (Fig. \ref{fig:bank2d}). So, in our trajectory planning algorithm, we use three discrete values of bank angles: 20°, 30° and 45°. All the scenarios and experiments in this paper have been done for an Airbus A320.

The rest of the paper is structured as follows: Section II discusses prior work done by us on avionics systems and related work done on trajectory generation along with novel aspects of the work presented in this paper; Section III describes the aerodynamic model used in this paper; Section IV describes our novel trajectory planning algorithm; Section V describes the dynamic data driven model for trajectory generation; Section VI contains details of the experiments done along with the results observed; Section VII contains conclusions and future work and Section VIII contains acknowledgements.
\section{Previous Work}
\begin{figure}
  \centering
  \includegraphics{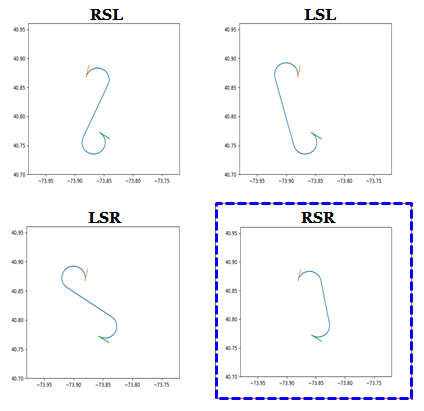}
  \caption{Four types of Dubins paths with a straight line segment: RSR, RSL, LSL and LSR. The shortest one (RSR in this case) is chosen.}
  \label{fig:dp}
\end{figure}
In prior work, we have developed a ProgrammIng Language for spatiO-Temporal data Streaming applications (PILOTS) to detect sensor failures from data and to estimate values for quantities of interest (e.g., airspeed, or fuel quantity) upon fault detection and isolation.  We have successfully applied PILOTS to data from actual aircraft accidents, including Air France 447 when pitot tubes icing resulted in wrong airspeed measurements, and Tuninter 1153, when a wrong fuel quantity indicator installation resulted in complete fuel exhaustion and engine failure~\cite{imai}. We have worked on data streaming application for avionics systems including development of failure models that can detect errors in sensor data, simulation of error detection and correction using real data from flights, development of error signatures to detect and classify abnormal conditions from sensor data, programming model to reason about spatio-temporal data streams, and design and implementation of a language for spatio-temporal data streaming applications~\cite{chen-pilots-dddas-2016,imai-galli-varela-pilots-dddas-2015,imai-klockr-varela-bdse-2013,imai-varela-pilots-dddas-2012,klockr-errorsignatures-dddas-2013,imai-varela-quest-2012}.

Dubins curves~\cite{dubins} are used to find the shortest paths, in two dimensions, between two configurations for a robot or a car that can move only in one direction (Fig. \ref{fig:dp}). Dubins 2D shortest car paths have been previously applied for generating shortest paths for aircraft in three dimensional space. Atkins \emph{et al} use Dubins paths to generate 3D trajectories for aircraft in the event of a loss of thrust emergency~\cite{atkins,atkins2}. They define worst case trajectory, direct trajectory and high altitude trajectory. Their high altitude trajectory has intermediate 'S' turns to lose excess altitude which might take the aircraft far away from the runway. Owen \emph{et al} propose Dubins airplane paths for fixed wing UAVs with power for maneuverability~\cite{owen}. They introduce three type of trajectories for low altitude, middle altitude and high altitude scenarios. The low altitude trajectory consists of a simple 3D Dubins airplane path, the high altitude trajectory consists of spiral turns preceeded by a Dubins airplane path and the middle altitude trajectory consists of an intermediate arc that allows the aircraft to lose excess altitude.

We try to improve upon prior work by proposing a trajectory generation algorithm that removes the need for \emph{'S' turns} and arcs and minimizes the number of different maneuvers in the trajectory. We do this by introducing an \textit{extended runway} in the final approach before landing that allows the aircraft to lose excess altitude. We also remove the need for a middle altitude scenario and incorporate it into the high altitude scenario. We evaluate generated trajectories using a set of \emph{trajectory safety metrics} that can be used to rank the trajectories depending on how safe they are. We use sensor data from the aircraft to recompute the baseline glide ratio after discrete intervals of time to regenerate new and more accurate trajectories that take into account the capabilities of the aircraft at that time. 
\section{Aircraft Model}
\begin{figure}[!htb]
  \centering
  \begin{subfigure}{.5\textwidth}
  \centering
  \includegraphics[width=.7\linewidth]{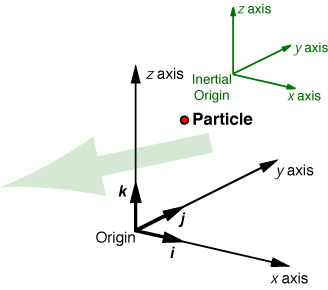}
  \caption{Position of a particle with respect to a 3D inertial frame\cite{stengel2015}.}
  \label{fig:ir}
\end{subfigure}%
\begin{subfigure}{.5\textwidth}
  \centering
  \includegraphics[width=.7\linewidth]{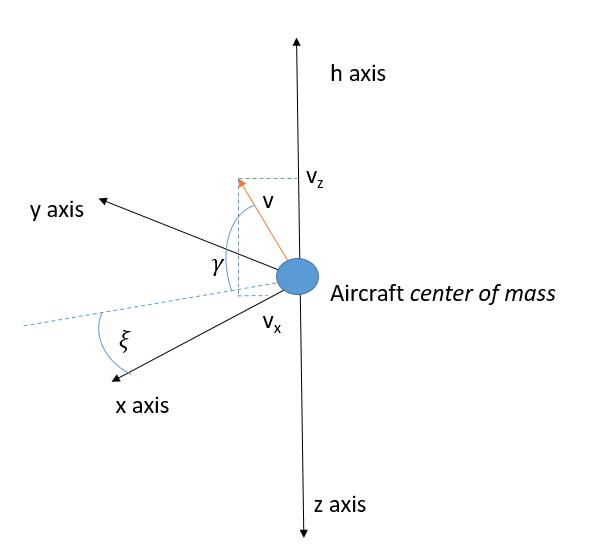}
  \caption{Inertial velocity expressed in polar coordinates.}
  \label{fig:pm}
\end{subfigure}
\caption{Position and inertial velocity of a point mass in a 3D frame.}
\label{}
\end{figure}
\begin{figure}[!htb]
  \centering
  \includegraphics[width=0.5\textwidth]{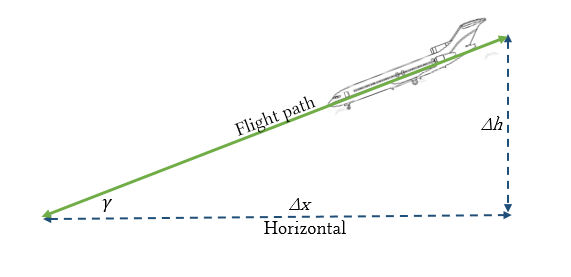}
  \caption{Glide ratio for a glider in straight line flight.}
  \label{fig:glide}
\end{figure}
\begin{figure}[!htb]
  \centering
  \begin{subfigure}{.5\textwidth}
  \centering
  \includegraphics[width=\linewidth]{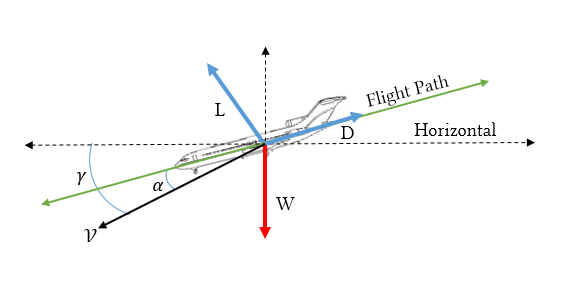}
  \caption{Forces on a gliding flight.}
  \label{fig:forces}
\end{subfigure}%
\begin{subfigure}{.5\textwidth}
  \centering
  \includegraphics[width=.7\linewidth]{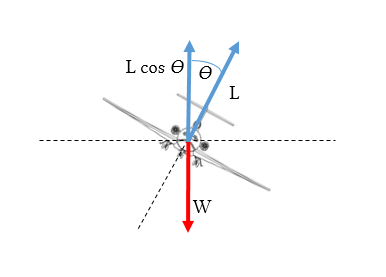}
  \caption{Weight vs lift in banked turns.}
  \label{fig:bank}
\end{subfigure}
\caption{Forces on a glider in straight line motion and banked turns.}
\label{}
\end{figure}

Stengel\cite{stengel2015} defines the position of a point with respect to a 3 dimensional inertial frame as follows:
\begin{equation}
r=\begin{bmatrix}
x\\ 
y\\
z\\
\end{bmatrix}
\end{equation}
Therefore, the velocity $(v)$ and linear momentum $(p)$ of a particle are given by: 
\begin{equation}
v=\frac{\mathrm{d} r}{\mathrm{d} t}=\dot{r}=\begin{bmatrix}
\dot{x}\\ 
\dot{y}\\
\dot{z}\\
\end{bmatrix}=\begin{bmatrix}
v_{x}\\ 
v_{y}\\
v_{z}\\
\end{bmatrix}
\end{equation}

\begin{equation}
p=mv=m\begin{bmatrix}
v_{x}\\ 
v_{y}\\
v_{z}\\
\end{bmatrix}
\end{equation}
where $m$ = mass of the particle.
Fig \ref{fig:pm} shows the inertial velocity of a point mass in polar coordinates, where $\gamma$ is the vertical flight path angle and $\xi$ is the horizontal flight path angle. Considering the motion to be a straight line motion in a particular direction, we can use $v_x$ to denote motion in the $xy$ horizontal plane. 2-dimensional equations for motion of a point mass, which coincides with center of mass of an aircraft, restricted to the vertical plane are given below:\\
\begin{equation}
\label{1}
 \begin{bmatrix}
\dot{x}\\ 
\dot{z}\\
\dot{v_x}\\
\dot{v_z} 
\end{bmatrix} = \begin{bmatrix}
v_x\\ 
v_z\\
f_x/m\\
f_z/m 
\end{bmatrix}
\end{equation}
Transforming velocity to polar coordinates:
\begin{equation}
 \begin{bmatrix}
x\\ 
z\\
\end{bmatrix} = \begin{bmatrix}
v_x\\ 
v_z\\
\end{bmatrix}= \begin{bmatrix}
v cos\gamma\\ 
-v sin\gamma\\
\end{bmatrix}\\
\implies \begin{bmatrix}
v\\ 
\gamma\\
\end{bmatrix}=\begin{bmatrix}
\sqrt{v_{x}^{2} + v_{z}^{2}}\\ 
-sin^{-1}(v_z/v)\\
\end{bmatrix}   
\end{equation}
Therefore, rates of change of velocity and flight path angle are given by:
\begin{equation}
    \begin{bmatrix}
\dot{v}\\ 
\dot{\gamma}\\
\end{bmatrix}=\begin{bmatrix}
\frac{\mathrm{d} }{\mathrm{d} t}\sqrt{v_{x}^{2} + v_{z}^{2}}\\ 
-\frac{\mathrm{d} }{\mathrm{d} t}sin^{-1}(v_z/v)\\
\end{bmatrix}
\end{equation}
Longitudinal equations of motion for a point mass are given by:
\begin{equation}
    \dot{x}(t)=v_x=v(t)cos\gamma(t)
\end{equation}
\begin{equation}
    \dot{z}(t)=v_z=-v(t)sin\gamma(t)
\end{equation}
\begin{equation}
\dot{v}(t)=\frac{(C_Tcos\alpha-C_D)\frac{1}{2}\rho(z)v^2(t)S-mg_0sin\gamma(t)}{m}    
\end{equation}
\begin{equation}
\dot{\gamma}(t)=\frac{(C_Tsin\alpha+C_L)\frac{1}{2}\rho(z)v^2(t)S-mg_0cos\gamma(t)}{mv(t)}    
\end{equation}
where $x$ is position projected on horizontal plane, $z$ is -height(altitude), $v$ is airspeed, $\gamma$ is flight path angle, $C_T$ is side force coefficient, $C_D$ is drag coefficient $\rho$ is density of air, $\alpha$ is the angle of attack, and $C_L$ is lift coefficient.\\
Lift and Drag are given by:
\begin{equation}
    L= C_L\frac{1}{2}\rho v_{air}^2S
\end{equation}
\begin{equation}
    D= C_D\frac{1}{2}\rho v_{air}^2S
\end{equation}
For a gliding flight, the condition of equilibrium is defined by the following equations:
\begin{equation}
    L=C_L\frac{1}{2} \rho v^2S=wcos\gamma
\end{equation}
\begin{equation}
    D=C_D\frac{1}{2} \rho v^2S=wsin\gamma
\end{equation}\\
where $h$ is the altitude vector, $S$ is the surface area of the wing and $w$ is the weight of the aircraft.\\
Therefore, the gliding flight path angle (Fig.~\ref{fig:forces}) can be found:\\
\begin{equation}
    \cot \gamma=\frac{L}{D}
    \label{cot1}
\end{equation}
From geometry, we have:
\begin{equation}
    \dot{x}=vcos\gamma
\end{equation}
\begin{equation}
    \dot{h}=vsin\gamma
\end{equation}
Therefore,
\begin{equation}
    \cot \gamma=\frac{\dot{x}}{\dot{h}}=\frac{\Delta x}{\Delta h}=g_0
    \label{cot2}
\end{equation}
where $g_0$ is the  \emph{glide ratio} for straight line glide at a constant airspeed.\\
Hence, from \ref{cot1} and \ref{cot2}, we can conclude that:
\begin{equation}
\label{m15}
g_{0}=\frac{L}{D}
\end{equation}
Glide range is maximum when $(L/D)$ is maximum (Fig.~\ref{fig:glide}).
\begin{equation}
    \left(\Delta x=\Delta h\cot \gamma\right) \to \Delta x_{max}\ \mbox{when}\ \frac{L}{D} \to \left(\frac{L}{D}\right)_{max}
\end{equation}
Corresponding airspeed is given by:
\begin{equation}
    v_{glide}=\sqrt{\frac{2w}{\rho S\sqrt{C_{D}^2+C_{L}^2}}}
\end{equation}
\\For banked turns, if the bank angle is $\theta$, the vertical component of lift, $L'=L\cos\theta$ (Fig.~\ref{fig:bank}). Hence the glide ratio $g_\theta$ is given by Eq. \ref{m16}, which forms the basis of our geometrical model of a gliding flight.
\begin{equation}
\label{m16}
\boxed{g_\theta=\frac{L'}{D}=\left(\frac{L}{D}\right)\cos\theta=g_{0} \cos \theta} 
\end{equation}

\section{Trajectory Planning Algorithm}
\begin{figure}[!htb]
  \centering
  \includegraphics[width=\textwidth]{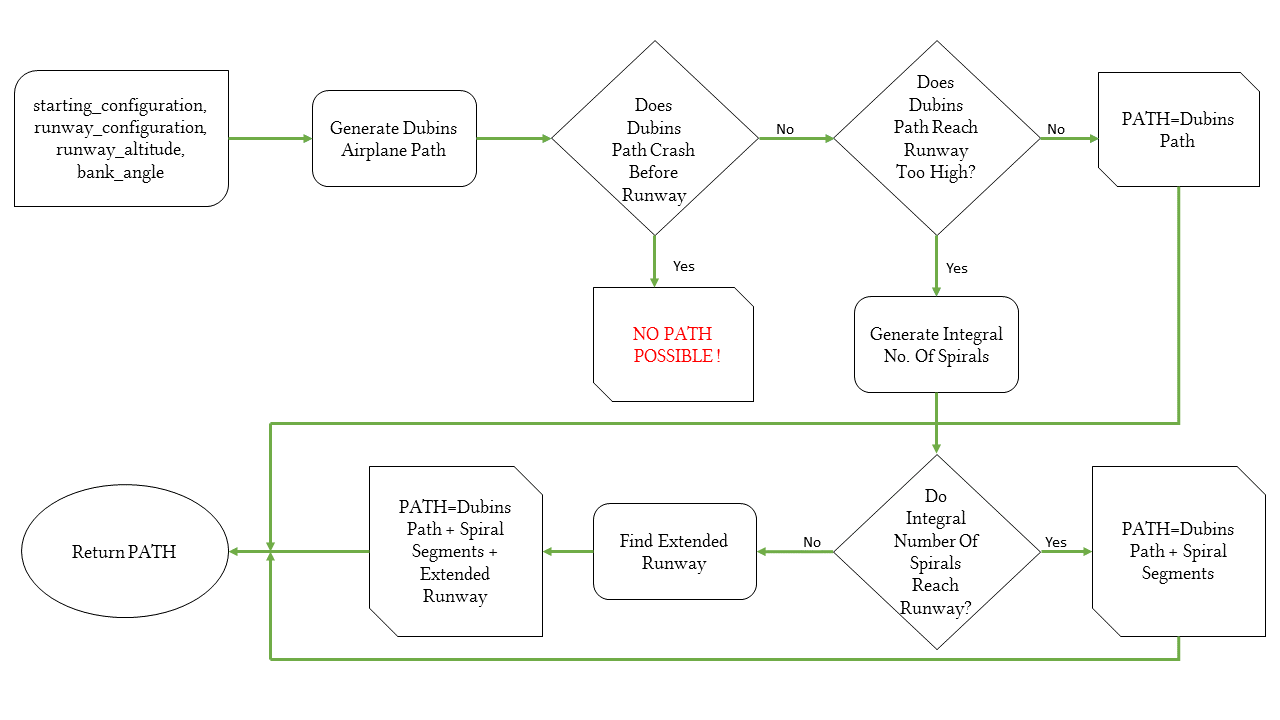}
  \caption{Flowchart of trajectory planning algorithm.}
  \label{fig:Flow}
\end{figure}

\
LOT emergencies can occur at any altitude. Depending on the altitude of the emergency, the type of possible trajectories to reachable runways varies. Owen \emph{et al}\cite{owen} describe three types of scenarios depending on the altitude of the aircraft with respect to the runway, namely, \emph{low altitude}, \emph{middle altitude} and \emph{high altitude} trajectories. We introduce the concept of an \textit{extended runway} when the altitude of the aircraft is too high for a simple Dubins airplane path to bring the aircraft to the runway and at the same time, the altitude of the aircraft is too low for spiralling down to the runway. Therefore, our trajectory planning algorithm (Fig.~\ref{fig:Flow}, Pseudocode \ref{pseudocode}) reduces the problem of finding trajectories to two scenarios: \textit{low altitude scenario} and \textit{high altitude scenario}.

\begin{lstfloat}
\begin{lstlisting}[language=Java, caption=Trajectory Planning Algorithm, label=pseudocode]
Trajectory_Generator(starting_configuration, runway_configuration, runway_altitude, bank_angle)
    dubins=Generate_Dubins(starting_configuration, runway_configuration, bank_angle)	
    dubins.end_altitude= altitude of last point of dubins 	
    If dubins.end_altitude < runway_altitude then		
        PATH = FAILURE 	
    End If	
    Else If dubins.end_altitude == runway_altitude then
        PATH = dubins 
    End If
    Else If dubins.end_altitude > runway_altitude then
         spiral=Generate_Spiral(dubins, end_altitude,runway_altitude)
         spiral.end_altitude= altitude of last point of spiral
         If spiral.end_altitude>runway_altitude then
             end_altitude=spiral.end_altitude
             distance=search_distance_parameter
             While end_altitude > runway_altitude do
                  new_point=Find_Point(distance) 
                  loss=loss of altitude in gliding from new_point to runway
                  extended_runway= new_point to runway
                  dubins_new=Generate_Dubins(starting_configuration, new_point, bank_angle)
                  spiral_new=Generate_Spiral(dubins_new, end_altitude,runway_altitude)
                  spiral_new.end_altitude= altitude of last point of spiral
                  end_altitude= spiral_new.end_altitude
                  If (end_altitude-loss) == runway_altitude then
                      PATH= dubins_new + spiral_new + extended_runway 
                      break
                  End If
                  Else
                      distance+=search_distance_parameter
              End While
         End If
         Else
             PATH= dubins + spiral
    End If
    Return PATH
End Trajectory_Generator
\end{lstlisting}
\end{lstfloat}
\subsection{Low Altitude Scenario}
\begin{figure}[!htb]
  \centering
  \begin{subfigure}{.5\textwidth}
  \centering
  \includegraphics[width=.8\linewidth]{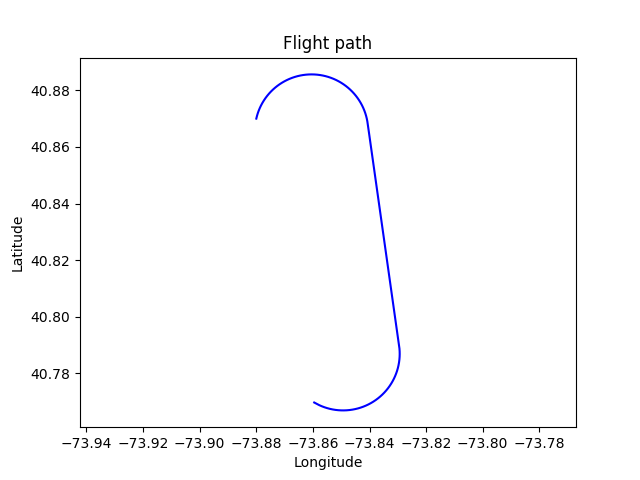}
  \caption{2D view.}
  \label{}
\end{subfigure}%
\begin{subfigure}{.5\textwidth}
  \centering
  \includegraphics[width=.8\linewidth]{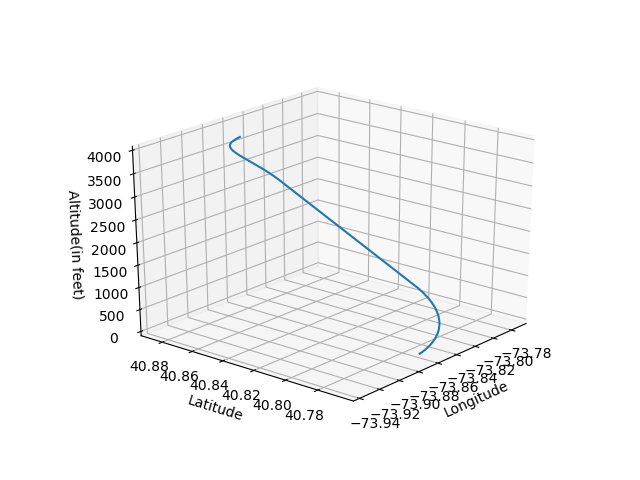}
  \caption{3D view}
  \label{}
\end{subfigure}
\caption{A low altitude trajectory.}
\label{fig:2D view of a low altitude trajectory.}
\end{figure}

We define a low altitude scenario as a case in which a simple 3D Dubins airplane path can bring the aircraft to the runway at the correct altitude. In this type of scenario, the generated trajectory consists of a Dubins airplane path from the starting point to the runway (Fig. \ref{fig:2D view of a low altitude trajectory.}).

\subsection{High Altitude Scenario}
\begin{figure}[!htb]
  \centering
  \begin{subfigure}{.5\textwidth}
  \centering
  \includegraphics[width=.8\linewidth]{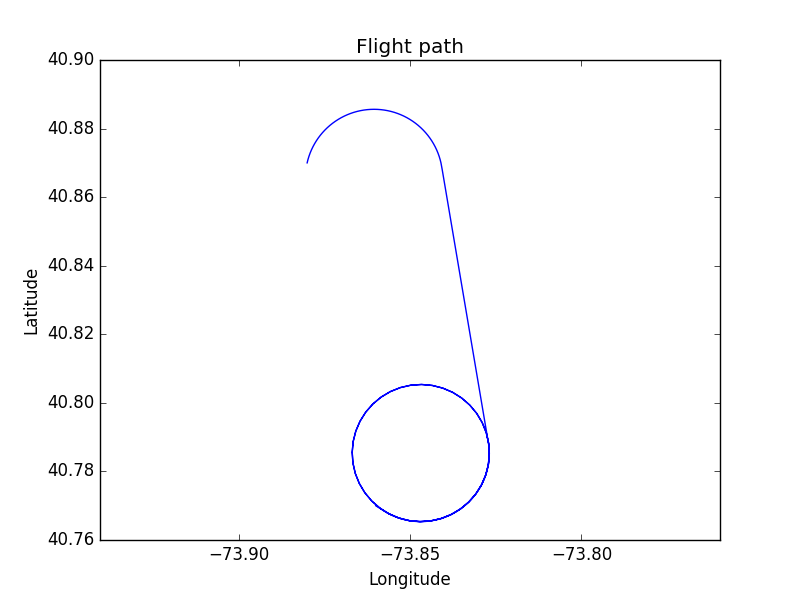}
  \caption{2D view.}
  \label{}
\end{subfigure}%
\begin{subfigure}{.5\textwidth}
  \centering
  \includegraphics[width=.8\linewidth]{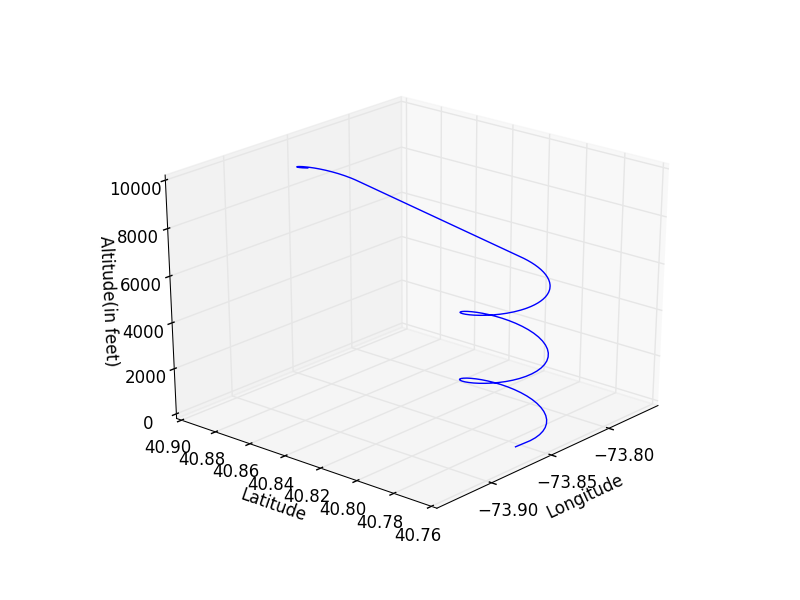}
  \caption{3D view}
  \label{}
\end{subfigure}
\caption{A high altitude trajectory.}
\label{fig:high2d1}
\end{figure}
Our model defines a high altitude scenario as any case in which a simple Dubins airplane path cannot bring the aircraft to the runway at the correct altitude, but brings it over the runway with an \textit{excess altitude}. In such a scenario, the excess altitude needs to be lost in the trajectory so that the pilot can safely land the aircraft. The value of the excess altitude has an impact on the type of trajectories that can be generated. There are two distinct types of cases that fall under the high altitude scenario:
\begin{itemize}
  \item The excess altitude is \emph{enough} for generating an integral number of spiral turns to bring the aircraft to the runway.
  \item The excess altitude is \emph{not enough} for generating an integral number of spiral turns to bring the aircraft to the runway.
\end{itemize}
In the first case, when the excess altitude allows for generating an integral number of spiral turns, the trajectory consists of a Dubins airplane path, followed by an integral number of spiral turns (Fig. \ref{fig:high2d1}).
\begin{figure}[!htb]
  \centering
  \begin{subfigure}{.5\textwidth}
  \centering
  \includegraphics[width=.8\linewidth]{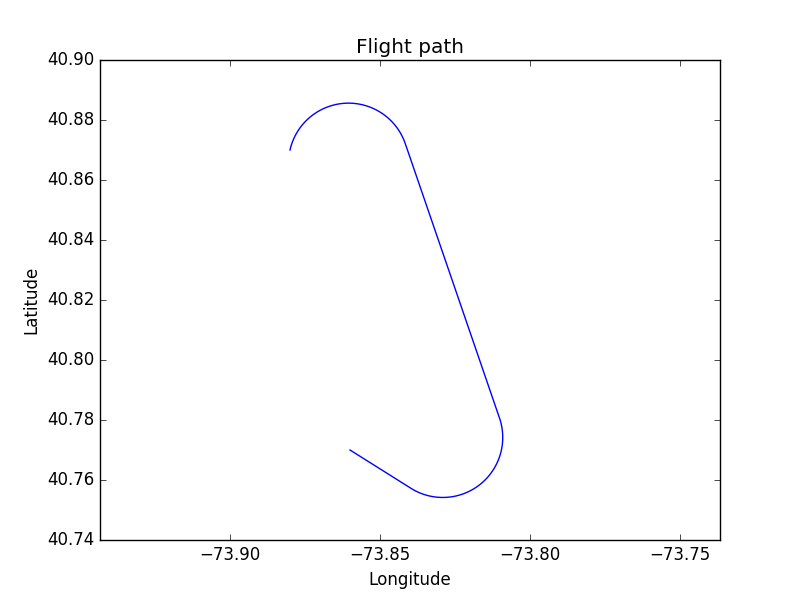}
  \caption{2D view.}
  \label{}
\end{subfigure}%
\begin{subfigure}{.5\textwidth}
  \centering
  \includegraphics[width=.8\linewidth]{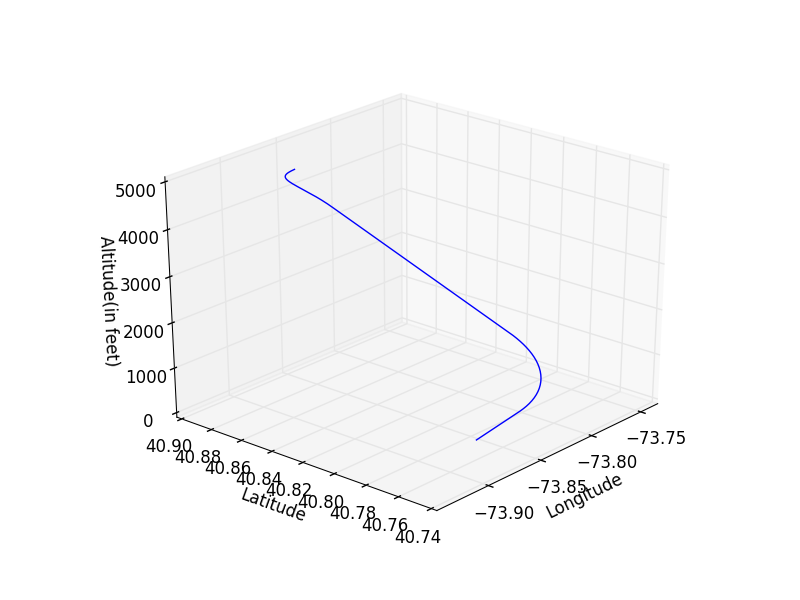}
  \caption{3D view}
  \label{}
\end{subfigure}
\caption{A middle altitude trajectory.}
\label{fig:middle2d}
\end{figure}
In the second case, when the excess altitude does not allow for generating an  integral number of spirals, our algorithm extends the final approach along the runway heading by adding an extended runway of length \textit{e}, where \textit{e} $ \in\left [  0,2\pi R g_{d,0}/g_{c,\theta}\right ) $ where $R$ is the radius of turn for the bank angle $\theta$ being used, $g_{d,0}$ is the \emph{dirty configuration glide ratio} for straight line gliding and $g_{c,\theta}$ is the clean configuration glide ratio for banked turns. The addition of this extended straight line approach in the final path allows the aircraft to lose the excess altitude before reaching the runway (Fig. \ref{fig:middle2d}). In our experiments, we used a \textit{search\_distance\_parameter} of 50 feet (Pseudocode~\ref{pseudocode}) while calculating the extended final segment. 
\begin{figure}[!htb]
  \centering
  \begin{subfigure}{.5\textwidth}
  \centering
  \includegraphics[width=.8\linewidth]{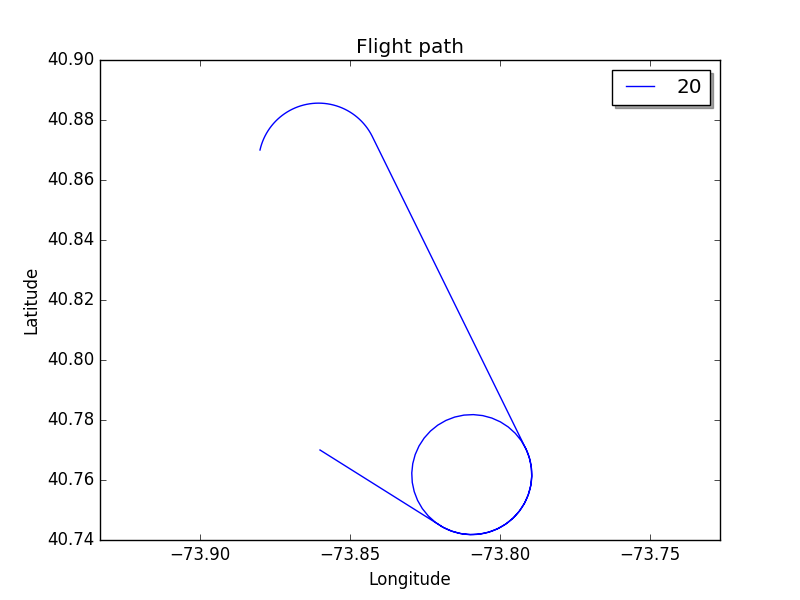}
  \caption{2D view.}
  \label{}
\end{subfigure}%
\begin{subfigure}{.5\textwidth}
  \centering
  \includegraphics[width=.8\linewidth]{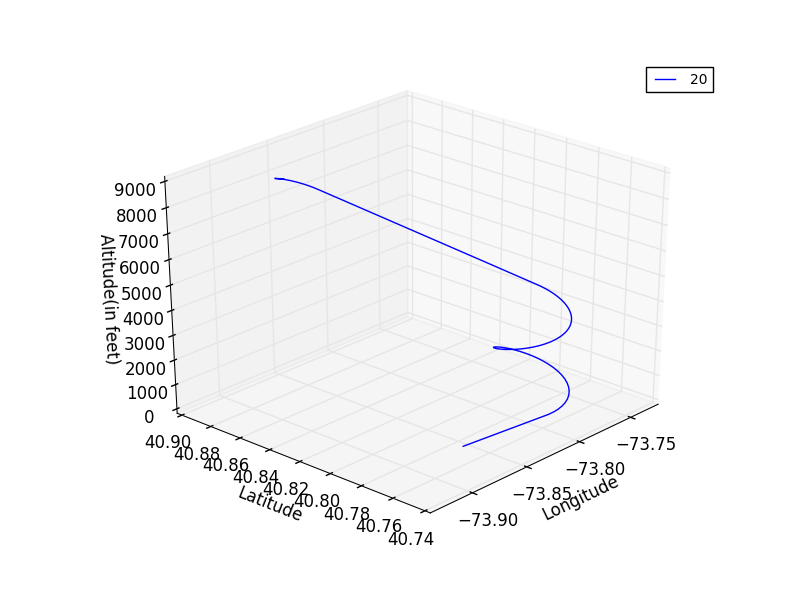}
  \caption{3D view}
  \label{}
\end{subfigure}
\caption{A high altitude trajectory with spiral segments and an extended final approach.}
\label{fig:higher2d}
\end{figure}

In certain cases, the value of the excess altitude might allow generation of integral number of spirals but not enough spirals to lose the excess altitude in entirety. Such special cases call for the need of using a combination of both spiral turns and an extended final straight line approach. Thus, in such scenarios, our trajectory planning algorithm generates trajectories that consist of a Dubins airplane path, followed by an integral number of spiral turns, followed by an extended final approach (Fig.~\ref{fig:higher2d}).

\begin{figure}[!htb]
\centering
\begin{subfigure}{.5\textwidth}
  \centering
  \includegraphics[width=\linewidth]{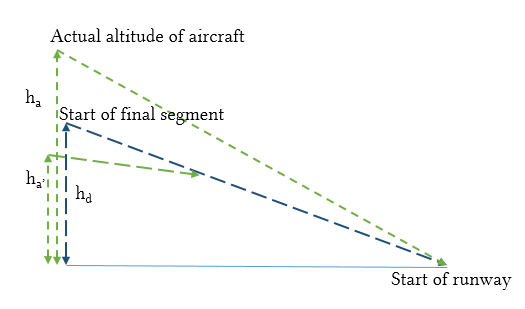}
  \caption{Using a dirty configuration glide ratio in final segment.}
  \label{difa}
\end{subfigure}%
\begin{subfigure}{.5\textwidth}
  \centering
  \includegraphics[width=\linewidth]{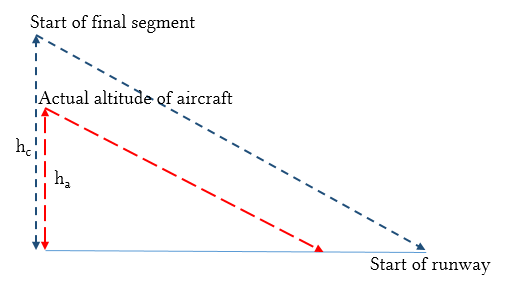}
  \caption{Using a clean configuration glide ratio in final segment.}
  \label{difb}
\end{subfigure}
\caption{Significance of using the dirty configuration glide ratio in final segment.}
\label{dif}
\end{figure}

We use the dirty configuration glide ratio in the extended final segments in order to ensure that the starting point of the extended final segment is at an altitude $h_d$ from which the flight can safely make it to the runway. The aircraft can reach the start of the final segment at an altitude $h_a\geq h_d$ and in case the aircraft reaches the starting point of the final segment at an altitude $h_a>h_d$, then the pilot will have the flexibility to lose the excess altitude $h_a-h_d$ by increasing drag by performing 'S' turns, or by increasing sideslip and forward slip, and still making it to the runway (Fig.~\ref{difa}). On the other hand, if the aircraft reaches the starting point of the final segment at an altitude $h_{a'}<h_d$, then the pilot can keep using clean configuration until it is safe to increase drag (Fig.~\ref{difa}) and make a successful landing. However, if we generate trajectories by using the clean configuration glide ratio in the final segment, then the final segment will start at an altitude $h_c$ that is too optimistic and in case the aircraft reaches the start of the final segment at an altitude $h_a<h_c$, then the aircraft will crash before reaching the runway (Fig.~\ref{difb}).

However, it should be noted that our current algorithm fails to return trajectories in certain cases, for example, when the aircraft is directly above the runway. It can generate trajectories \emph{only} for those cases where the 2D Dubins path from initial to final configurations consists of a curve, followed by a straight line segment, followed by a curve.

Our trajectory generation software generates space discretized trajectories of the form $\mathbb{T}:\left[r_0,r_1,r_2,r_3,....,r_N\right]$, where $r_i=\begin{bmatrix}
x_i\\ 
y_i\\ 
z_i
\end{bmatrix}$ and $x_i$, $y_i$ and $z_i$ define the position of a point $i$ with respect to a 3 dimensional inertial frame.
\section{Trajectory Safety Metrics}
We introduce a set of \emph{trajectory safety metrics} to evaluate our trajectories. Each trajectory $\mathbb{T}$ generates a value for each metric and each of these values are then normalized relative to the minimum (Eq. ~\ref{minnorm}) or maximum (Eq. ~\ref{maxnorm}) value (whichever is desired).  
\begin{equation}
\label{minnorm}
 \left \| x \right \|=\frac{x-x_{max}}{x_{min}-x_{max}}
\end{equation}
\begin{equation}
\label{maxnorm}
 \left \| x \right \|=\frac{x-x_{min}}{x_{max}-x_{min}}
\end{equation}
The safety metrics that have been considered for this paper are:
\begin{itemize}
\item \emph{Average Altitude} ($\bar{z}$)- This metric computes the average of the altitude ($z$) for all $N$ points in $\mathbb{T}$. This metric is normalized against the maximum value because for a given trajectory, a higher average altitude from the ground is desirable.
\begin{equation}
\label{m1}
\bar{z} = \frac{\sum_{i=0}^{N}z_i}{N}
\end{equation}
\item \emph{Average Distance From Runway} ($\bar{d}$)- This metric computes the average of the distance ($d$) from the runway for all $N$ points in $\mathbb{T}$. This metric is normalized against the minimum value because for a given trajectory, minimum distance from the runway at all time is desirable.
\begin{equation}
\label{m2}
\bar{d} = \frac{\sum_{i=0}^{N}d(r_i,r_{R})}{N}
\end{equation}
where $r_R$ is the position vector of the runway and $d(r_1,r_2)$ is given by Eq.~\ref{ex}.
\begin{equation}
\label{ex}
d(r_1,r_2)=\sqrt{(x_1-x_2)^2+(y_1-y_2)^2+(z_1-z_2)^2}
\end{equation}
\item \emph{Average Bank Angle Over Height} ($\bar{\left(\frac{\theta}{z}\right)}$)- This metric measures the occurrence of steep turns near the ground. This is computed by taking an average of the ratio between bank angle ($\theta$) and altitude ($z$) for all $N$ points in $\mathbb{T}$. Since it is not desirable to have steep turns very close to the ground, this metric is normalized against the minimum value.
\begin{equation}
\label{m3}
\bar{\left(\frac{\theta}{z}\right)} = \frac{\sum_{i=0}^{N}\left(\frac{\theta_i}{z_i}\right)}{N}
\end{equation}
\item \emph{Number of turns} ($n$)- This metric counts the number of turns ($n$) in $\mathbb{T}$. It is desirable to have as minimum number of turns as possible, hence it is normalized against the minimum value.
\begin{equation}
\label{m4}
n  = \mbox{Number\ of\ turns\ in\ Dubins\ airplane\ path\ +\ number\ of\ 360\degree\ turns\ in\ spiral\ segment}
\end{equation}
\item \emph{Total length} ($l$)- This metric measures the total length ($l$) of the trajectory $\mathbb{T}$. It is normalized against the minimum as shorter trajectories are more desirable than longer ones.
\begin{equation}
\label{m5}
l = \sum_{i=1}^{N} d(r_i,r_{i-1})
\end{equation}
\item \emph{Extended final runway distance} ($e$)- This metric measures the length ($e$) of the extended final straight line approach. Trajectories with longer final straight line approaches are desirable as a longer final approach allows for the pilot to make adjustments to drag and speed easily just before landing. Thus, this metric is normalized against the maximum value. 
\begin{equation}
\label{m5}
e = d(r_e,r_R)
\end{equation}
where $r_e$ is the position vector of the starting point of the extended final segment and $r_R$ is the position vector of the runway.
\end{itemize}
 We introduce an \emph{utility function} ($u$), which is computed by taking the average of the normalized values of the above metrics, and used to rank the trajectories. The higher the value of the utility function, the better is the rank of the trajectory.
\begin{equation}
\label{m7}
\boxed{u = \frac{\left \| \bar{z} \right \| + \left \| \bar{d} \right \|+ \left \| \bar{\left(\frac{\theta}{h}\right)}\right \| +\left \| n\right \| +\left \| l \right \|+\left \| e\right \| }{6}}
\end{equation} 
This utility function can be easily modified to account for other safety factors like wind, terrain; \emph{etc} in future work.

 \section{Dynamic Data Driven Flight Trajectory Generation}
\begin{figure}[H]
  \centering
  \includegraphics[width=0.8\textwidth]{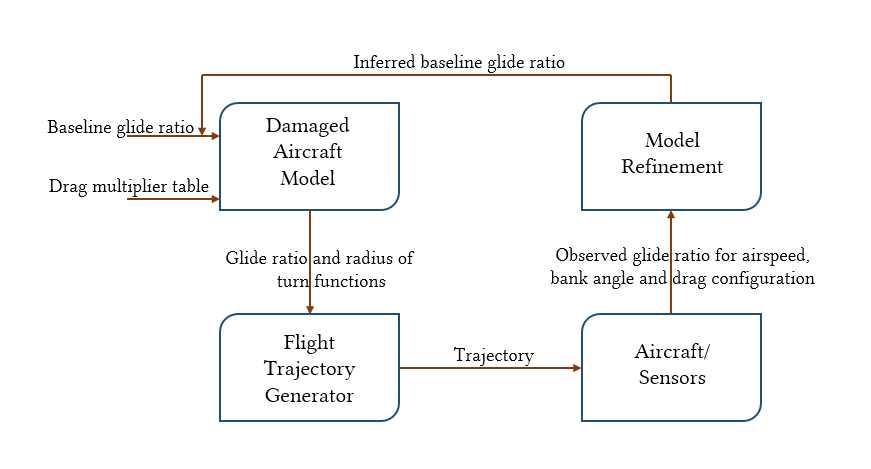}
  \caption{Dynamic data driven flight trajectory generation.}
  \label{fig:model1}
\end{figure}

\begin{figure}
  \centering
  \includegraphics[width=0.75\textwidth]{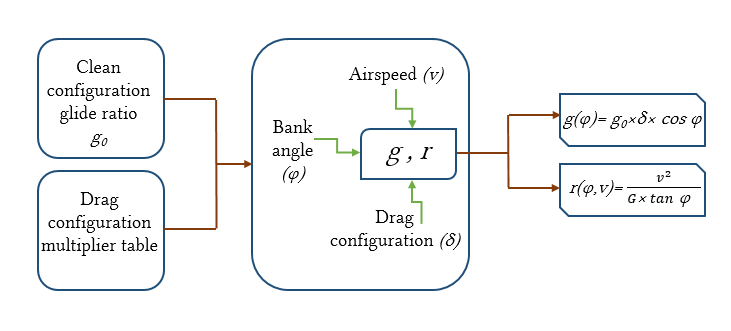}
  \caption{Damaged aircraft model.}
  \label{fig:model2}
\end{figure}

\begin{figure}
  \centering
  \includegraphics[width=0.8\textwidth]{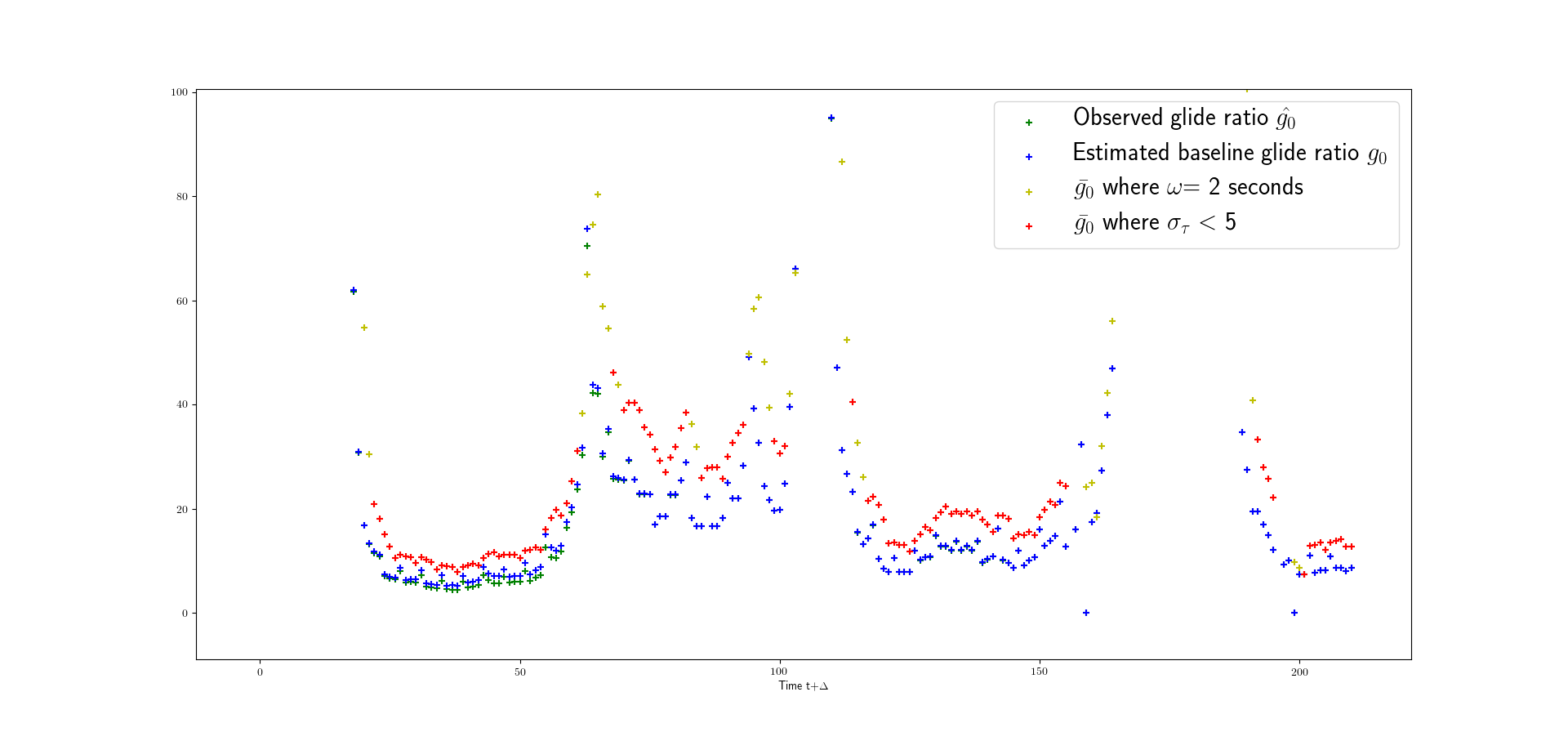}
  \caption{Effect of too short size of $\omega$.}
  \label{dur1}
\end{figure}

\begin{figure}
  \centering
  \includegraphics[width=0.8\textwidth]{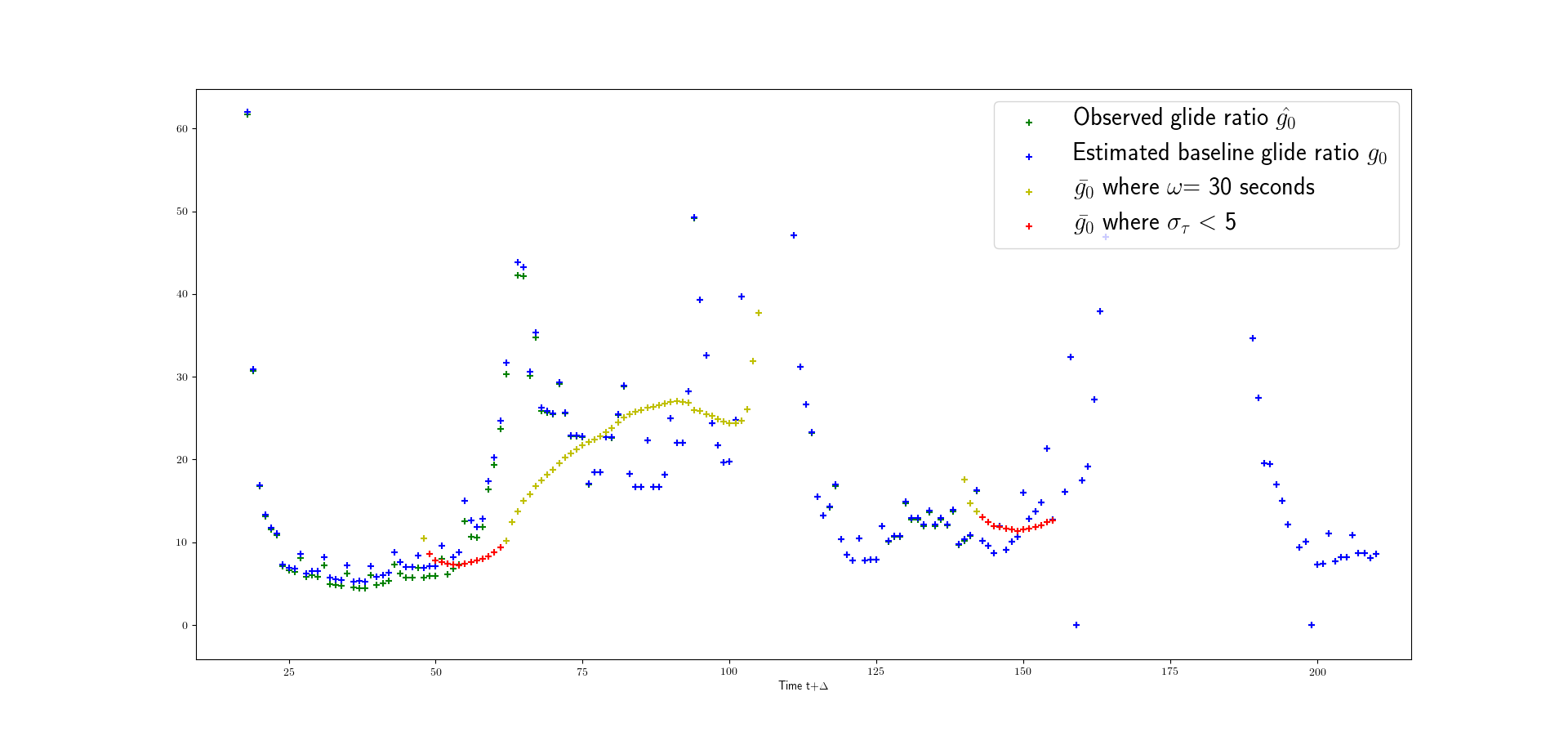}
  \caption{Effect of too large size of $\omega$.}
  \label{dur2}
\end{figure}

\begin{figure}
  \centering
  \includegraphics[width=0.8\textwidth]{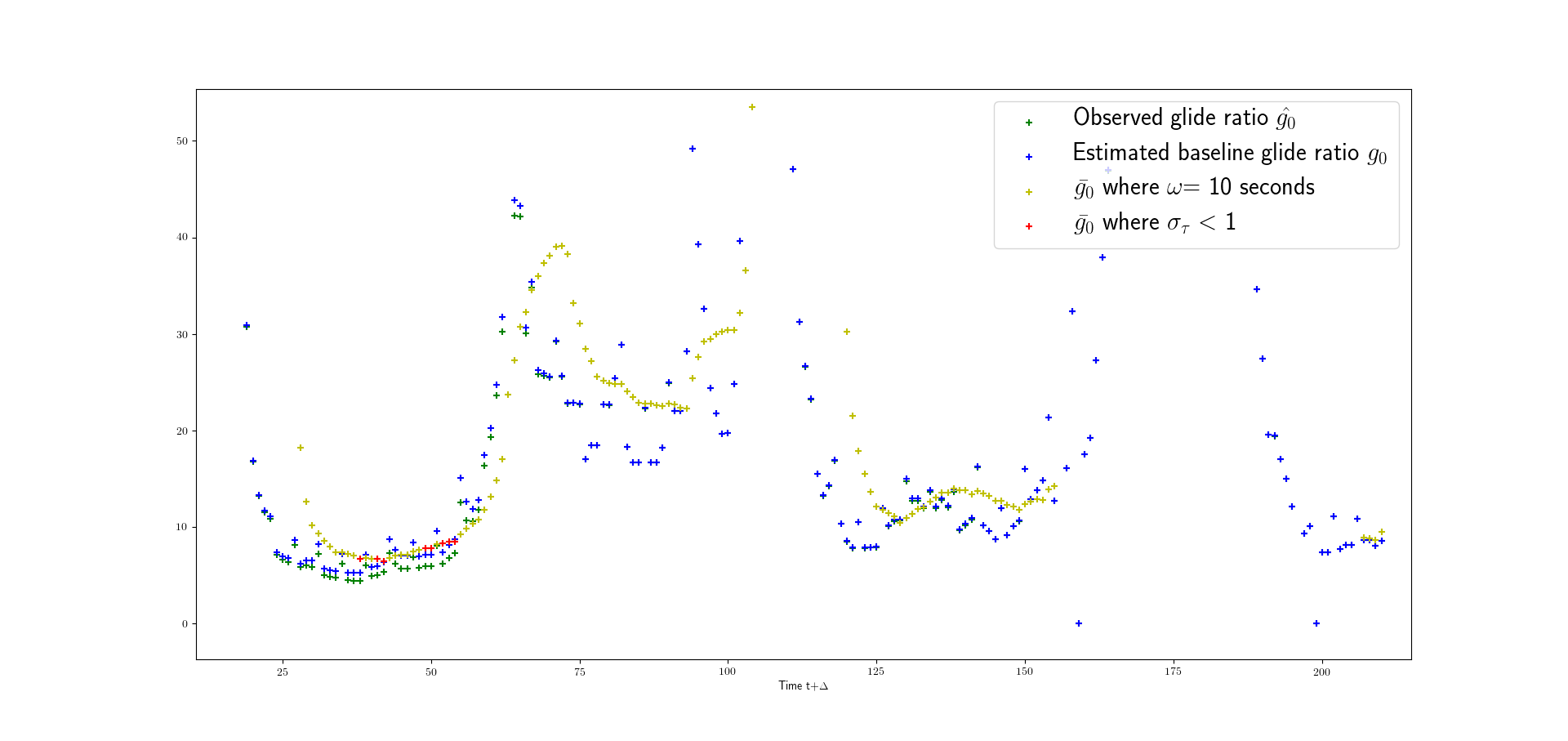}
  \caption{Effect of too small value of $\sigma_\tau$.}
  \label{dev1}
\end{figure}

\begin{figure}
  \centering
  \includegraphics[width=0.8\textwidth]{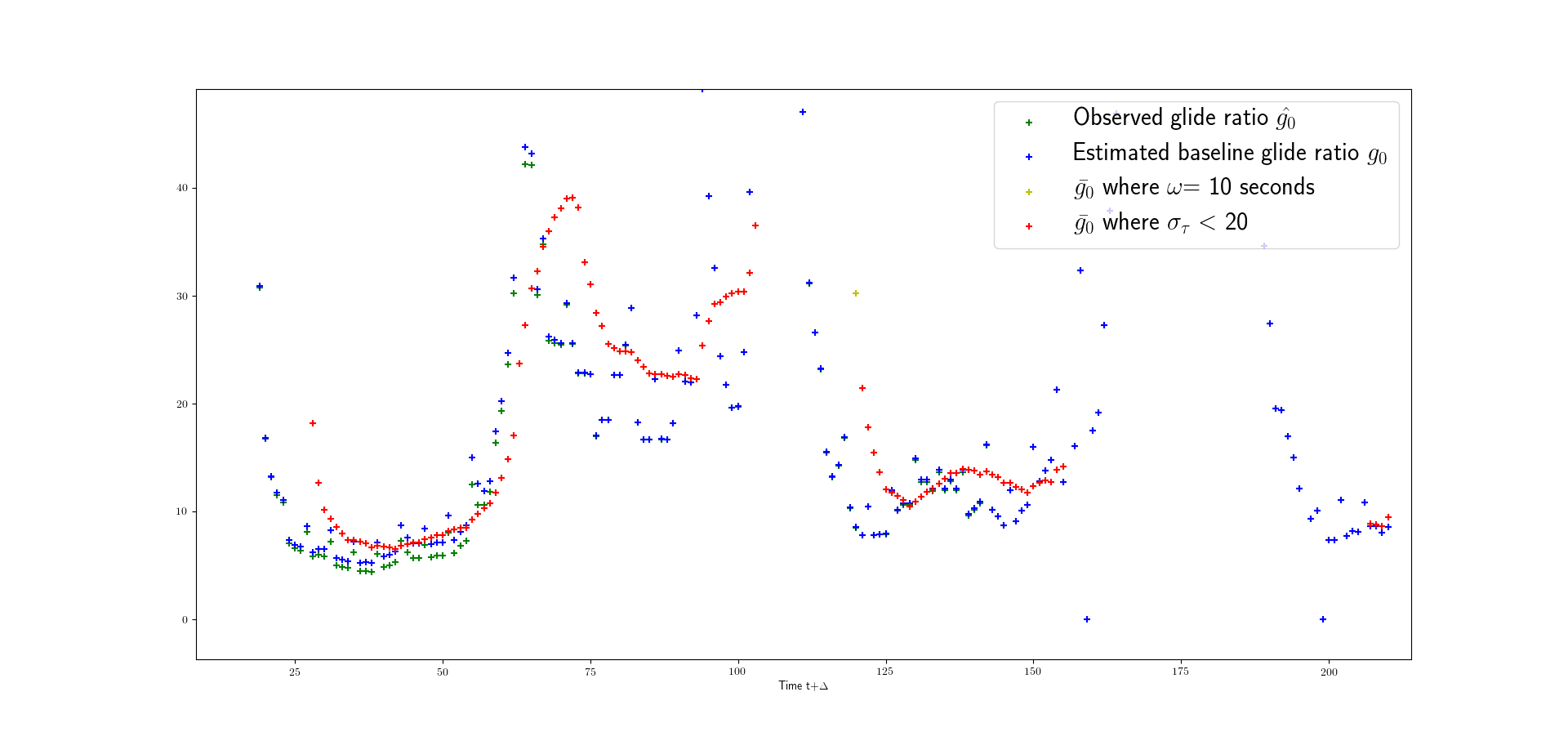}
  \caption{Effect of too large value of $\sigma_\tau$.}
  \label{dev2}
\end{figure}

In order to generate trajectories that are as faithful to the current performance of the aircraft as possible, after discrete time intervals, we take data from the aircraft sensors and estimate the correct \emph{baseline glide ratio} (the glide ratio for a clean configuration with best gliding airspeed in a straight flight), $g_{0}$. A flowchart of this approach is given in Fig.~\ref{fig:model1}. The DDDAS approach has four components: the \emph{damaged aircraft model}, the \emph{flight trajectory generator}, \emph{aircraft/sensors} and the \emph{model refinement} component.
The \emph{damaged aircraft model} (Fig.~\ref{fig:model2}) takes as inputs the new baseline glide ratio ($g_{0}$) and a \emph{drag multiplier table} to compute the glide ratio (\(g(\phi)\)) for every bank angle (\(\phi\)) and the corresponding radius of turn (\(r(\phi,v)\)) for a given bank angle ($\phi$) and airspeed ($v$). The drag multiplier table contains the ratios (\(\delta\)) that can be used to obtain the baseline glide ratio ($g_{0}(c)$) for a drag configuration \emph{c} from the baseline glide ratio ($g_{0}$) of a clean configuration. 
\begin{equation}
\label{m8}
g_{0}(c)=\delta \times g_{0}
\end{equation}
Given the baseline glide ratio ($g_{0}$) and drag configuration ratio (\(\delta\)), the glide ratio for a bank angle (\(\phi\)) can be obtained from Eq. \ref{m9}. 
\begin{equation}
\label{m9}
g(\phi)=g_{0}\times \delta \times \cos\phi
\end{equation}
Given the \emph{airspeed} ($v$) and \emph{gravitational constant} ($G=11.29\ \mbox{kn}^2\mbox{ft}^{-1}$), the radius of turn (\(r(\phi,v)\)) for a bank angle ($\phi$) and airspeed ($v$) can be obtained from Eq. \ref{m10} .
\begin{equation}
\label{m10}
r(\phi,v)=\frac{v^{2}}{G\times\tan \phi}
\end{equation}
The functions for calculating the glide ratio (\(g(\phi)\)) for every bank angle (\(\phi\)) and the corresponding radius of turn (\(r(\phi,v)\)) are sent from the damaged aircraft model to the flight trajectory generator to be used in the trajectory planning algorithm for computing trajectories.\\
\begin{figure}
  \centering
  \includegraphics[width=0.8\textwidth]{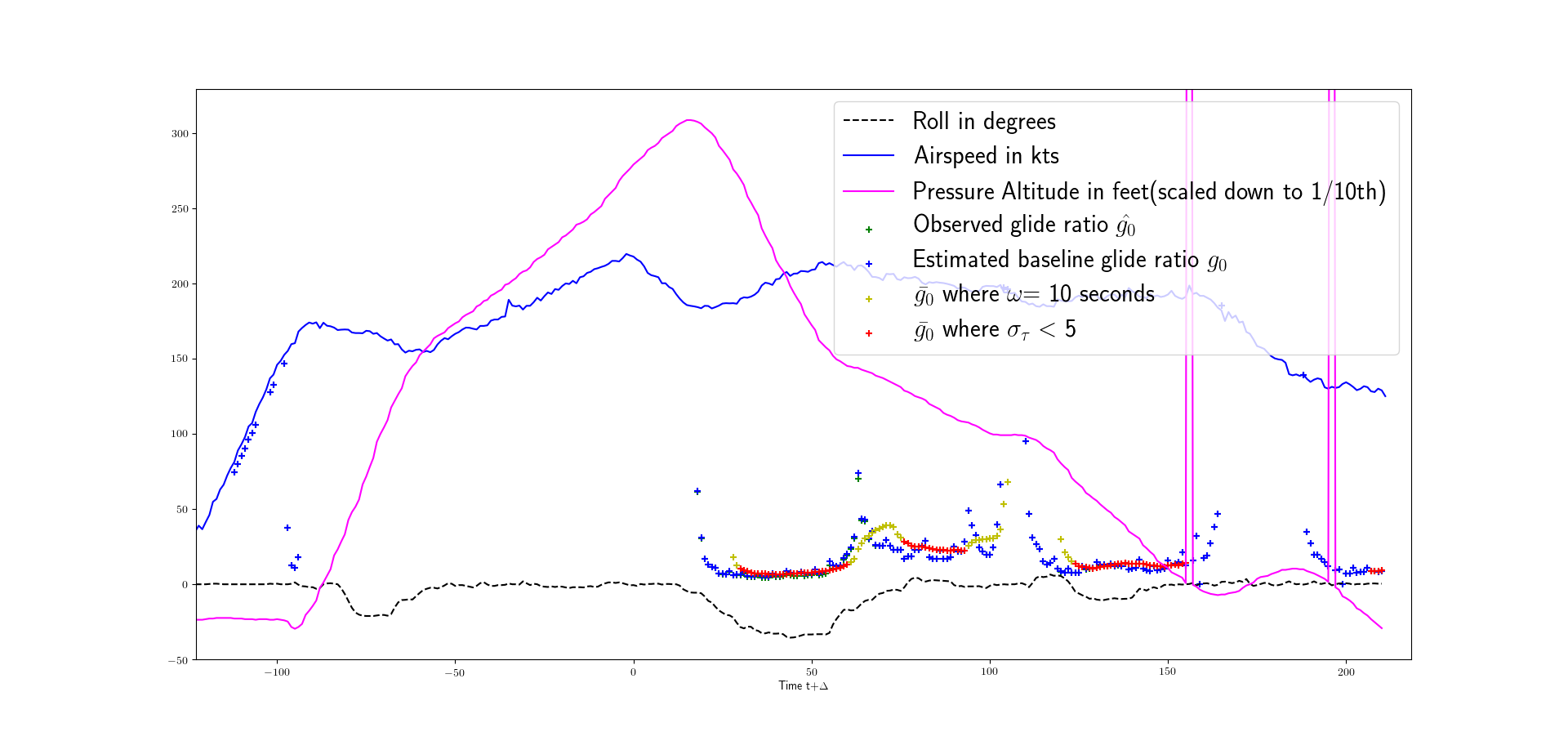}
  \caption{Estimation of glide ratio from sensors (FDR data of US Airways 1549).}
  \label{egr}
\end{figure}
We constantly read data from the aircraft sensors to estimate the observed glide ratio $\hat{g_{\theta}}$. This is done every second by using the pressure altitude ($A_p$) and the airspeed ($v$) that are available from the sensor data. For a given instant $t_i$, we can compute the observed glide ratio $\hat{g_{\theta}}_{t_i}$ by taking a ratio of the horizontal distance travelled to the altitude lost in the preceding $\eta$ seconds ~(Eq.~\ref{egr1}). This is followed by two steps: first, it is checked if this instant $t_i$ was preceded by a window $\omega$ of steady descent with no rise in altitude; then it is checked if the distribution of $\hat{g_{\theta}}_{t_i}$ in the period $\omega$ had a standard deviation within a threshold $\sigma_\tau$. This is done to detect \emph{stable windows} of flight in order to make sure that aberrations and large deviations in the glide ratio are not taken into account while calculating new trajectories.
\begin{equation}
\hat{g_{\theta}}_{t_i}= \frac{\mbox{Horizontal distance covered in preceding $\eta$ seconds}}{\mbox{Loss in altitude in preceding $\eta$ seconds}} \\ = \frac{\sum_{i-3}^{i}v_{t_i}}{A_{p_{t_{i-3}}}-A_{p_{t_i}}} 
\label{egr1}
\end{equation}
If the window $\omega$ preceding an instant $t_i$ is a stable window, we compute the observed glide ratio $\hat{g_{\theta}}$ for the observed bank angle $\theta$ by taking a mean of the glide ratios in $\omega$. 
\begin{equation}
\hat{g_{\theta}}= \frac{\sum_{i=1}^{n}\hat{g_{\theta}}_{t_i}}{n}\ \mbox{for all}\ t_i\in \omega
\label{egr2}
\end{equation}
In our experiments, we take $\eta=4$, $\omega$ of duration 10 seconds and $\sigma_\tau=5$ (Fig.~\ref{egr}). The choice of these values have a major impact on the estimation of $\hat{g_{\theta}}$ because too short $\omega$ does not have enough data to give a proper estimate of $\hat{g_{\theta}}$ while too long $\omega$ makes it difficult to find a proper estimate of $\hat{g_{\theta}}$ (Fig.~\ref{dur1}, Fig.~\ref{dur2}) and too small value of $\sigma_\tau$ makes the conditions too strict to find suitable values of $\hat{g_{\theta}}$ while too large value of $\sigma_\tau$ makes it difficult to filter out noisy values of $\hat{g_{\theta}}$ (Fig.~\ref{dev1}, Fig.~\ref{dev2}).

The observed glide ratio for airspeed, bank angle and drag configuration is then sent to the \emph{model refinement} component along with the bank angle and drag configuration.

When it receives the data from the sensors, the model refinement component calculates the new baseline glide ratio ${g_0}^{'}$ using the observed bank angle $\theta$, drag configuration $\delta$ from the sensor data and the observed glide ratio $\hat{g_{\theta}}$ for that bank angle.\\ 
\begin{equation}
\hat{g_\theta}={g_0}^{'}\delta\cos\theta
\implies \boxed{{g_0}^{'}=\frac{\hat{g_\theta}}{\delta\cos\theta}}
\end{equation}
This is done assuming that the aircraft is maintaining the best gliding airspeed for clean configuration.\\
\begin{figure}[!htb]
\centering
\begin{subfigure}{.5\textwidth}
  \centering
  \includegraphics[width=.8\linewidth]{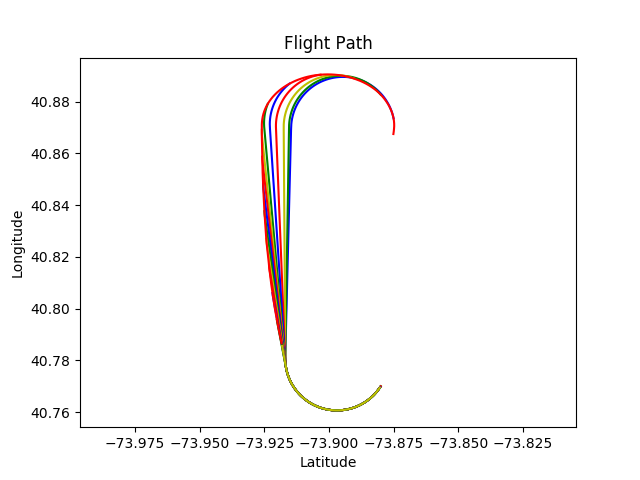}
  \caption{2D View.}
  \label{fig:sub1}
\end{subfigure}%
\begin{subfigure}{.5\textwidth}
  \centering
  \includegraphics[width=.8\linewidth]{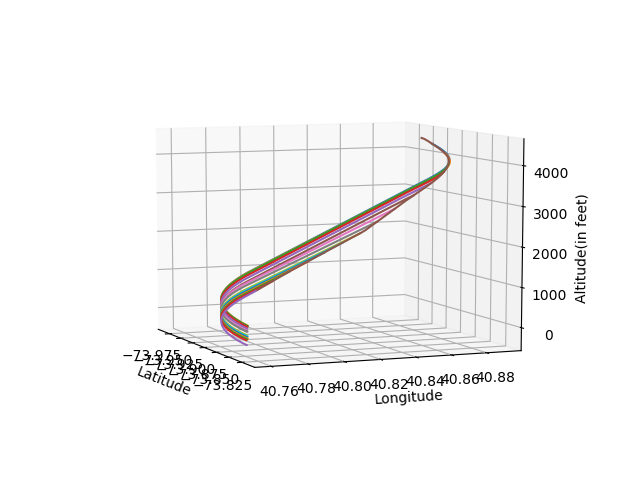}
  \caption{3D View.}
  \label{fig:sub2}
\end{subfigure}
\caption{Trajectories to LGA4 from an altitude of 4551 feet. New trajectories computed using data from flight simulator.}
\label{fig:dsim2}
\end{figure}

Fig.~\ref{fig:dsim2} shows dynamic data driven approach being used to generate trajectories to LGA4 from an altitude of 4451 feet. The image clearly shows that correcting the baseline glide ratio from flight data can have major impact on new trajectories that are generated. It allows us to account for dynamic factors such as partial power, wind and effects of surface damage and compute a trajectory corresponding to the current performance capabilities.
\FloatBarrier
 \section{Experimentation and Results}
 \subsection{Simulations With La Guardia Airport}
\begin{figure}[!htb]
\centering
\begin{subfigure}{.5\textwidth}
  \centering
  \includegraphics[width=.8\linewidth]{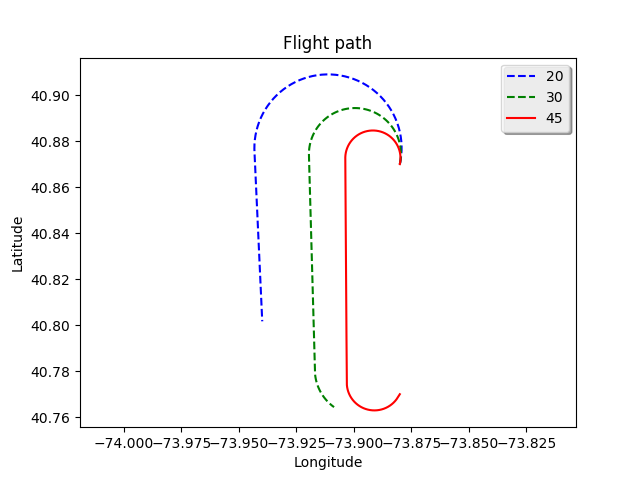}
  \caption{2D View.}
  \label{fig:sub1}
\end{subfigure}%
\begin{subfigure}{.5\textwidth}
  \centering
  \includegraphics[width=.8\linewidth]{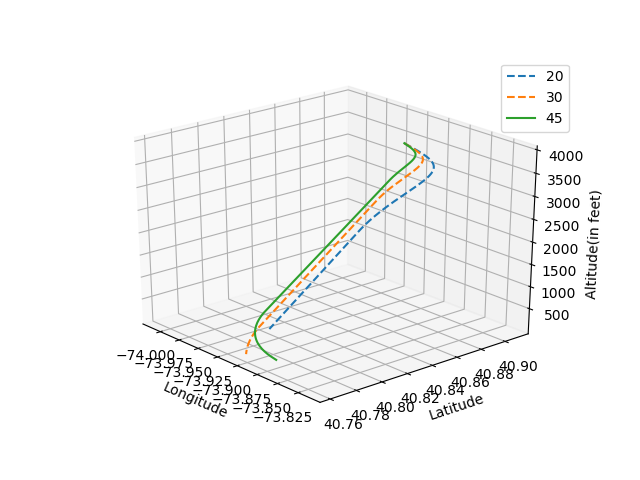}
  \caption{3D View.}
  \label{fig:sub2}
\end{subfigure}
\caption{Trajectory to LGA4 from 4000 feet.}
\label{fig:22t41}
\centering
\begin{subfigure}{.5\textwidth}
  \centering
  \includegraphics[width=.8\linewidth]{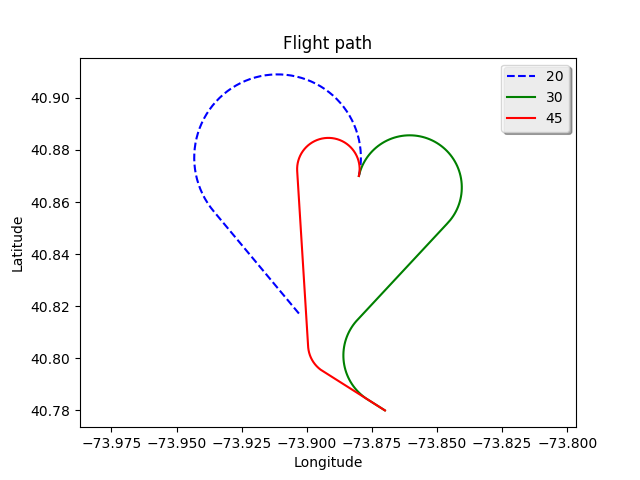}
  \caption{2D View.}
  \label{fig:sub1}
\end{subfigure}%
\begin{subfigure}{.5\textwidth}
  \centering
  \includegraphics[width=.8\linewidth]{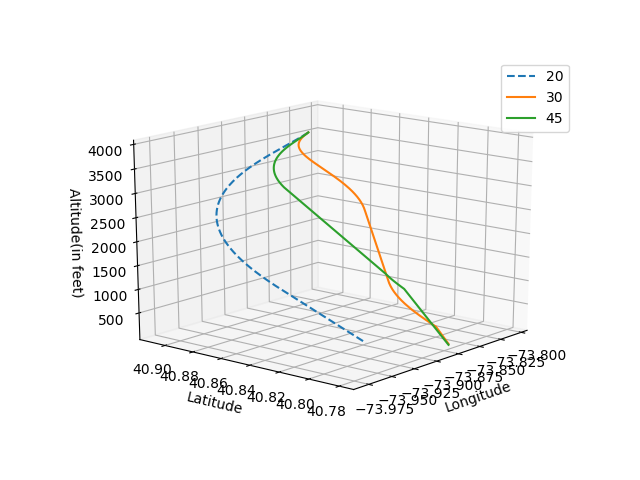}
  \caption{3D View.}
  \label{fig:sub2}
\end{subfigure}
\caption{Trajectory to LGA13 from 4000 feet.}
\label{fig:22t44}
\centering
\begin{subfigure}{.5\textwidth}
  \centering
  \includegraphics[width=.8\linewidth]{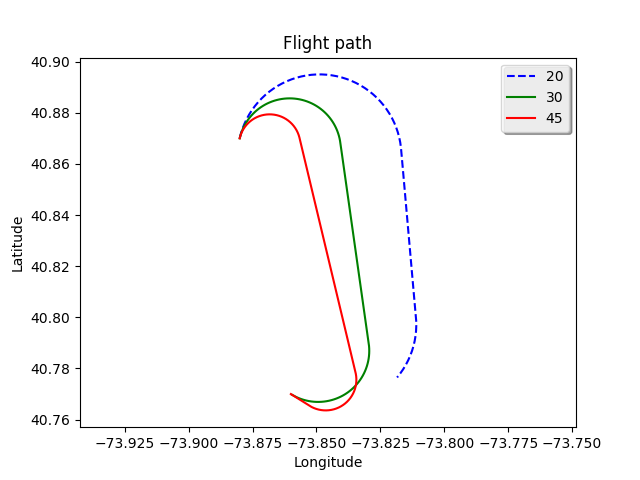}
  \caption{2D View.}
  \label{fig:sub1}
\end{subfigure}%
\begin{subfigure}{.5\textwidth}
  \centering
  \includegraphics[width=.8\linewidth]{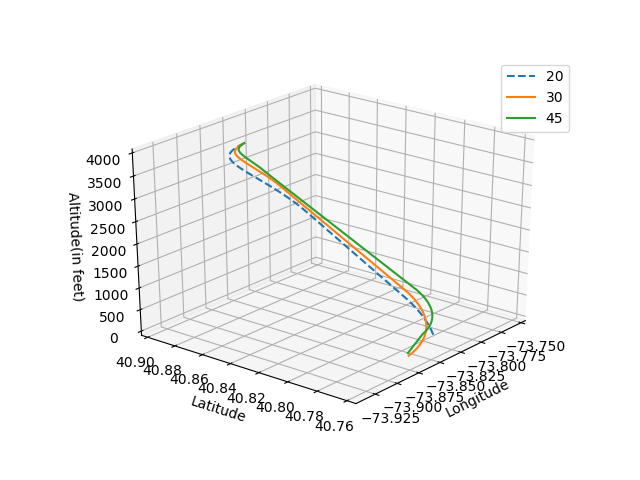}
  \caption{3D View.}
  \label{fig:sub2}
\end{subfigure}
\caption{Trajectory to LGA31 from 4000 feet.}
\label{fig:22t46}
\end{figure}

\begin{figure}[!htb]
\centering
\begin{subfigure}{.5\textwidth}
  \centering
  \includegraphics[width=.8\linewidth]{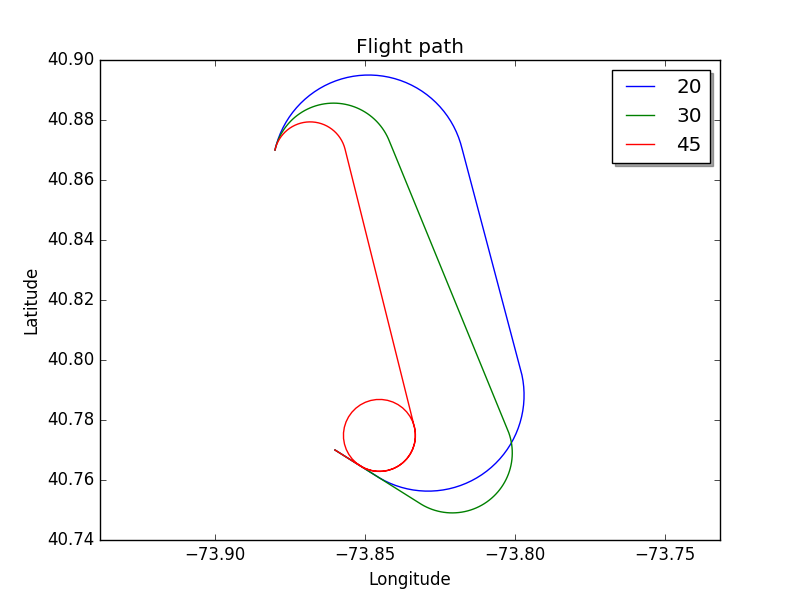}
  \caption{2D View.}
  \label{fig:sub1}
\end{subfigure}%
\begin{subfigure}{.5\textwidth}
  \centering
  \includegraphics[width=.8\linewidth]{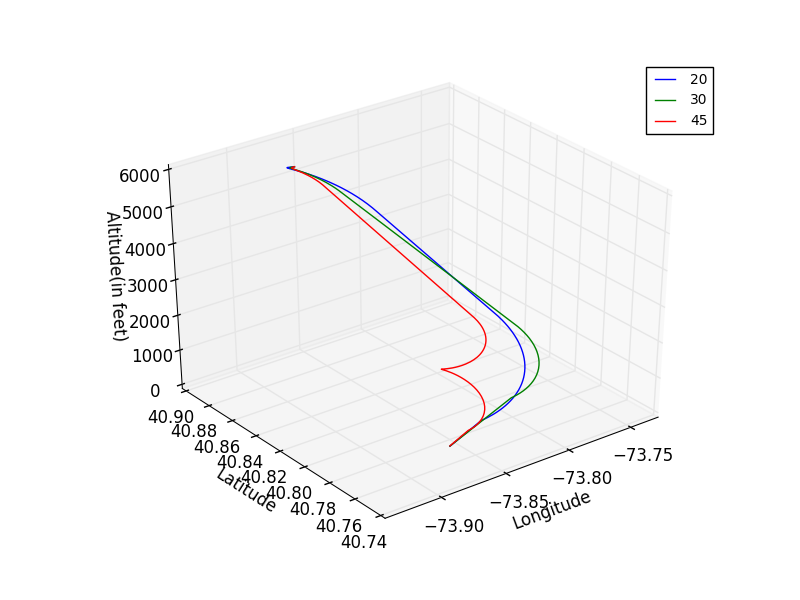}
  \caption{3D View.}
  \label{fig:sub2}
\end{subfigure}
\caption{Trajectory to LGA31 from 6000 feet.}
\label{fig:22t412}
\end{figure}

\begin{figure}[!htb]
\centering
\begin{subfigure}{.5\textwidth}
  \centering
  \includegraphics[width=.8\linewidth]{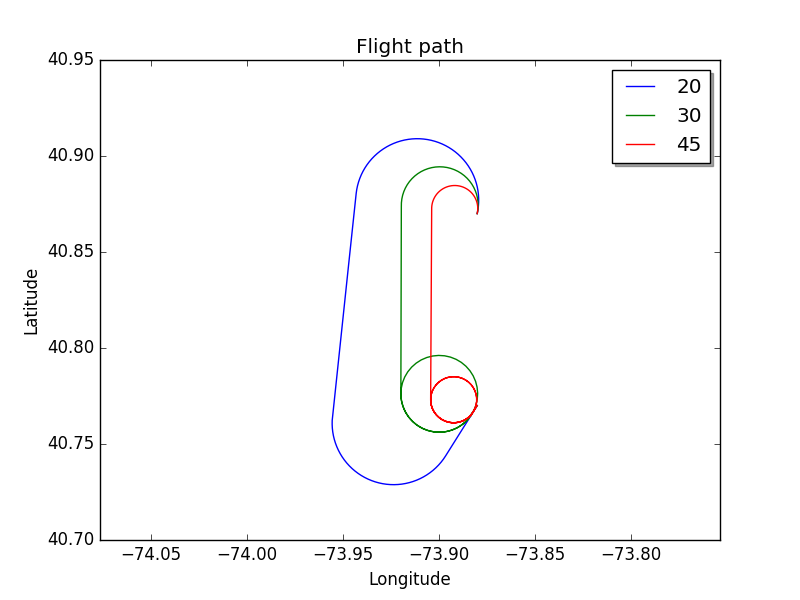}
  \caption{2D View.}
  \label{fig:sub1}
\end{subfigure}%
\begin{subfigure}{.5\textwidth}
  \centering
  \includegraphics[width=.8\linewidth]{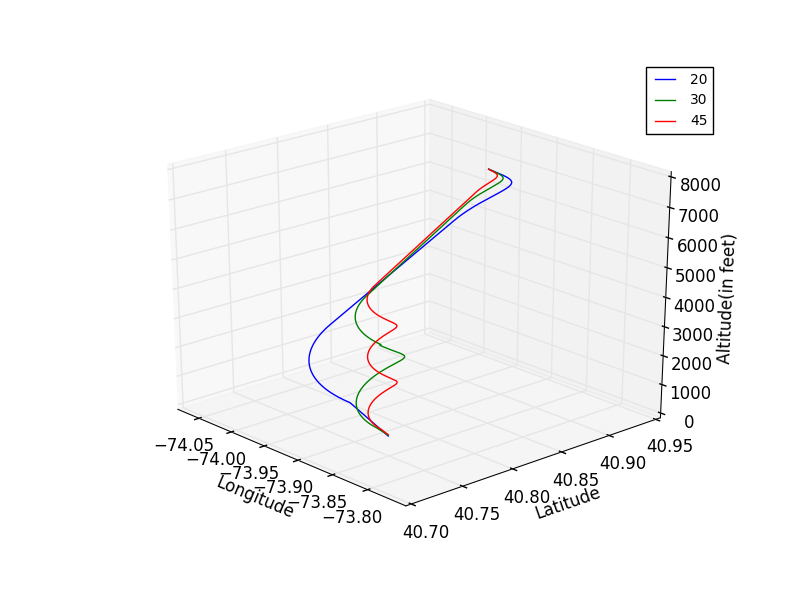}
  \caption{3D View.}
  \label{fig:sub2}
\end{subfigure}
\caption{Trajectory to LGA4 from 8000 feet.}
\label{fig:22t414}
\end{figure}

\begin{figure}[!htb]
\centering
\begin{subfigure}{.5\textwidth}
  \centering
  \includegraphics[width=.8\linewidth]{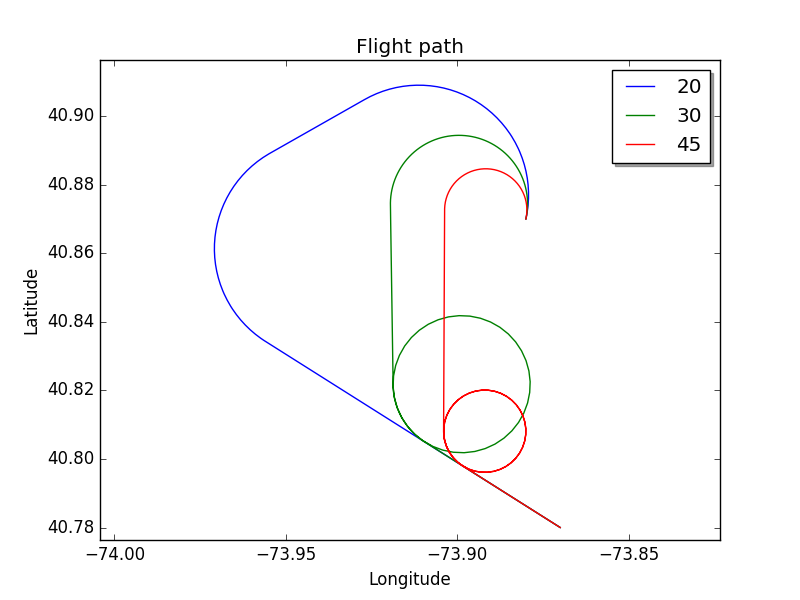}
  \caption{2D View.}
  \label{fig:sub1}
\end{subfigure}%
\begin{subfigure}{.5\textwidth}
  \centering
  \includegraphics[width=.8\linewidth]{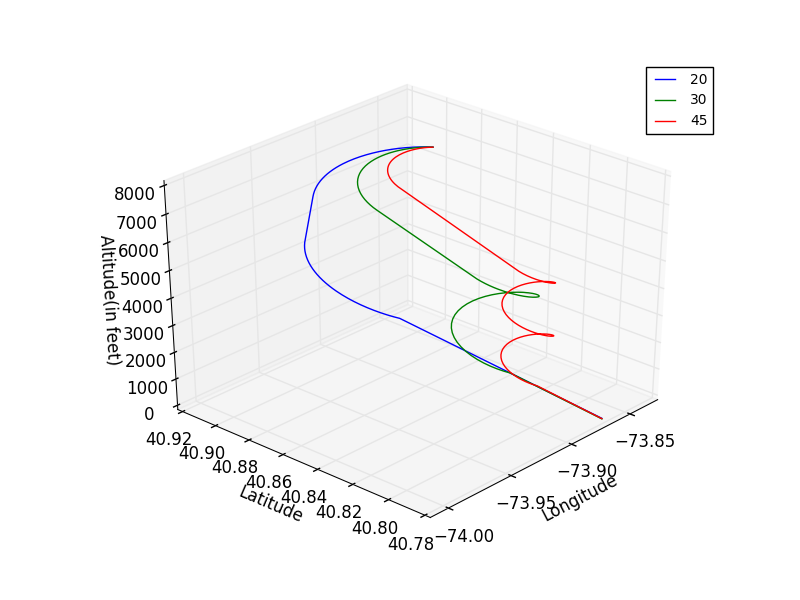}
  \caption{3D View.}
  \label{fig:sub2}
\end{subfigure}
\caption{Trajectory to LGA13 from 8000 feet.}
\label{fig:22t416}
\end{figure}

\begin{figure}[!htb]
\centering
\begin{subfigure}{.5\textwidth}
  \centering
  \includegraphics[width=.8\linewidth]{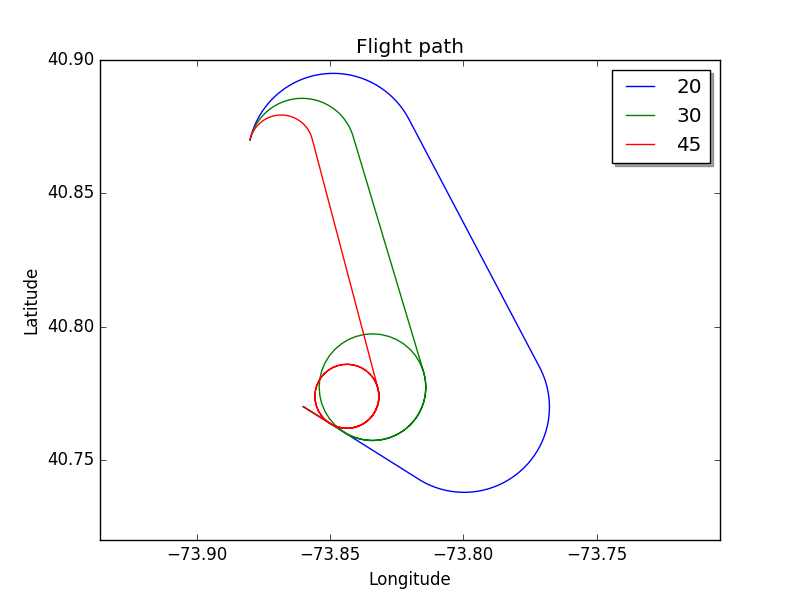}
  \caption{2D View.}
  \label{fig:sub1}
\end{subfigure}%
\begin{subfigure}{.5\textwidth}
  \centering
  \includegraphics[width=.8\linewidth]{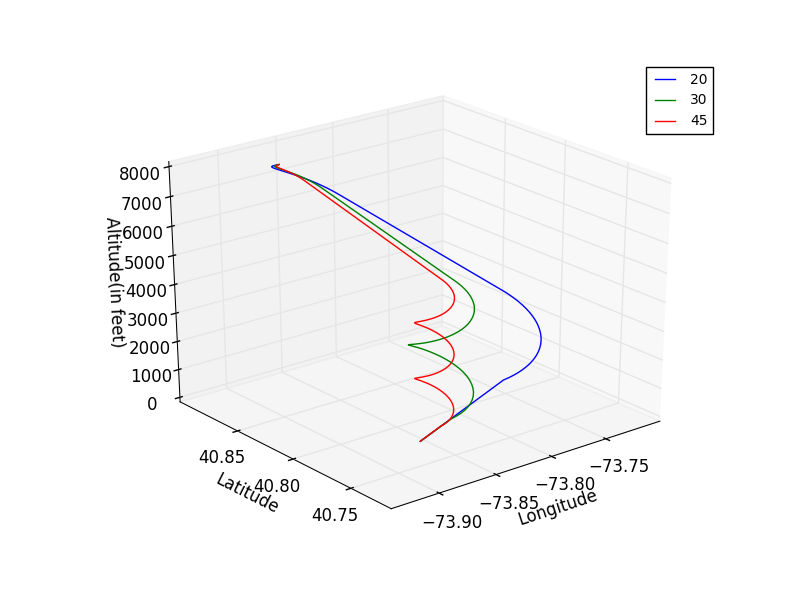}
  \caption{3D View.}
  \label{fig:sub2}
\end{subfigure}
\caption{Trajectory to LGA31 from 8000 feet.}
\label{fig:22t418}
\end{figure}

\begin{figure}[!htb]
\centering
\begin{subfigure}{.5\textwidth}
  \centering
  \includegraphics[width=.8\linewidth]{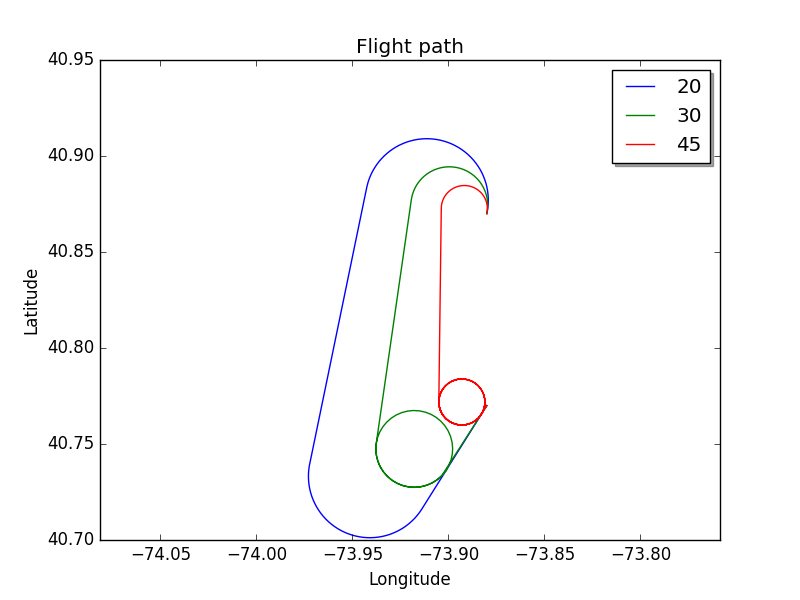}
  \caption{2D View.}
  \label{fig:sub1}
\end{subfigure}%
\begin{subfigure}{.5\textwidth}
  \centering
  \includegraphics[width=.8\linewidth]{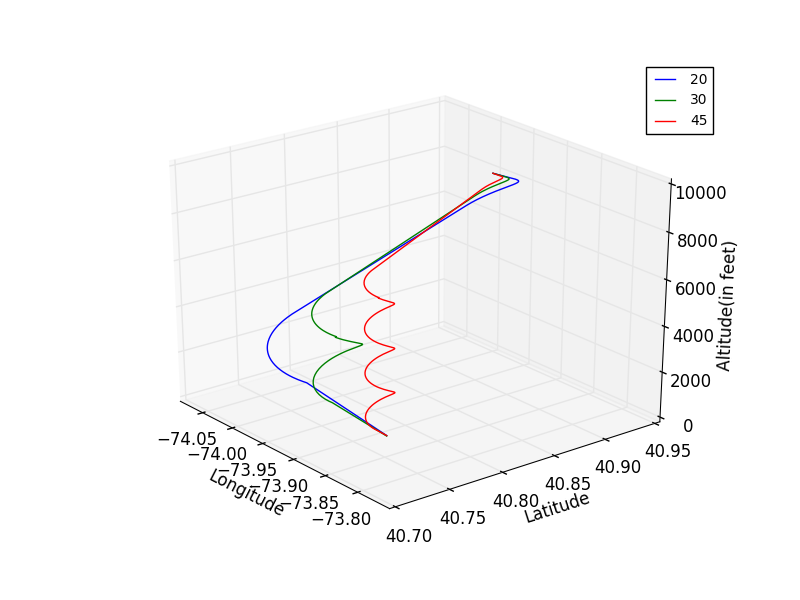}
  \caption{3D View.}
  \label{fig:sub2}
\end{subfigure}
\caption{Trajectory to LGA4 from 10000 feet.}
\label{fig:22t420}
\end{figure}

\begin{figure}[!htb]
\centering
\begin{subfigure}{.5\textwidth}
  \centering
  \includegraphics[width=.8\linewidth]{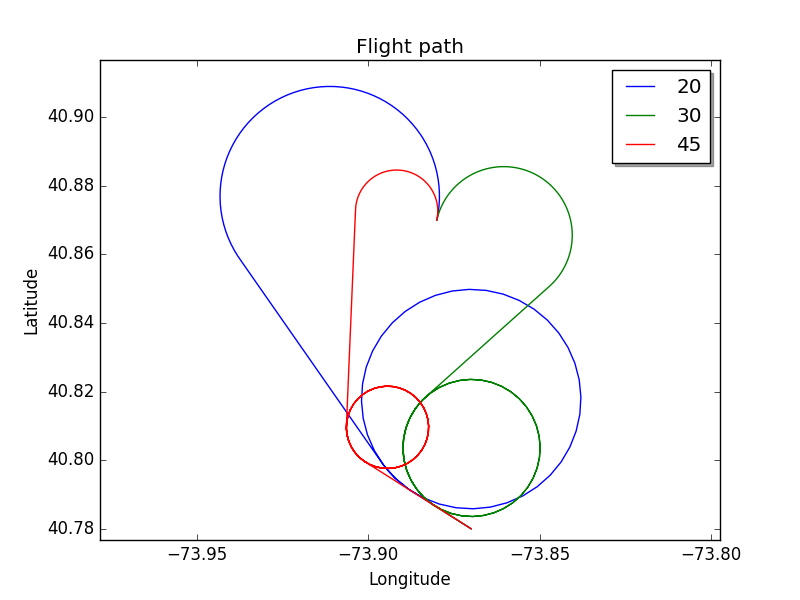}
  \caption{2D View.}
  \label{fig:sub1}
\end{subfigure}%
\begin{subfigure}{.5\textwidth}
  \centering
  \includegraphics[width=.8\linewidth]{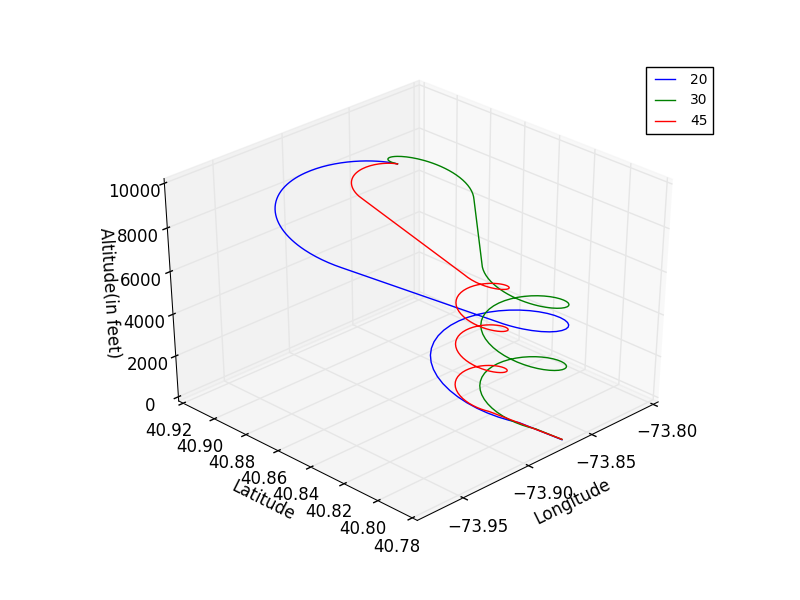}
  \caption{3D View.}
  \label{fig:sub2}
\end{subfigure}
\caption{Trajectory to LGA13 from 10000 feet.}
\label{fig:22t422}
\end{figure}

\begin{figure}[!htb]
\centering
\begin{subfigure}{.5\textwidth}
  \centering
  \includegraphics[width=.8\linewidth]{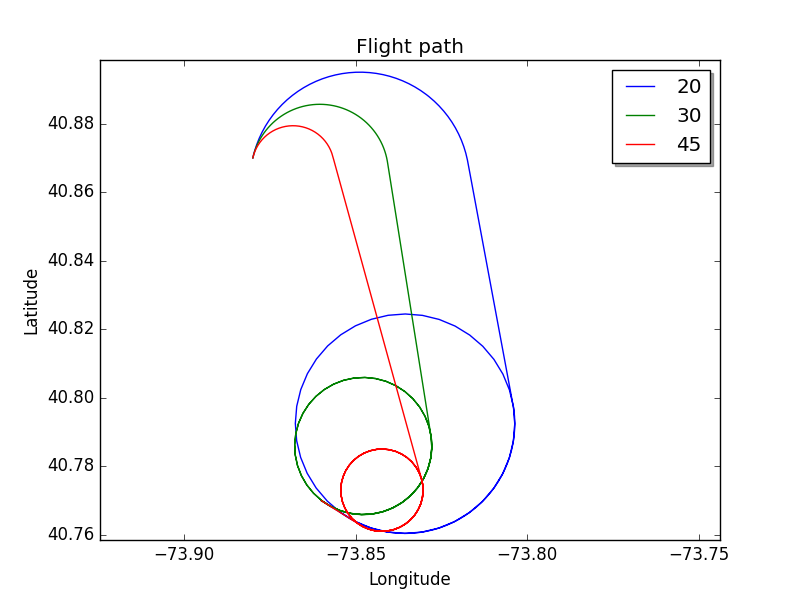}
  \caption{2D View.}
  \label{fig:sub1}
\end{subfigure}%
\begin{subfigure}{.5\textwidth}
  \centering
  \includegraphics[width=.8\linewidth]{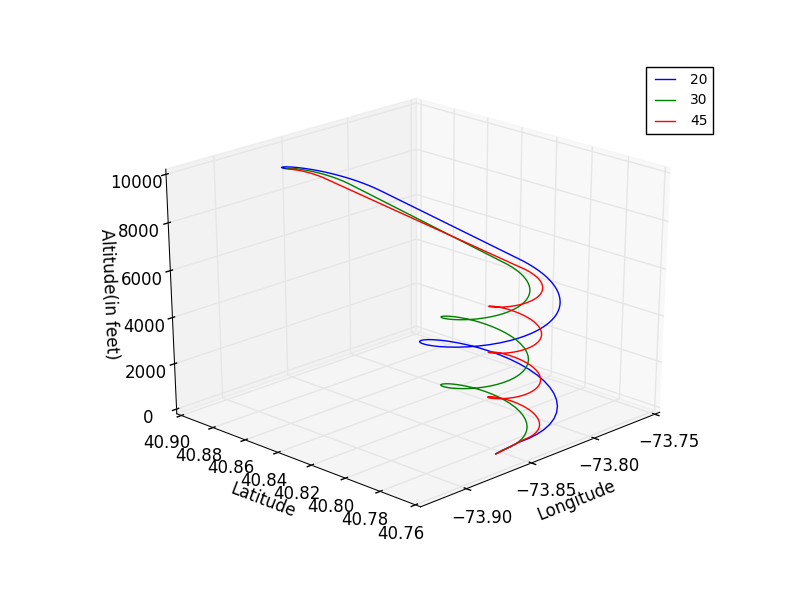}
  \caption{3D View.}
  \label{fig:sub2}
\end{subfigure}
\caption{Trajectory to LGA31 from 10000 feet.}
\label{fig:22t424}
\end{figure}

 We ran simulations with several hypothetical scenarios from different altitudes. Figures~ \ref{fig:22t41} to \ref{fig:22t424} show the 2D and 3D views of the trajectories generated for LGA4, LGA13 and LGA31 from altitudes of 4000, 6000, 8000 and 10000 feet. The starting point had the configuration: \{longitude:-73.88000°, latitude:40.86500°, heading: 12.9°\}. In all cases, a clean configuration glide ratio of 17.25:1 was used for generating the trajectories with a dirty configuration glide ratio of 9:1 in the extended final segments. 
\begin{figure}[!htb]
\centering
\begin{subfigure}{.5\textwidth}
  \centering
  \includegraphics[width=.8\linewidth]{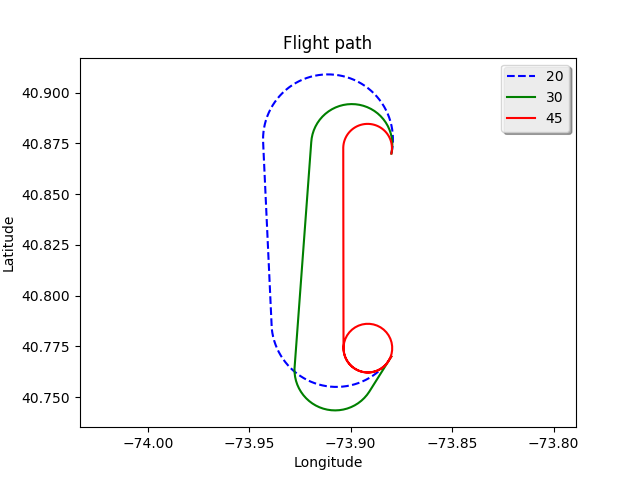}
  \caption{2D View.}
  \label{fig:sub1}
\end{subfigure}%
\begin{subfigure}{.5\textwidth}
  \centering
  \includegraphics[width=.8\linewidth]{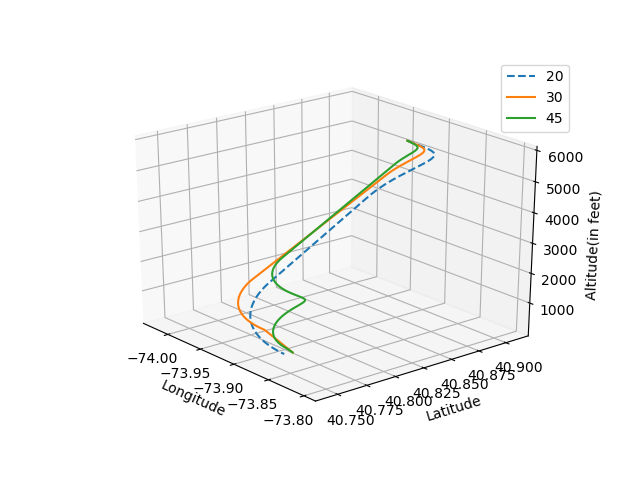}
  \caption{3D View.}
  \label{fig:sub2}
\end{subfigure}
\caption{Trajectory to LGA4 from 6000 feet.}
\label{fig:22t49}
\end{figure}
\begin{figure}[!htb]
\centering
\begin{subfigure}{.5\textwidth}
  \centering
  \includegraphics[width=.8\linewidth]{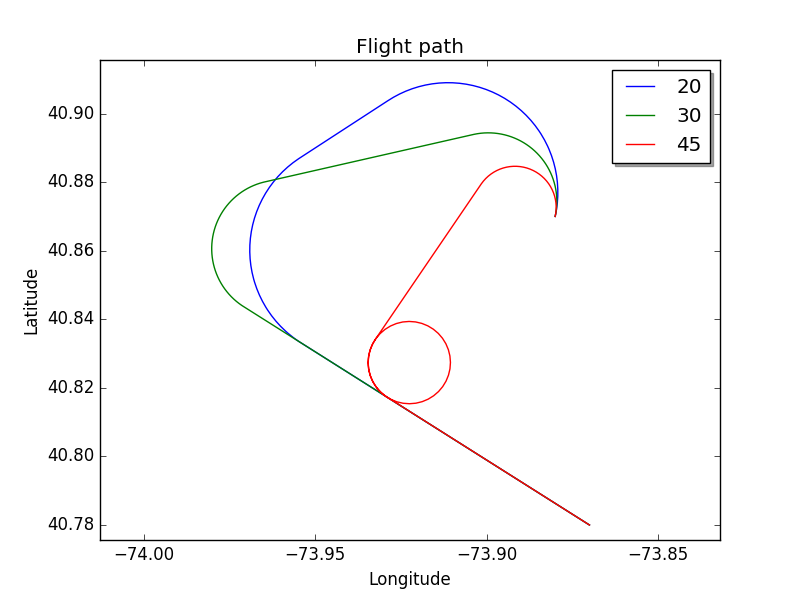}
  \caption{2D View.}
  \label{fig:sub1}
\end{subfigure}%
\begin{subfigure}{.5\textwidth}
  \centering
  \includegraphics[width=.8\linewidth]{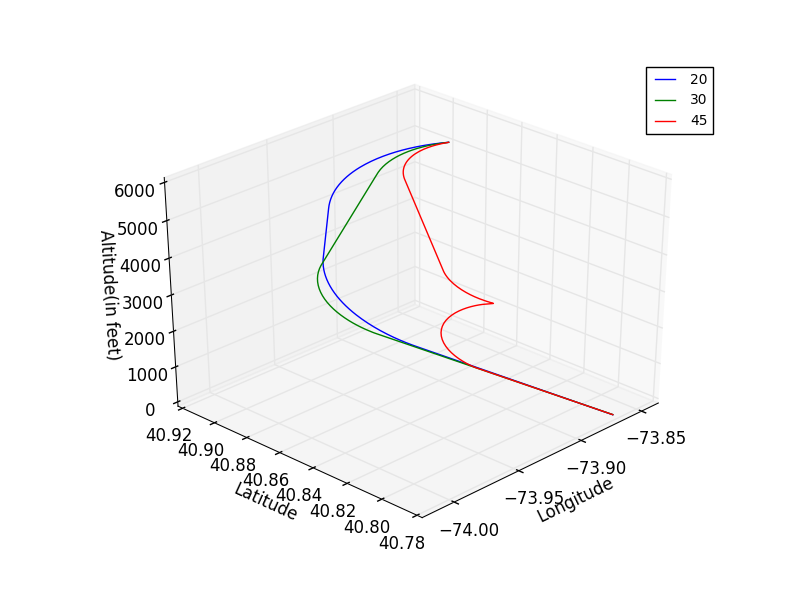}
  \caption{3D View.}
  \label{fig:sub2}
\end{subfigure}
\caption{Trajectory to LGA13 from 6000 feet.}
\label{fig:22t49}
\end{figure}
\FloatBarrier
\subsection{US Airways Flight 1549}
Us Airways flight 1549 was a flight that took off from New York City's La Guardia Airport on January 15, 2009 and lost power in both engines when it struck a flock of Canada geese a couple of minutes after takeoff. The pilots managed to land the Airbus A320-214 successfully in the Hudson river and save the lives of everyone on board the flight. In order to analyze the other possible options that may have been available to the pilot instead of landing the aircraft in the Hudson, we used our trajectory planning algorithm to recreate the conditions of the particular incident of flight 1549.\\
\begin{figure*}[!htb]
  \centering
  \includegraphics[width=\textwidth]{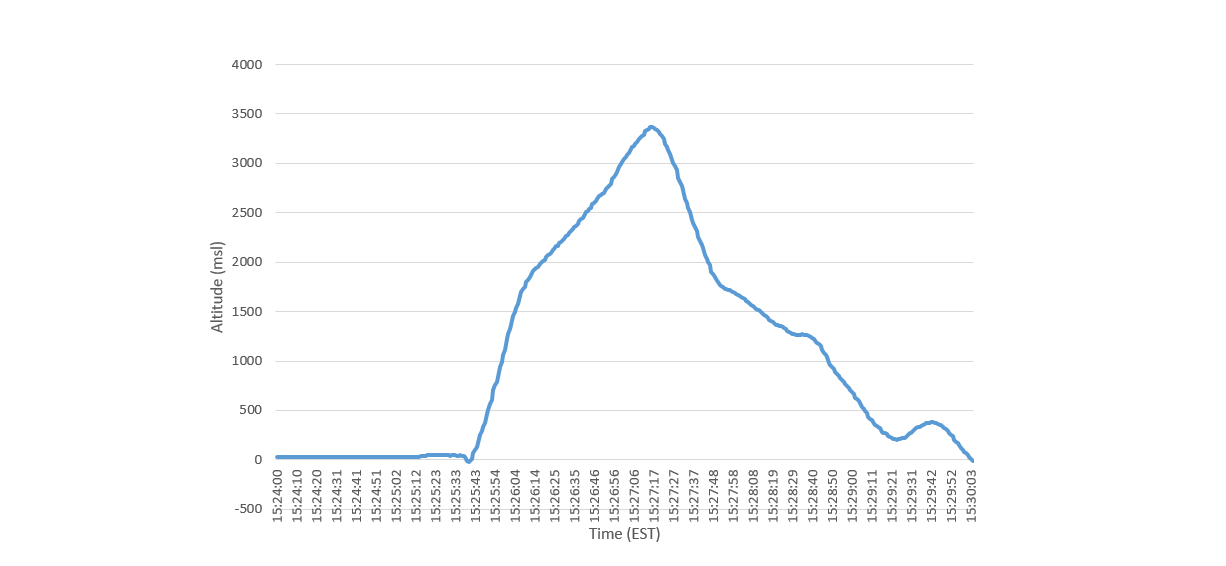}
  \caption{Flight 1549 Time vs Altitude graph.}
  \label{1549tag}
\end{figure*}
\begin{table*}
\caption {US Airways 1549 Flight Data Recorder data.} \label{tab:table2}
\centering
\begin{tabular}{c c c c c c c c} 
Time Delay & Latitude(decimal) & Longitude(decimal) & Pressure Altitude(feet) &  true altitude(feet) & magnetic heading(degrees) & Airspeed(kts)\\ [1ex] 
t & 40.8477 & -73.8758 & 2792 & 3056 & 0 & 218 \\ [1ex] 
t+4 & 40.8513 & -73.8767 & 2888 & 3152 & 0.7 & 207.125 \\ [1ex] 
t+8 & 40.8547 & -73.8781 & 2968 & 3232 & 0 & 200.25 \\ [1ex]  
t+12 & 40.8581 & -73.8786 & 3048 & 3312 & 0.4 & 193 \\ [1ex]
t+16 & 40.8617 & -73.8794 & 3088 & 3352 & 358.9 & 185.25 \\ [1ex] 
t+20 & 40.865 & -73.88 & 3040 & 3304 & 357.2 & 185.25 \\ [1ex] 
t+24 & 40.8678 & -73.8806 & 2916 & 3180 & 352.6 & 185.375 \\ [1ex]
t+28 & 40.8711 & -73.8819 & 2760 & 3024 & 344.5 & 187 \\ [1ex]  
t+32 & 40.8739 & -73.8842 & 2580 & 2844 & 333.3 & 190.625 \\ [1ex]
t+36 & 40.761 & -73.8861 & 2368 & 2632 & 320.6 & 198.75 \\ [1ex] 
t+40 & 40.8789 & -73.8897 & 2156 & 2420 & 305.5 & 202.875 \\ [1ex]
\end{tabular}
\end{table*}
\begin{figure}[!htb]
\centering
\begin{subfigure}{.5\textwidth}
  \centering
  \includegraphics[width=.8\linewidth]{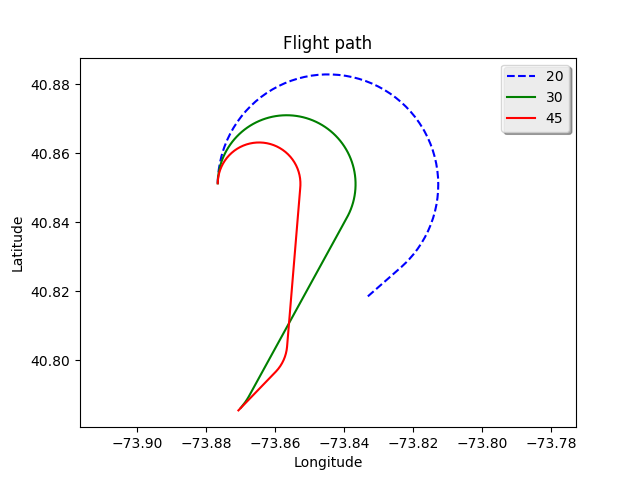}
  \caption{2D View.}
  \label{fig:sub1}
\end{subfigure}%
\begin{subfigure}{.5\textwidth}
  \centering
  \includegraphics[width=.8\linewidth]{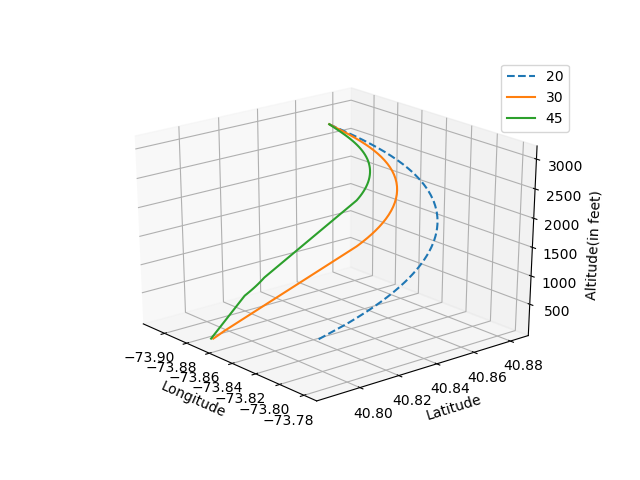}
  \caption{3D View.}
  \label{fig:sub2}
\end{subfigure}
\caption{Trajectory to LGA22 with a glide ratio of 17.25:1 at time t+4 .}
\label{fig:22t4}
\end{figure}
\begin{figure}[!htb]
\centering
\begin{subfigure}{.5\textwidth}
  \centering
  \includegraphics[width=.8\linewidth]{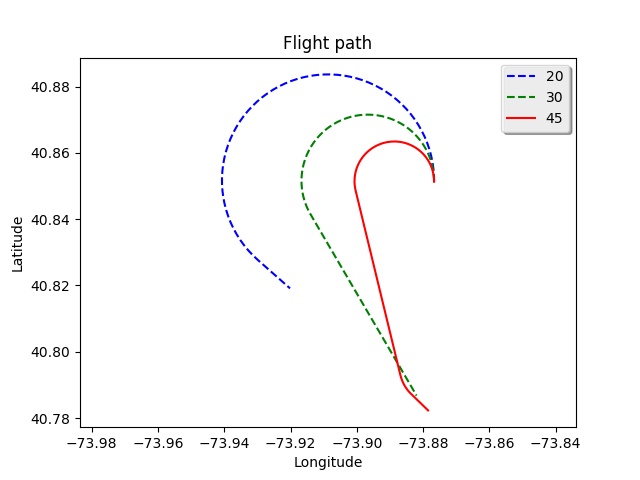}
  \caption{2D View.}
  \label{fig:sub1}
\end{subfigure}%
\begin{subfigure}{.5\textwidth}
  \centering
  \includegraphics[width=.8\linewidth]{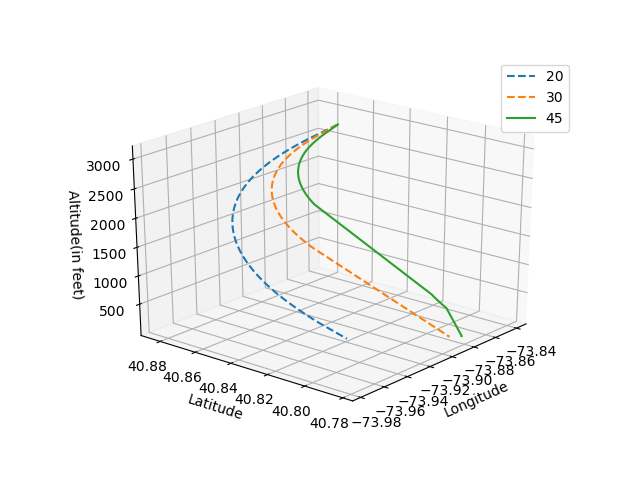}
  \caption{3D View.}
  \label{fig:sub2}
\end{subfigure}
\caption{Trajectory to LGA13 with a glide ratio of 17.25:1 at time t+4 .}
\label{fig:13t4}
\end{figure}
\begin{figure}[!htb]
\centering
\begin{subfigure}{.5\textwidth}
  \centering
  \includegraphics[width=.8\linewidth]{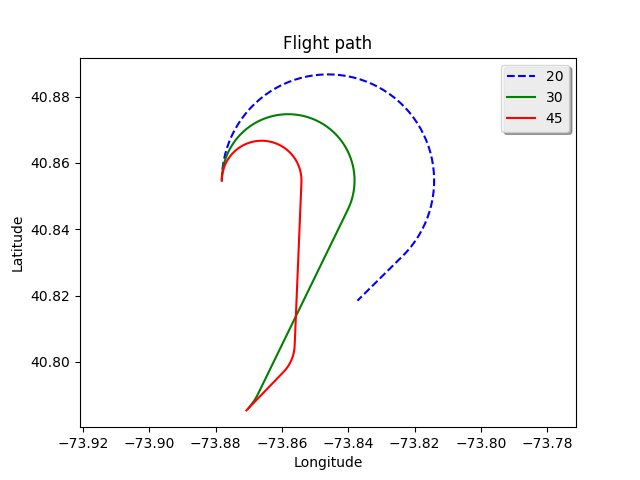}
  \caption{2D View.}
  \label{fig:sub1}
\end{subfigure}%
\begin{subfigure}{.5\textwidth}
  \centering
  \includegraphics[width=.8\linewidth]{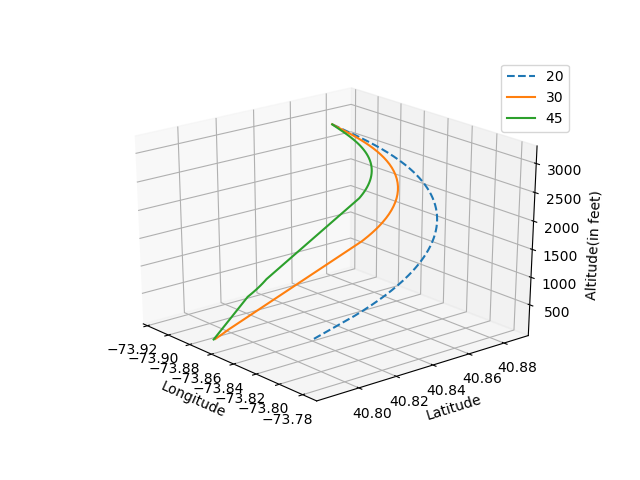}
  \caption{3D View.}
  \label{fig:sub2}
\end{subfigure}
\caption{Trajectory to LGA22 with a glide ratio of 17.25:1 at time t+8.}
\label{fig:22t8}
\end{figure}
\begin{figure}[!htb]
\centering
\begin{subfigure}{.5\textwidth}
  \centering
  \includegraphics[width=.8\linewidth]{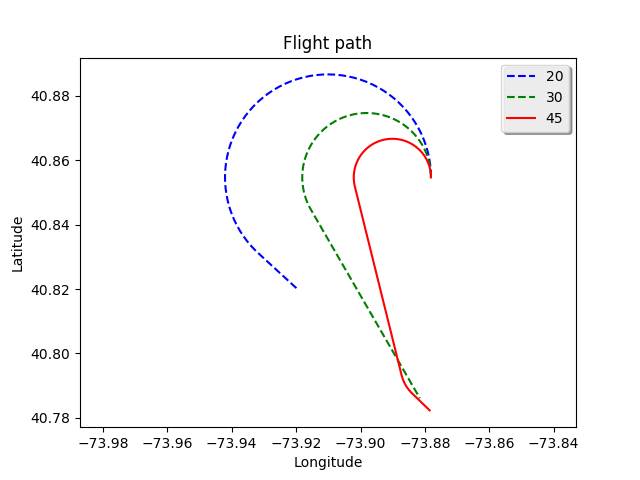}
  \caption{2D View.}
  \label{fig:sub1}
\end{subfigure}%
\begin{subfigure}{.5\textwidth}
  \centering
  \includegraphics[width=.8\linewidth]{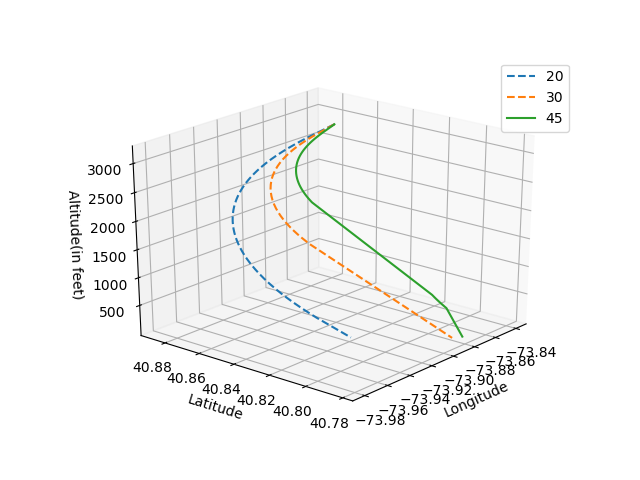}
  \caption{3D View.}
  \label{fig:sub2}
\end{subfigure}
\caption{Trajectory to LGA13 with a glide ratio of 17.25:1 at time t+8.}
\label{fig:13t8}
\end{figure}
\begin{figure}[!htb]
\centering
\begin{subfigure}{.5\textwidth}
  \centering
  \includegraphics[width=.8\linewidth]{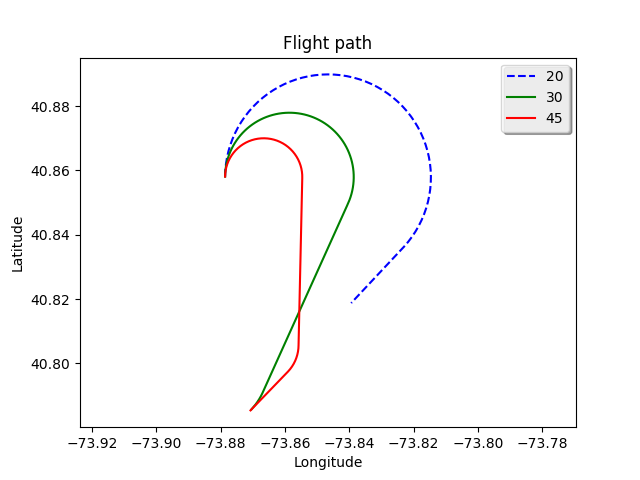}
  \caption{2D View.}
  \label{fig:sub1}
\end{subfigure}%
\begin{subfigure}{.5\textwidth}
  \centering
  \includegraphics[width=.8\linewidth]{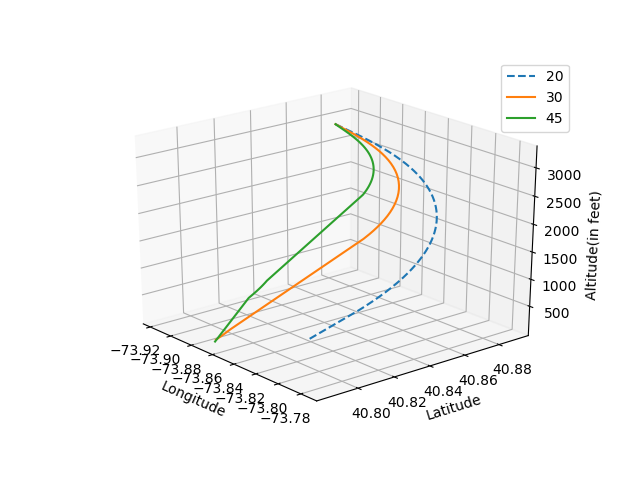}
  \caption{3D View.}
  \label{fig:sub2}
\end{subfigure}
\caption{Trajectory to LGA22 with a glide ratio of 17.25:1 at time t+12.}
\label{fig:22t12}
\end{figure}
\begin{figure}[!htb]
\centering
\begin{subfigure}{.5\textwidth}
  \centering
  \includegraphics[width=.8\linewidth]{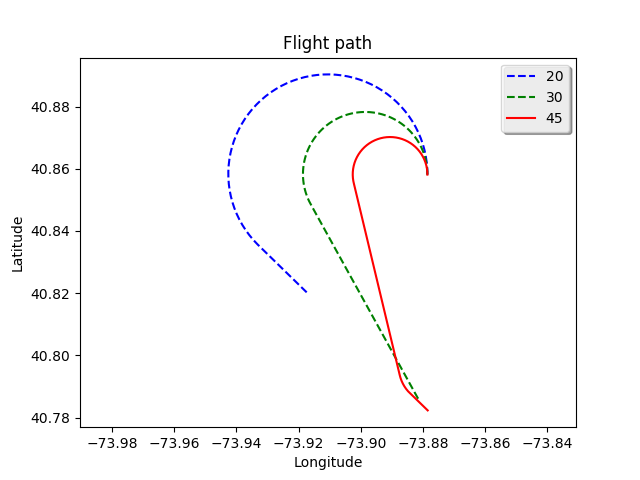}
  \caption{2D View.}
  \label{fig:sub1}
\end{subfigure}%
\begin{subfigure}{.5\textwidth}
  \centering
  \includegraphics[width=.8\linewidth]{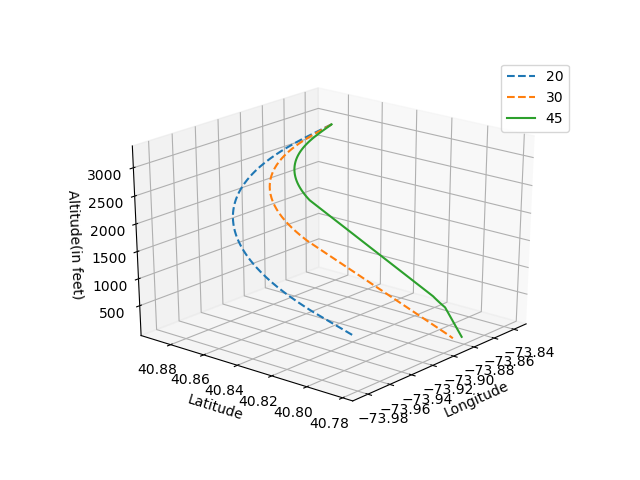}
  \caption{3D View.}
  \label{fig:sub2}
\end{subfigure}
\caption{Trajectory to LGA13 with a glide ratio of 17.25:1 at time t+12.}
\label{fig:13t12}
\end{figure}
\begin{figure}[!htb]
\centering
\begin{subfigure}{.5\textwidth}
  \centering
  \includegraphics[width=.8\linewidth]{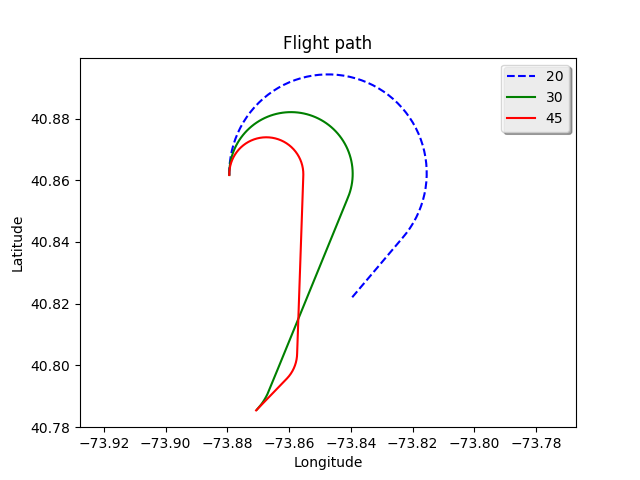}
  \caption{2D View.}
  \label{fig:sub1}
\end{subfigure}%
\begin{subfigure}{.5\textwidth}
  \centering
  \includegraphics[width=.8\linewidth]{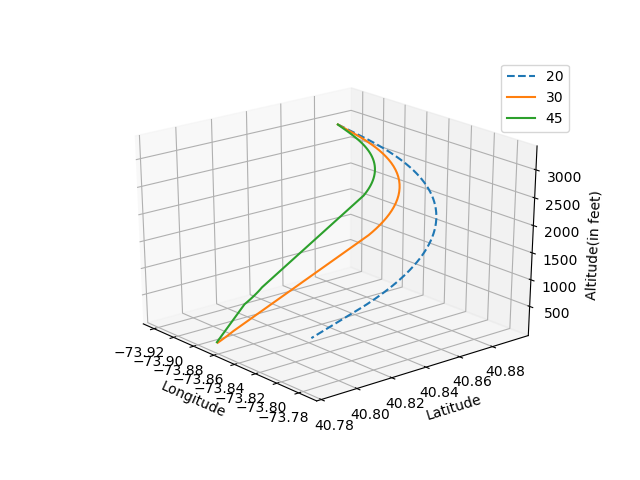}
  \caption{3D View.}
  \label{fig:sub2}
\end{subfigure}
\caption{Trajectory to LGA22 with a glide ratio of 17.25:1 at time t+16.}
\label{fig:22t16}
\end{figure}
\begin{figure}[!htb]
\centering
\begin{subfigure}{.5\textwidth}
  \centering
  \includegraphics[width=.8\linewidth]{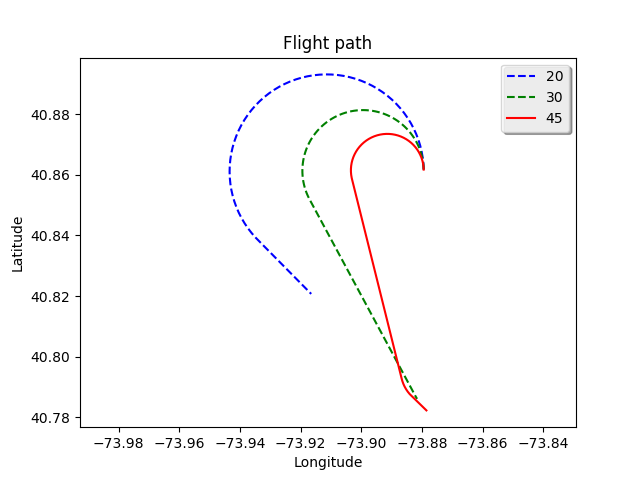}
  \caption{2D View.}
  \label{fig:sub1}
\end{subfigure}%
\begin{subfigure}{.5\textwidth}
  \centering
  \includegraphics[width=.8\linewidth]{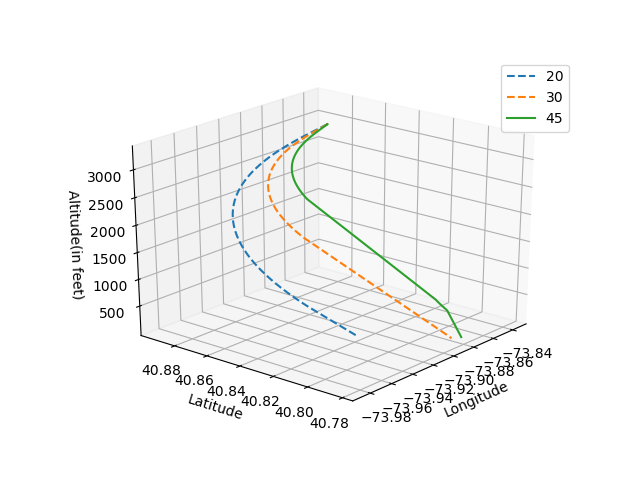}
  \caption{3D View.}
  \label{fig:sub2}
\end{subfigure}
\caption{Trajectory to LGA13 with a glide ratio of 17.25:1 at time t+16.}
\label{fig:13t16}
\end{figure}
\begin{figure}[!htb]
\centering
\begin{subfigure}{.5\textwidth}
  \centering
  \includegraphics[width=.8\linewidth]{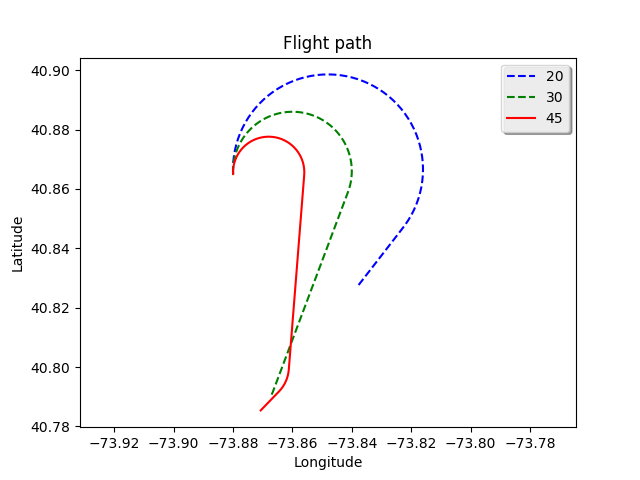}
  \caption{2D View.}
  \label{fig:sub1}
\end{subfigure}%
\begin{subfigure}{.5\textwidth}
  \centering
  \includegraphics[width=.8\linewidth]{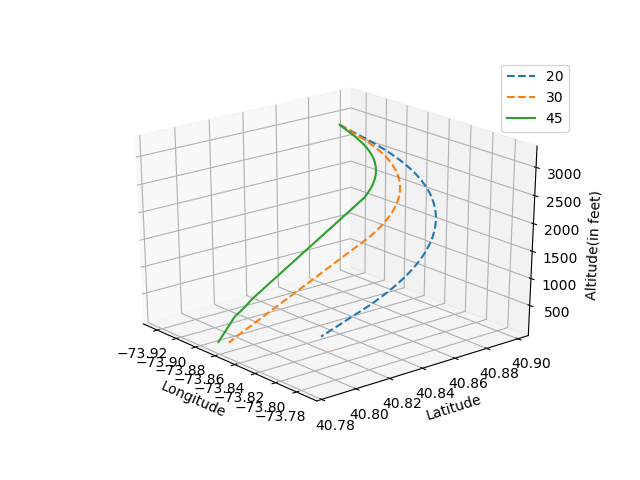}
  \caption{3D View.}
  \label{fig:sub2}
\end{subfigure}
\caption{Trajectory to LGA22 with a glide ratio of 17.25:1 at time t+20.}
\label{fig:22t20}
\end{figure}
\begin{figure}[!htb]
\centering
\begin{subfigure}{.5\textwidth}
  \centering
  \includegraphics[width=.8\linewidth]{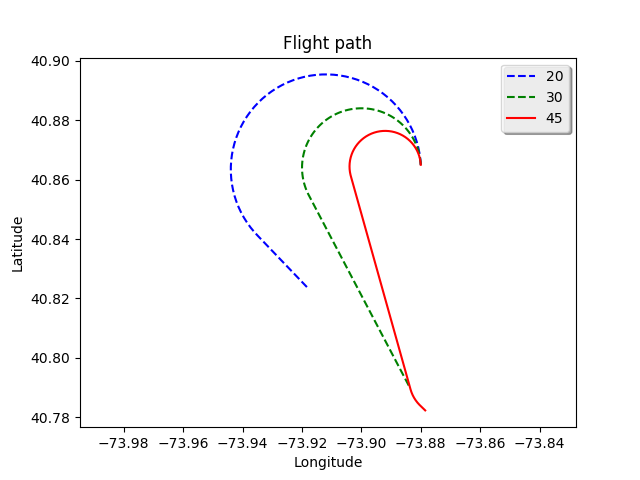}
  \caption{2D View.}
  \label{fig:sub1}
\end{subfigure}%
\begin{subfigure}{.5\textwidth}
  \centering
  \includegraphics[width=.8\linewidth]{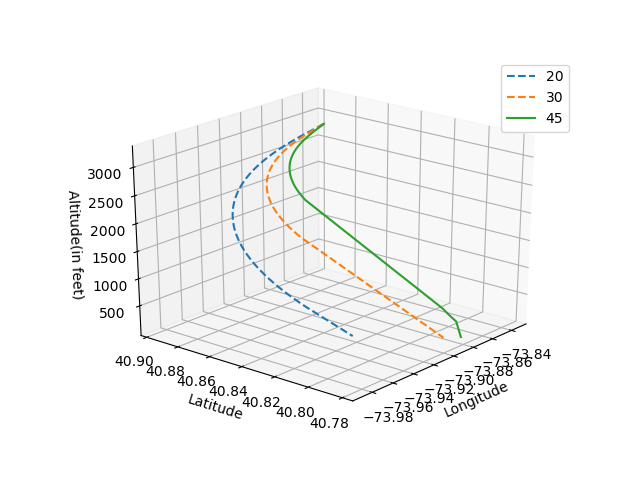}
  \caption{3D View.}
  \label{fig:sub2}
\end{subfigure}
\caption{Trajectory to LGA13 with a glide ratio of 17.25:1 at time t+20.}
\label{fig:13t20}
\end{figure}
\begin{figure}[!htb]
\centering
\begin{subfigure}{.5\textwidth}
  \centering
  \includegraphics[width=.8\linewidth]{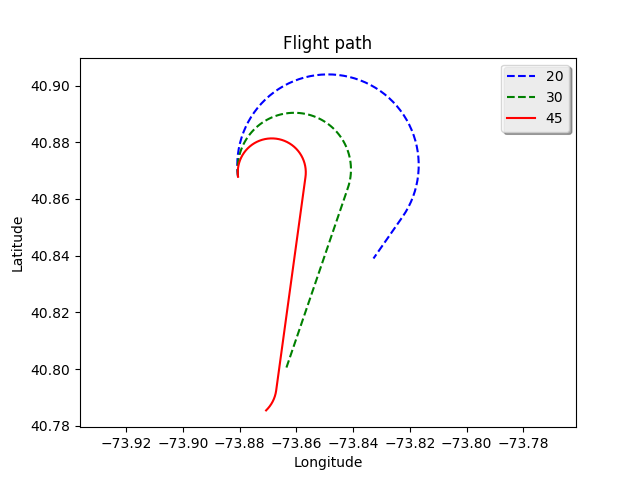}
  \caption{2D View.}
  \label{fig:sub1}
\end{subfigure}%
\begin{subfigure}{.5\textwidth}
  \centering
  \includegraphics[width=.8\linewidth]{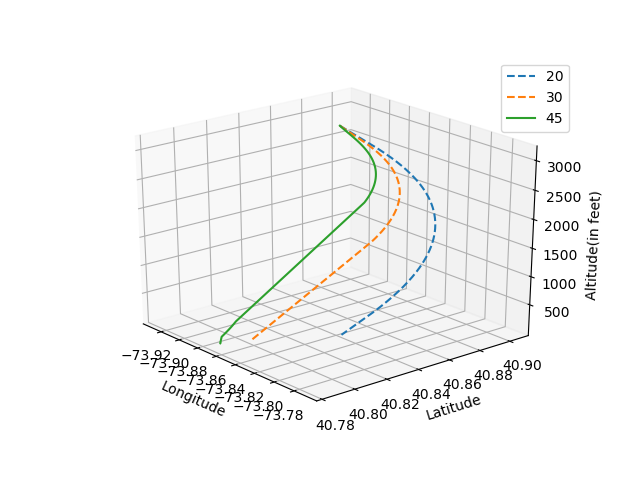}
  \caption{3D View.}
  \label{fig:sub2}
\end{subfigure}
\caption{Trajectory to LGA22 with a glide ratio of 17.25:1 at time t+24.}
\label{fig:22t24}
\end{figure}
\begin{figure}[!htb]
\centering
\begin{subfigure}{.5\textwidth}
  \centering
  \includegraphics[width=.8\linewidth]{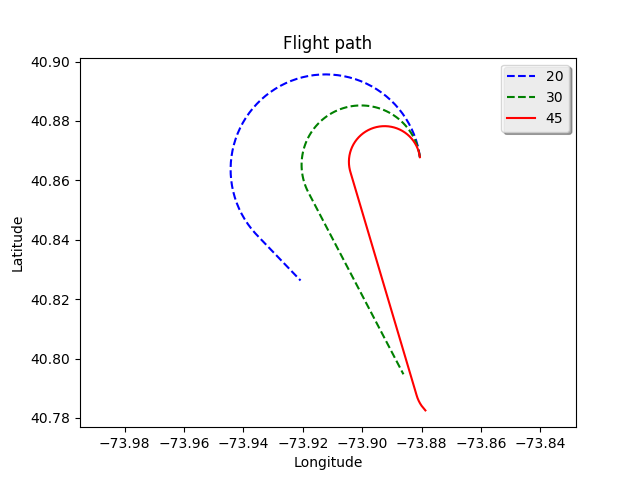}
  \caption{2D View.}
  \label{fig:sub1}
\end{subfigure}%
\begin{subfigure}{.5\textwidth}
  \centering
  \includegraphics[width=.8\linewidth]{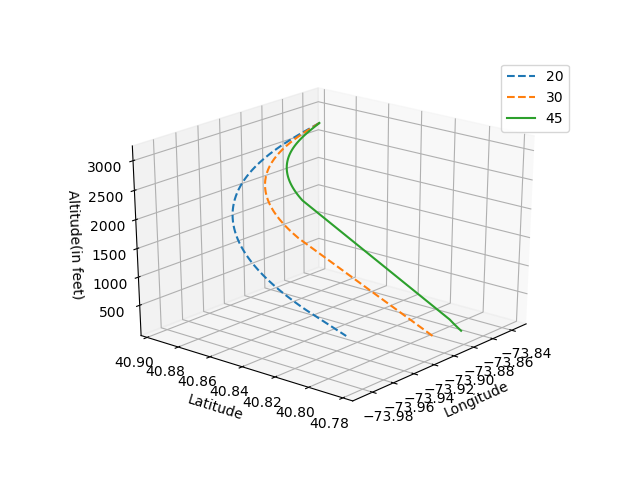}
  \caption{3D View.}
  \label{fig:sub2}
\end{subfigure}
\caption{Trajectory to LGA13 with a glide ratio of 17.25:1 at time t+24.}
\label{fig:13t24}
\end{figure}
\begin{figure}[!htb]
\centering
\begin{subfigure}{.5\textwidth}
  \centering
  \includegraphics[width=.8\linewidth]{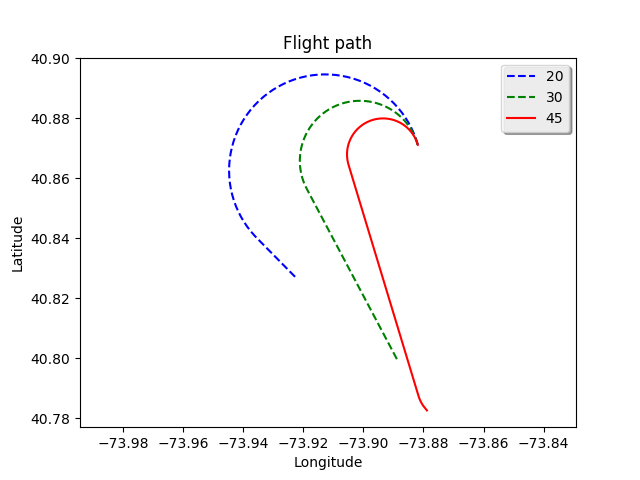}
  \caption{2D View.}
  \label{fig:sub1}
\end{subfigure}%
\begin{subfigure}{.5\textwidth}
  \centering
  \includegraphics[width=.8\linewidth]{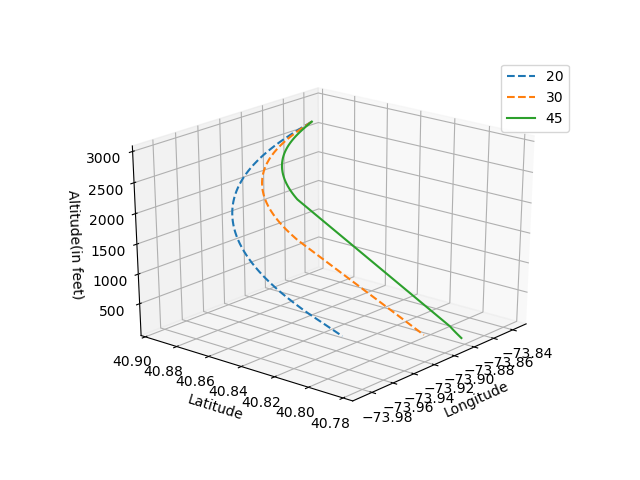}
  \caption{3D View.}
  \label{fig:sub2}
\end{subfigure}
\caption{Trajectory to LGA13 with a glide ratio of 17.25:1 at time t+28.}
\label{fig:13t28}
\end{figure}

\begin{table}
\caption {Rank of trajectories generated for US Airways 1549 for t+4 seconds with glide ratio 17.25:1.}
\label{tab:table3}
\centering
\begin{tabular}{c c c c c c c c c c} 
Runway  & Bank angle & $\left \| \bar{d} \right \|$ &$ \left \| \bar{z} \right \|$&$ \left \| l \right \|$&$ \left \| n\right \|$&$ \left \| \bar{\left(\frac{\theta}{h}\right)}\right \|$&$ \left \| e\right \|$&$ u $& Rank\\ [1ex]
 LGA22 & 30 & 0.55 & 0.00 & 0.00 & 1.00 & 1.00 & 0.00 & 0.43 & 3\\
 LGA22 & 45 & 1.00 & 0.78 & 1.00 & 0.00 & 0.00 & 1.00 & 0.63 & 1\\
 LGA13 & 45 & 0.00 & 1.00 & 0.66 & 0.00 & 0.77 & 0.67 & 0.52 & 2\\
\end{tabular}
\end{table}
\begin{table}
\caption {Rank of trajectories generated for US Airways 1549 for t+8 seconds with glide ratio 17.25:1.}
\label{tab:table4}
\centering
\begin{tabular}{c c c c c c c c c c} 
Runway  & Bank angle & $\left \| \bar{d} \right \|$ &$ \left \| \bar{z} \right \|$&$ \left \| l \right \|$&$ \left \| n\right \|$&$ \left \| \bar{\left(\frac{\theta}{h}\right)}\right \|$&$ \left \| e\right \|$&$ u $& Rank\\ [1ex]
LGA22  & 30 & 0.61 & 0.00 & 0.00 & 1.00 & 1.00 & 0.00 & 0.43 & 3\\
LGA22  & 45 & 1.00 & 0.76 & 1.00 & 0.00 & 0.00 & 1.00 & 0.63 & 1 \\
LGA13  & 45 & 0.00 & 1.00 & 0.66 & 0.00 & 0.50 & 0.68 & 0.47 & 2\\
\end{tabular}
\end{table}
\begin{table}
\caption {Rank of trajectories generated for US Airways 1549 for t+12 seconds with glide ratio 17.25:1.}
\label{tab:table5}
\centering
\begin{tabular}{c c c c c c c c c c} 
Runway  & Bank angle & $\left \| \bar{d} \right \|$ &$ \left \| \bar{z} \right \|$&$ \left \| l \right \|$&$ \left \| n\right \|$&$ \left \| \bar{\left(\frac{\theta}{h}\right)}\right \|$&$ \left \| e\right \|$&$ u $& Rank\\ [1ex]
 LGA22 & 30 & 0.66 & 0.00 & 0.00 & 1.00 & 1.00 & 0.00 & 0.44 & 3\\
 LGA22 & 45 & 1.00 & 0.77 & 1.00 & 0.00 & 0.00 & 1.00 & 0.63 & 1\\
 LGA13 & 45 & 0.00 & 1.00 & 0.63 & 0.00 & 0.75 & 0.69 & 0.52 & 2\\  
\end{tabular}
\end{table}
\begin{table}
\caption {Rank of trajectories generated for US Airways 1549 for t+16 seconds with glide ratio 17.25:1.}
\label{tab:table6}
\centering
\begin{tabular}{c c c c c c c c c c} 
Runway  & Bank angle & $\left \| \bar{d} \right \|$ &$ \left \| \bar{z} \right \|$&$ \left \| l \right \|$&$ \left \| n\right \|$&$ \left \| \bar{\left(\frac{\theta}{h}\right)}\right \|$&$ \left \| e\right \|$&$ u $& Rank\\ [1ex]
 LGA22 & 30 & 0.55 & 0.00 & 0.00 & 1.00 & 0.00 & 0.00   & 0.26 & 3 \\
 LGA22 & 45 & 1.00 & 0.70 & 1.00 & 0.00 & 0.56 & 1.00   & 0.71 & 1\\
 LGA13 & 45 & 0.00 & 1.00 & 0.69 & 0.00 & 1.00 & 0.72   & 0.56 & 2\\
\end{tabular}
\end{table}
\begin{table}
\caption {Rank of trajectories generated for US Airways 1549 for t+20 seconds with glide ratio 17.25:1.}
\label{tab:table7}
\centering
\begin{tabular}{c c c c c c c c c c} 
Runway  & Bank angle & $\left \| \bar{d} \right \|$ &$ \left \| \bar{z} \right \|$&$ \left \| l \right \|$&$ \left \| n\right \|$&$ \left \| \bar{\left(\frac{\theta}{h}\right)}\right \|$&$ \left \| e\right \|$&$ u $& Rank\\ [1ex]
LGA22  & 45 & 0.00 & 1.00 & 1.00 & 1.00 & 0.00 & 1.00 & 0.67 & 1\\
LGA13  & 45 & 1.00 & 0.00 & 0.00 & 1.00 & 1.00 & 0.00 & 0.50 & 2\\
\end{tabular}
\end{table}
\begin{table}
\caption {Rank of trajectories generated for US Airways 1549 for t+24 seconds with glide ratio 17.25:1.}
\label{tab:table8}
\centering
\begin{tabular}{c c c c c c c c c c} 
Runway  & Bank angle & $\left \| \bar{d} \right \|$ &$ \left \| \bar{z} \right \|$&$ \left \| l \right \|$&$ \left \| n\right \|$&$ \left \| \bar{\left(\frac{\theta}{h}\right)}\right \|$&$ \left \| e\right \|$&$ u $& Rank\\ [1ex]
 LGA22 & 45 & 1.00 & 0.00 & 0.00 & 1.00 & 0.00 & 1.00 & 0.50 & 2\\
 LGA13 & 45 & 0.00 & 1.00 & 1.00 & 1.00 & 1.00 & 1.00 & 0.83 & 1\\
\end{tabular}
\end{table}
\begin{table}
\caption {Rank of trajectories generated for US Airways 1549 for t+4 seconds with glide ratio 19:1.}
\label{tab:table9}
\centering
\begin{tabular}{c c c c c c c c c c} 
Runway  & Bank angle & $\left \| \bar{d} \right \|$ &$ \left \| \bar{z} \right \|$&$ \left \| l \right \|$&$ \left \| n\right \|$&$ \left \| \bar{\left(\frac{\theta}{h}\right)}\right \|$&$ \left \| e\right \|$&$ u $& Rank\\ [1ex]
LGA22 & 30 & 0.20 & 0.30 & 0.16 & 1.00 & 0.88 & 0.33 & 0.47 & 3\\
LGA22 & 45 & 0.38 & 1.00 & 1.00 & 1.00 & 0.71 & 1.00 & 0.85 & 1\\
LGA13 & 30 & 0.00 & 0.12 & 0.00 & 1.00 & 1.00 & 0.16 & 0.38 & 4\\
LGA13 & 45 & 0.35 & 0.81 & 0.91 & 1.00 & 0.70 & 0.86 & 0.77 & 2\\
LGA31 & 45 & 1.00 & 0.00 & 0.16 & 1.00 & 0.00 & 0.00 & 0.36 & 5\\
\end{tabular}
\end{table}
\begin{table}
\caption {Rank of trajectories generated for US Airways 1549 for t+8 seconds with glide ratio 19:1.}
\label{tab:table10}
\centering
\begin{tabular}{c c c c c c c c c c} 
Runway  & Bank angle & $\left \| \bar{d} \right \|$ &$ \left \| \bar{z} \right \|$&$ \left \| l \right \|$&$ \left \| n\right \|$&$ \left \| \bar{\left(\frac{\theta}{h}\right)}\right \|$&$ \left \| e\right \|$&$ u $& Rank\\ [1ex]
LGA22 & 30 & 0.12 & 0.34 & 0.16 & 1.00 & 0.87 & 0.41 & 0.48 & 3\\
LGA22 & 45 & 0.35 & 1.00 & 1.00 & 1.00 & 0.68 & 1.00 & 0.81 & 1\\
LGA13 & 30 & 0.00 & 0.14 & 0.00 & 1.00 & 1.00 & 0.17 & 0.38 & 4\\
LGA13 & 45 & 0.33 & 0.82 & 0.93 & 1.00 & 0.80 & 0.87 & 0.74 & 2\\
LGA31 & 45 & 1.00 & 0.00 & 0.13 & 1.00 & 0.00 & 0.00 & 0.36 & 5\\
\end{tabular}
\end{table}
\begin{table}
\caption {Rank of trajectories generated for US Airways 1549 for t+12 seconds with glide ratio 19:1.}
\label{tab:table11}
\centering
\begin{tabular}{c c c c c c c c c c} 
Runway  & Bank angle & $\left \| \bar{d} \right \|$ &$ \left \| \bar{z} \right \|$&$ \left \| l \right \|$&$ \left \| n\right \|$&$ \left \| \bar{\left(\frac{\theta}{h}\right)}\right \|$&$ \left \| e\right \|$&$ u $& Rank\\ [1ex]
LGA22 & 30 & 0.09 & 0.36 & 0.20 & 1.00 & 0.81 & 0.41 & 0.47 & 3\\
LGA22 & 45 & 0.33 & 1.00 & 1.00 & 1.00 & 0.54 & 1.00 & 0.81 & 1\\
LGA13 & 30 & 0.00 & 0.11 & 0.00 & 1.00 & 1.00 & 0.17 & 0.38 & 4\\
LGA13 & 45 & 0.28 & 0.81 & 0.92 & 1.00 & 0.56 & 0.87 & 0.74 & 2\\
LGA31 & 45 & 1.00 & 0.00 & 0.17 & 1.00 & 0.00 & 0.00 & 0.36 & 5\\
\end{tabular}
\end{table}
\begin{table}
\caption {Rank of trajectories generated for US Airways 1549 for t+16 seconds with glide ratio 19:1.}
\label{tab:table12}
\centering
\begin{tabular}{c c c c c c c c c c} 
Runway  & Bank angle & $\left \| \bar{d} \right \|$ &$ \left \| \bar{z} \right \|$&$ \left \| l \right \|$&$ \left \| n\right \|$&$ \left \| \bar{\left(\frac{\theta}{h}\right)}\right \|$&$ \left \| e\right \|$&$ u $& Rank\\ [1ex]
LGA22 & 30 & 0.43 & 0.20 & 0.14 & 1.00 & 0.62 & 0.22 & 0.44 & 3\\
LGA22 & 45 & 1.00 & 1.00 & 1.00 & 1.00 & 0.00 & 1.00 & 0.83 & 2\\
LGA13 & 30 & 0.00 & 0.00 & 0.00 & 1.00 & 1.00 & 0.00 & 0.33 & 4\\
LGA13 & 45 & 0.96 & 0.80 & 0.95 & 1.00 & 0.51 & 0.83 & 0.84 & 1\\
\end{tabular}
\end{table}
\begin{table}
\caption {Rank of trajectories generated for US Airways 1549 for t+20 seconds with glide ratio 19:1.}
\label{tab:table13}
\centering
\begin{tabular}{c c c c c c c c c c} 
Runway  & Bank angle & $\left \| \bar{d} \right \|$ &$ \left \| \bar{z} \right \|$&$ \left \| l \right \|$&$ \left \| n\right \|$&$ \left \| \bar{\left(\frac{\theta}{h}\right)}\right \|$&$ \left \| e\right \|$&$ u $& Rank\\ [1ex]
LGA22 & 30 & 0.65 & 0.03 & 0.09 & 1.00 & 0.26 & 0.06 & 0.35 & 3\\
LGA22 & 45 & 1.00 & 1.00 & 1.00 & 1.00 & 0.00 & 1.00 & 0.83 & 1\\
LGA13 & 30 & 0.00 & 0.00 & 0.00 & 1.00 & 1.00 & 0.00 & 0.33 & 4\\
LGA13 & 45 & 0.94 & 0.84 & 0.97 & 1.00 & 0.12 & 0.86 & 0.79 & 2\\
\end{tabular}
\end{table}
\begin{table}
\caption {Rank of trajectories generated for US Airways 1549 for t+24 seconds with glide ratio 19:1.}
\label{tab:table14}
\centering
\begin{tabular}{c c c c c c c c c c} 
Runway  & Bank angle & $\left \| \bar{d} \right \|$ &$ \left \| \bar{z} \right \|$&$ \left \| l \right \|$&$ \left \| n\right \|$&$ \left \| \bar{\left(\frac{\theta}{h}\right)}\right \|$&$ \left \| e\right \|$&$ u $& Rank\\ [1ex]
LGA22 & 45 & 0.99 & 0.97 & 0.94 & 0.00 & 0.44 & 0.98 & 0.72 & 2\\
LGA13 & 30 & 0.00 & 0.00 & 0.00 & 1.00 & 0.00 & 0.00 & 0.17 & 3\\
LGA13 & 45 & 1.00 & 1.00 & 1.00 & 0.00 & 1.00 & 1.00 & 0.83 & 1\\
\end{tabular}
\end{table}
\begin{table}
\caption {Rank of trajectories generated for US Airways 1549 for t+28 seconds with glide ratio 19:1.}
\label{tab:table15}
\centering
\begin{tabular}{c c c c c c c c c c} 
Runway  & Bank angle & $\left \| \bar{d} \right \|$ &$ \left \| \bar{z} \right \|$&$ \left \| l \right \|$&$ \left \| n\right \|$&$ \left \| \bar{\left(\frac{\theta}{h}\right)}\right \|$&$ \left \| e\right \|$&$ u $& Rank\\ [1ex]
LGA22 & 45 & 1.00 & 0.00 & 0.00 & 1.00 & 0.00 & 0.00 & 0.33 & 2\\
LGA13 & 45 & 0.00 & 1.00 & 1.00 & 1.00 & 1.00 & 1.00 & 0.83 & 1\\
\end{tabular}
\end{table}
\begin{table}
\caption {Rank of trajectories generated for US Airways 1549 for t+32 seconds with glide ratio 19:1.}
\label{tab:table16}
\centering
\begin{tabular}{c c c c c c c c c c} 
Runway  & Bank angle & $\left \| \bar{d} \right \|$ &$ \left \| \bar{z} \right \|$&$ \left \| l \right \|$&$ \left \| n\right \|$&$ \left \| \bar{\left(\frac{\theta}{h}\right)}\right \|$&$ \left \| e\right \|$&$ u $& Rank\\ [1ex]
LGA22 & 45 & 1.00 & 0.00 & 0.00 & 1.00 & 0.00 & 1.00 & 0.50 & 2\\
LGA13 & 45 & 0.00 & 1.00 & 1.00 & 1.00 & 1.00 & 1.00 & 0.83 & 1\\
\end{tabular}
\end{table}
\begin{figure}[!htb]
\centering
\begin{subfigure}{.5\textwidth}
  \centering
  \includegraphics[width=.8\linewidth]{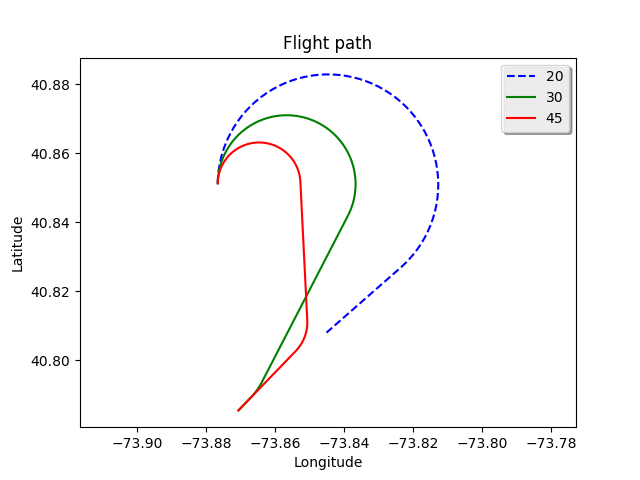}
  \caption{2D View.}
  \label{fig:sub1}
\end{subfigure}%
\begin{subfigure}{.5\textwidth}
  \centering
  \includegraphics[width=.8\linewidth]{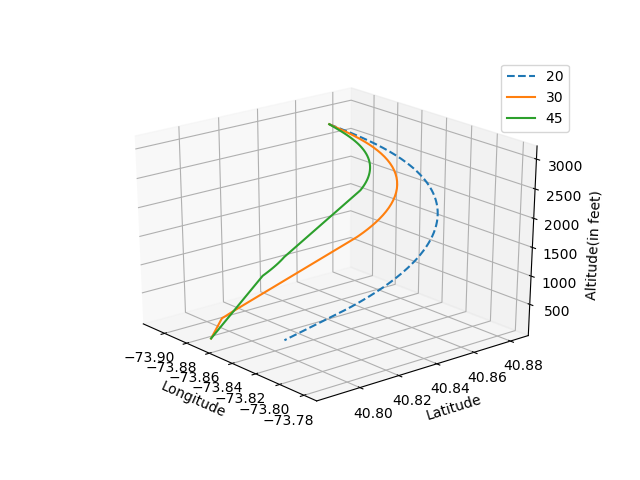}
  \caption{3D View.}
  \label{fig:sub2}
\end{subfigure}
\caption{Trajectory to LGA22 with a glide ratio of 19:1 at time t+4 .}
\label{fig:22_4}
\end{figure}
\begin{figure}[!htb]
\centering
\begin{subfigure}{.5\textwidth}
  \centering
  \includegraphics[width=.8\linewidth]{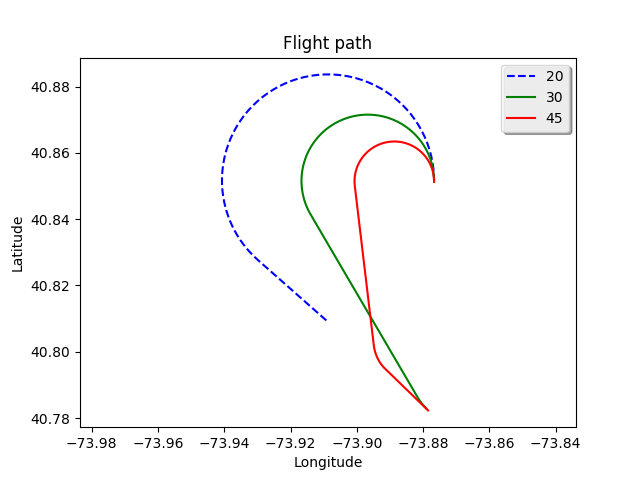}
  \caption{2D View.}
  \label{fig:sub1}
\end{subfigure}%
\begin{subfigure}{.5\textwidth}
  \centering
  \includegraphics[width=.8\linewidth]{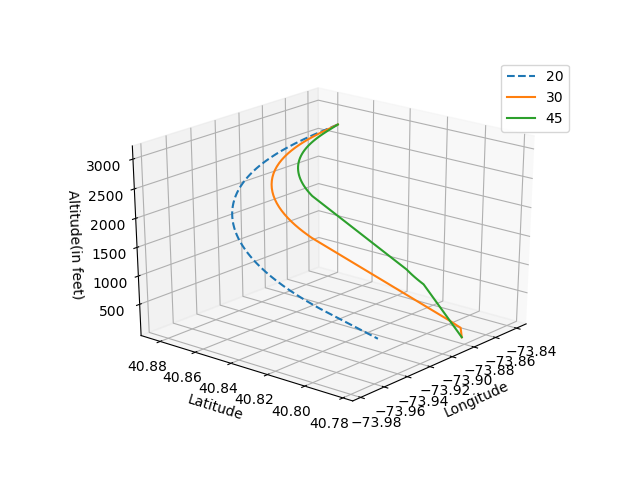}
  \caption{3D View.}
  \label{fig:sub2}
\end{subfigure}
\caption{Trajectory to LGA13 with a glide ratio of 19:1 at time t+4 .}
\label{fig:13_4}
\end{figure}
\begin{figure}[!htb]
\centering
\begin{subfigure}{.5\textwidth}
  \centering
  \includegraphics[width=.8\linewidth]{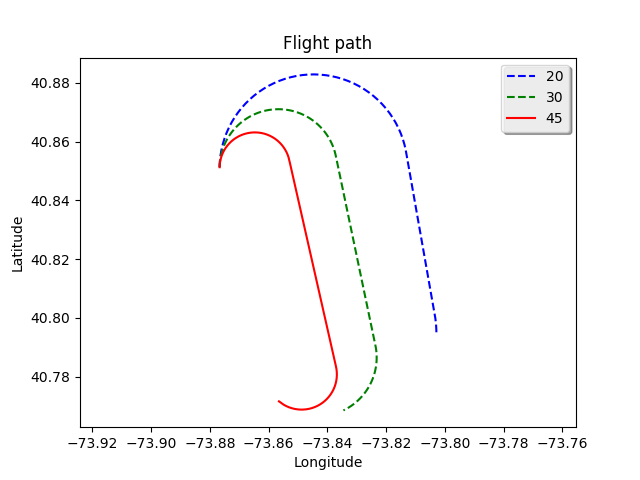}
  \caption{2D View.}
  \label{fig:sub1}
\end{subfigure}%
\begin{subfigure}{.5\textwidth}
  \centering
  \includegraphics[width=.8\linewidth]{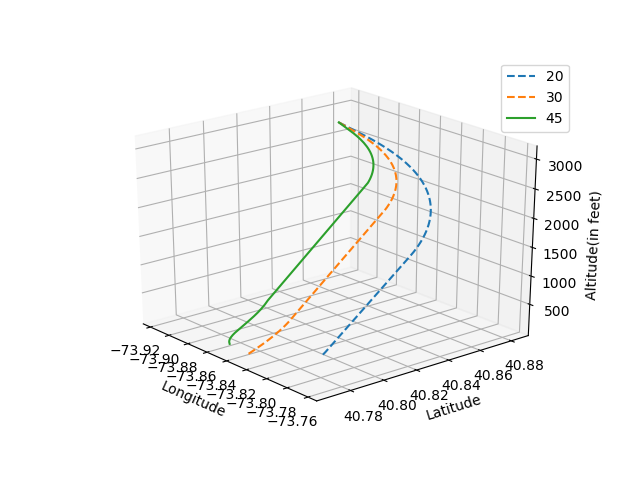}
  \caption{3D View.}
  \label{fig:sub2}
\end{subfigure}
\caption{Trajectory to LGA31 with a glide ratio of 19:1 at time t+4 .}
\label{fig:31_4}
\end{figure}
\begin{figure}[!htb]
\centering
\begin{subfigure}{.5\textwidth}
  \centering
  \includegraphics[width=.8\linewidth]{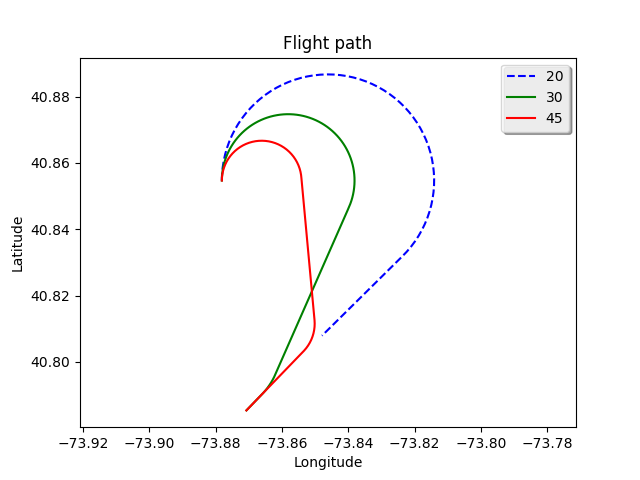}
  \caption{2D View.}
  \label{fig:sub1}
\end{subfigure}%
\begin{subfigure}{.5\textwidth}
  \centering
  \includegraphics[width=.8\linewidth]{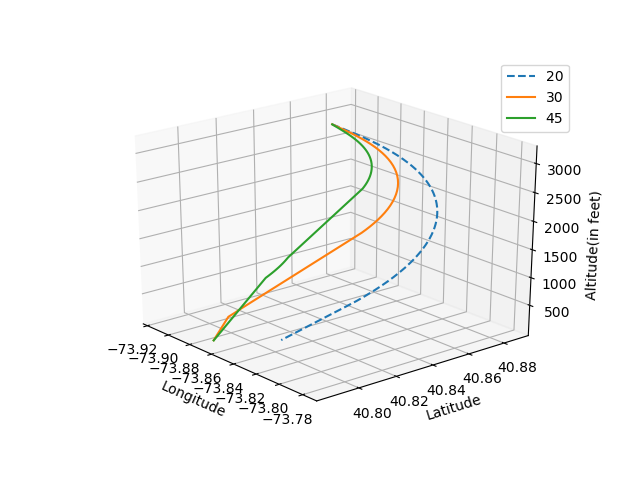}
  \caption{3D View.}
  \label{fig:sub2}
\end{subfigure}
\caption{Trajectory to LGA22 with a glide ratio of 19:1 at time t+8.}
\label{fig:22_8}
\end{figure}
\begin{figure}[!htb]
\centering
\begin{subfigure}{.5\textwidth}
  \centering
  \includegraphics[width=.8\linewidth]{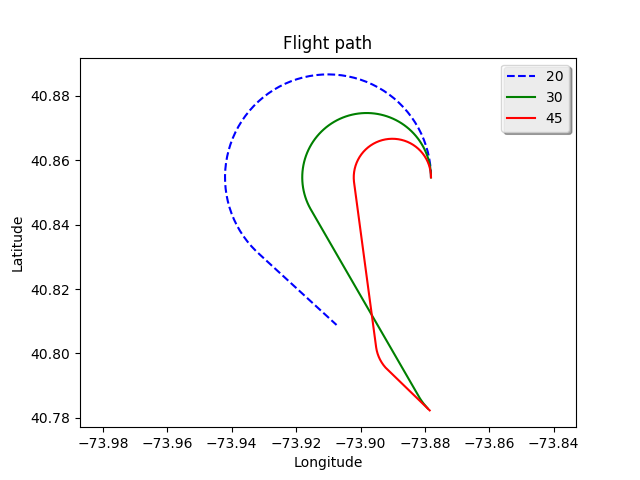}
  \caption{2D View.}
  \label{fig:sub1}
\end{subfigure}%
\begin{subfigure}{.5\textwidth}
  \centering
  \includegraphics[width=.8\linewidth]{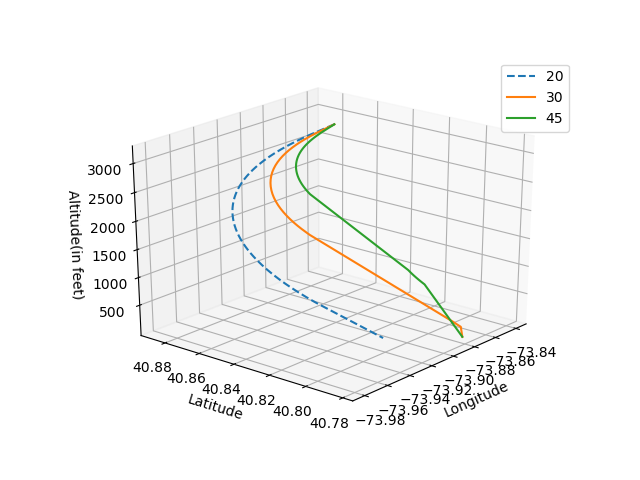}
  \caption{3D View.}
  \label{fig:sub2}
\end{subfigure}
\caption{Trajectory to LGA13 with a glide ratio of 19:1 at time t+8.}
\label{fig:13_8}
\end{figure}

\begin{figure}[!htb]
\centering
\begin{subfigure}{.5\textwidth}
  \centering
  \includegraphics[width=.8\linewidth]{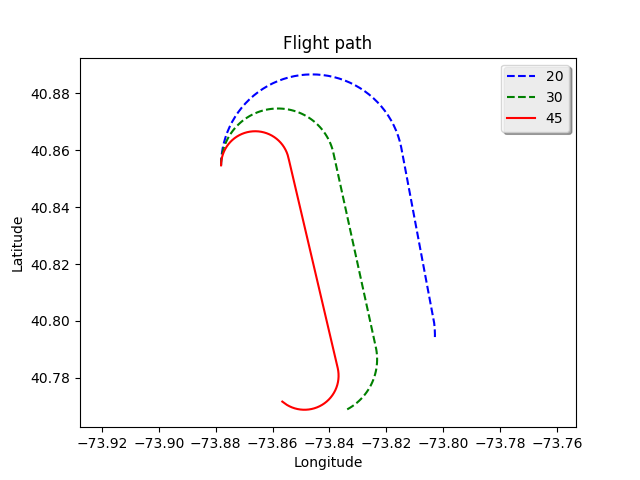}
  \caption{2D View.}
  \label{fig:sub1}
\end{subfigure}%
\begin{subfigure}{.5\textwidth}
  \centering
  \includegraphics[width=.8\linewidth]{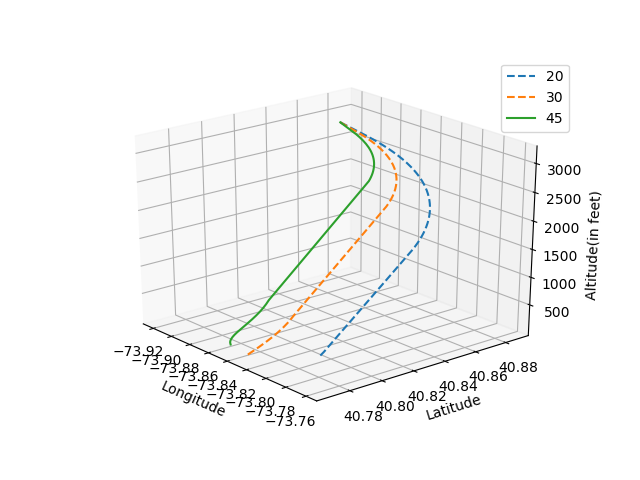}
  \caption{3D View.}
  \label{fig:sub2}
\end{subfigure}
\caption{Trajectory to LGA31 with a glide ratio of 19:1 at time t+8.}
\label{fig:31_8}
\end{figure}
\begin{figure}[!htb]
\centering
\begin{subfigure}{.5\textwidth}
  \centering
  \includegraphics[width=.8\linewidth]{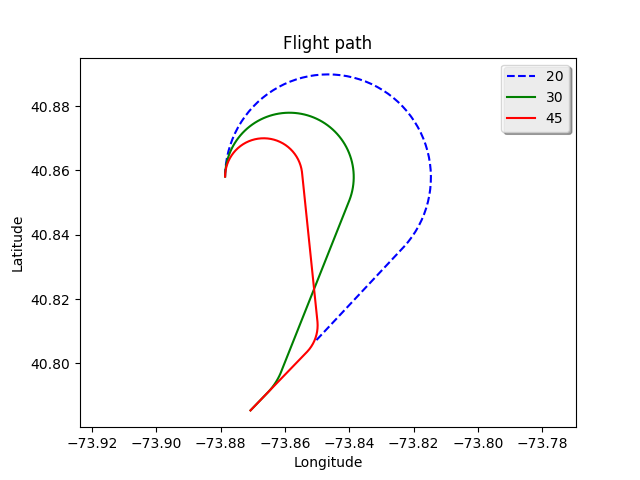}
  \caption{2D View.}
  \label{fig:sub1}
\end{subfigure}%
\begin{subfigure}{.5\textwidth}
  \centering
  \includegraphics[width=.8\linewidth]{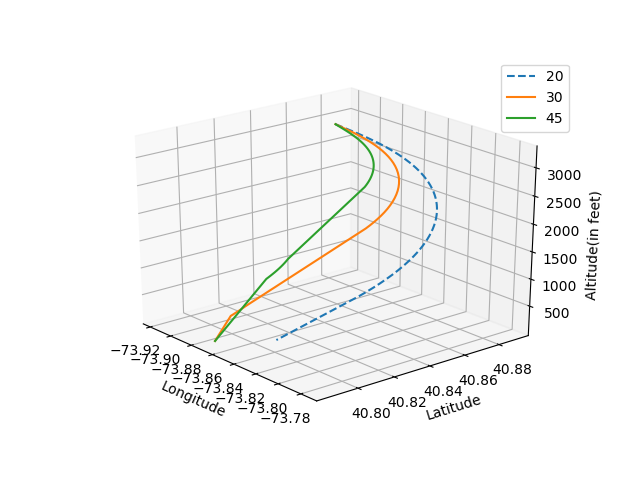}
  \caption{3D View.}
  \label{fig:sub2}
\end{subfigure}
\caption{Trajectory to LGA22 with a glide ratio of 19:1 at time t+12.}
\label{fig:22_12}
\end{figure}

\begin{figure}[!htb]
\centering
\begin{subfigure}{.5\textwidth}
  \centering
  \includegraphics[width=.8\linewidth]{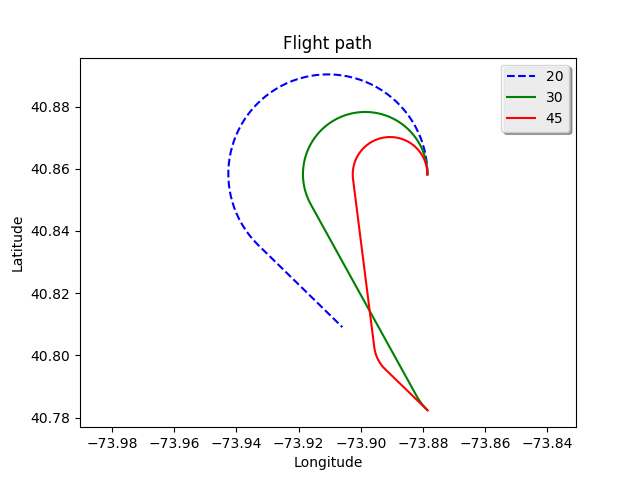}
  \caption{2D View.}
  \label{fig:sub1}
\end{subfigure}%
\begin{subfigure}{.5\textwidth}
  \centering
  \includegraphics[width=.8\linewidth]{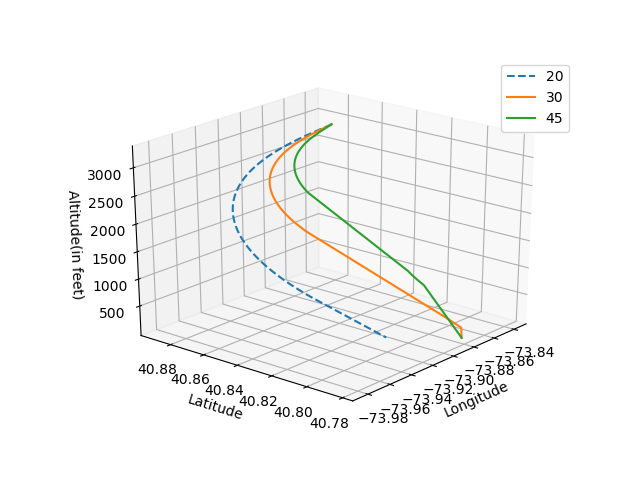}
  \caption{3D View.}
  \label{fig:sub2}
\end{subfigure}
\caption{Trajectory to LGA13 with a glide ratio of 19:1 at time t+12.}
\label{fig:13_12}
\end{figure}
\begin{figure}[!htb]
\centering
\begin{subfigure}{.5\textwidth}
  \centering
  \includegraphics[width=.8\linewidth]{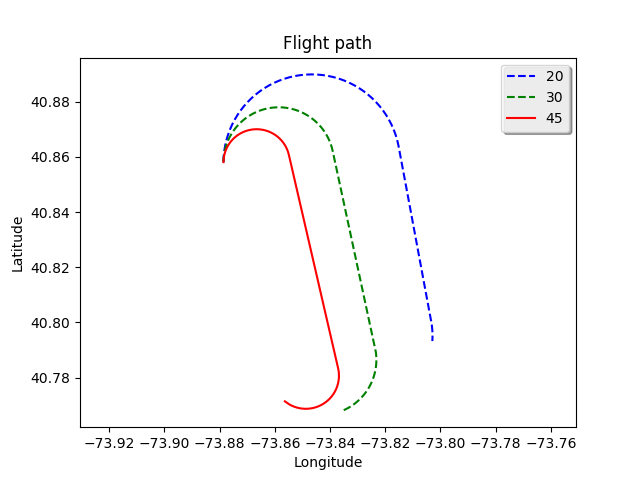}
  \caption{2D View.}
  \label{fig:sub1}
\end{subfigure}%
\begin{subfigure}{.5\textwidth}
  \centering
  \includegraphics[width=.8\linewidth]{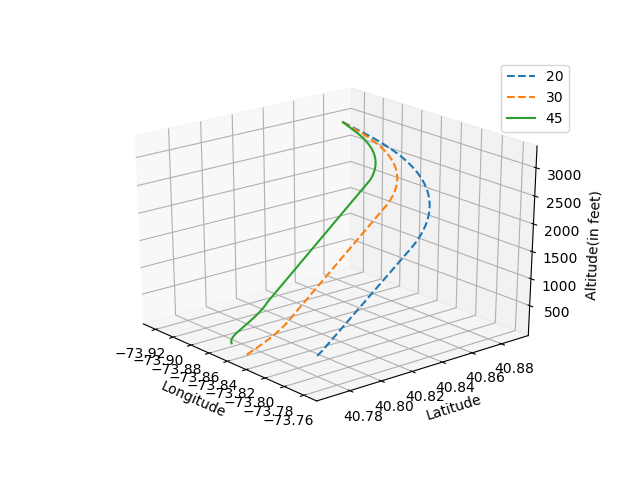}
  \caption{3D View.}
  \label{fig:sub2}
\end{subfigure}
\caption{Trajectory to LGA31 with a glide ratio of 19:1 at time t+12.}
\label{fig:31_12}
\end{figure}

\begin{figure}[!htb]
\centering
\begin{subfigure}{.5\textwidth}
  \centering
  \includegraphics[width=.8\linewidth]{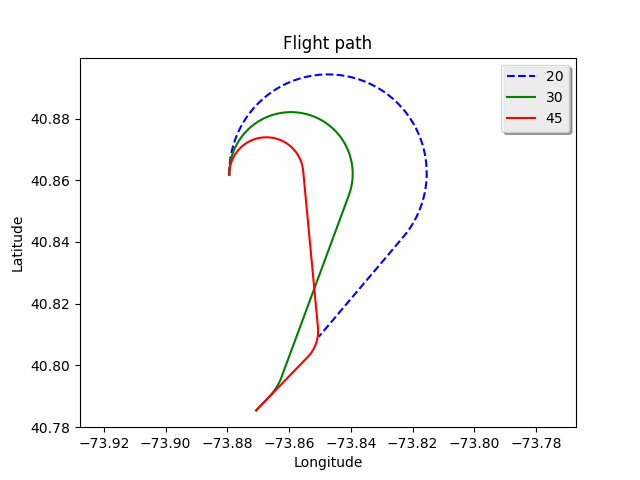}
  \caption{2D View.}
  \label{fig:sub1}
\end{subfigure}%
\begin{subfigure}{.5\textwidth}
  \centering
  \includegraphics[width=.8\linewidth]{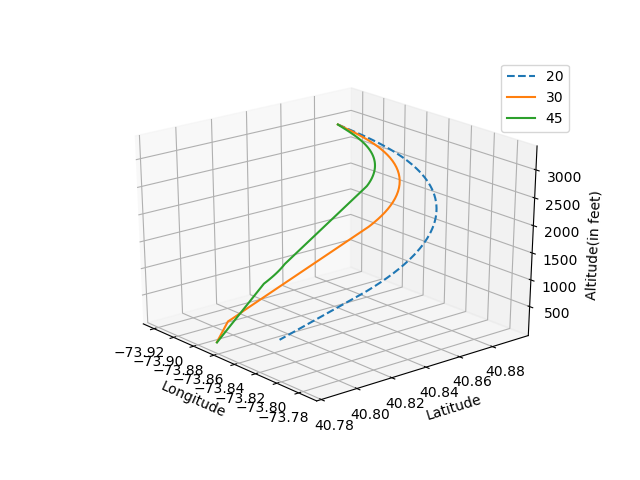}
  \caption{3D View.}
  \label{fig:sub2}
\end{subfigure}
\caption{Trajectory to LGA22 with a glide ratio of 19:1 at time t+16.}
\label{fig:22_16}
\end{figure}

\begin{figure}[!htb]
\centering
\begin{subfigure}{.5\textwidth}
  \centering
  \includegraphics[width=.8\linewidth]{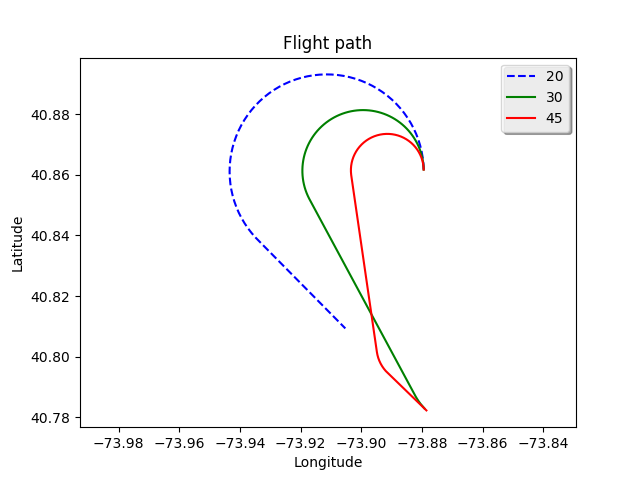}
  \caption{2D View.}
  \label{fig:sub1}
\end{subfigure}%
\begin{subfigure}{.5\textwidth}
  \centering
  \includegraphics[width=.8\linewidth]{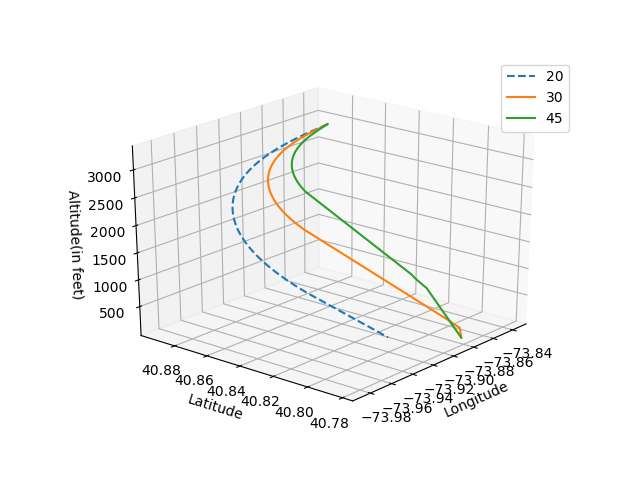}
  \caption{3D View.}
  \label{fig:sub2}
\end{subfigure}
\caption{Trajectory to LGA13 with a glide ratio of 19:1 at time t+16.}
\label{fig:13_16}
\end{figure}

\begin{figure}[!htb]
\centering
\begin{subfigure}{.5\textwidth}
  \centering
  \includegraphics[width=.8\linewidth]{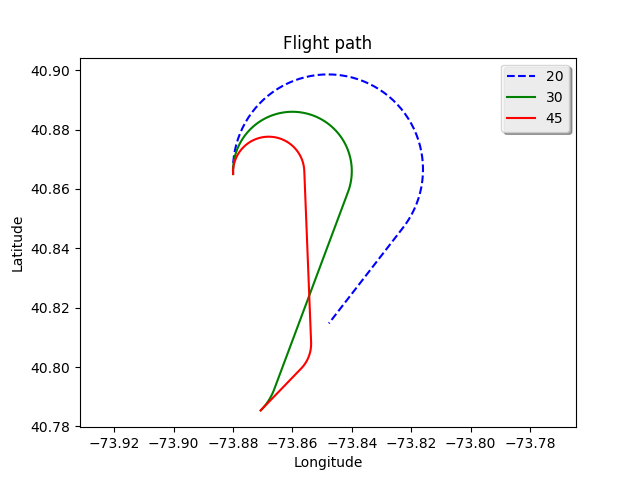}
  \caption{2D View.}
  \label{fig:sub1}
\end{subfigure}%
\begin{subfigure}{.5\textwidth}
  \centering
  \includegraphics[width=.8\linewidth]{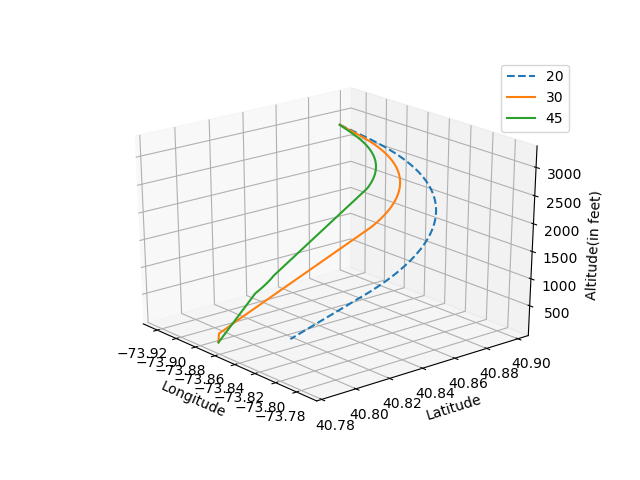}
  \caption{3D View.}
  \label{fig:sub2}
\end{subfigure}
\caption{Trajectory to LGA22 with a glide ratio of 19:1 at time t+20.}
\label{fig:22_20}
\end{figure}

\begin{figure}[!htb]
\centering
\begin{subfigure}{.5\textwidth}
  \centering
  \includegraphics[width=.8\linewidth]{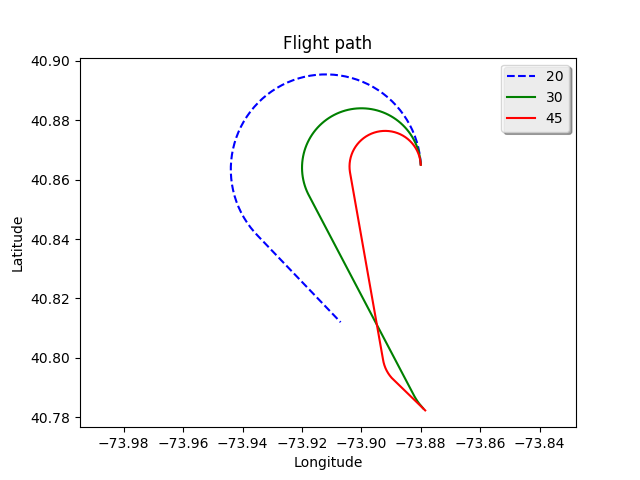}
  \caption{2D View.}
  \label{fig:sub1}
\end{subfigure}%
\begin{subfigure}{.5\textwidth}
  \centering
  \includegraphics[width=.8\linewidth]{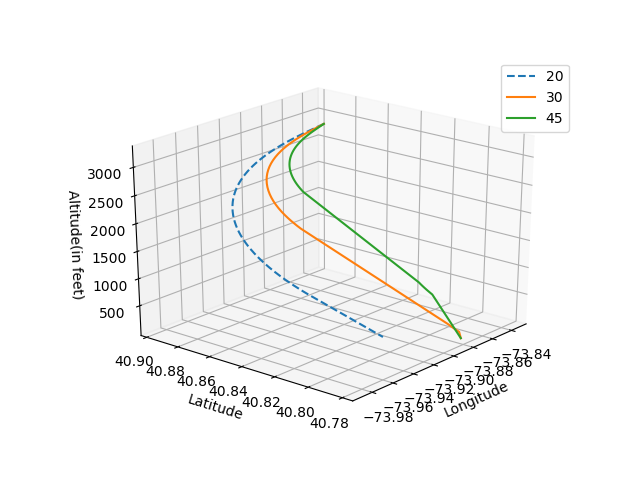}
  \caption{3D View.}
  \label{fig:sub2}
\end{subfigure}
\caption{Trajectory to LGA13 with a glide ratio of 19:1 at time t+20.}
\label{fig:13_20}
\end{figure}

\begin{figure}[!htb]
\centering
\begin{subfigure}{.5\textwidth}
  \centering
  \includegraphics[width=.8\linewidth]{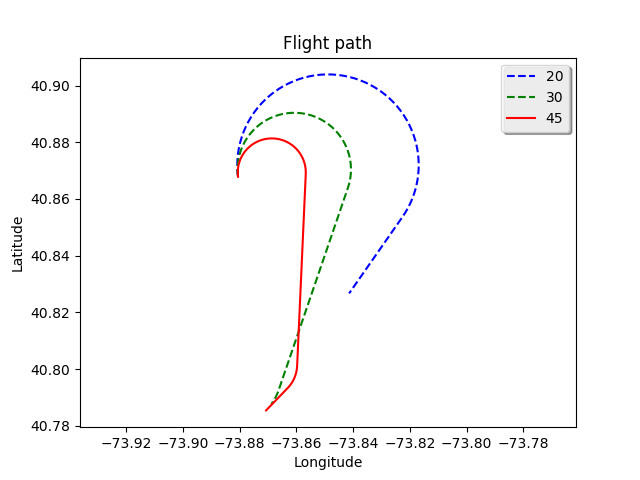}
  \caption{2D View.}
  \label{fig:sub1}
\end{subfigure}%
\begin{subfigure}{.5\textwidth}
  \centering
  \includegraphics[width=.8\linewidth]{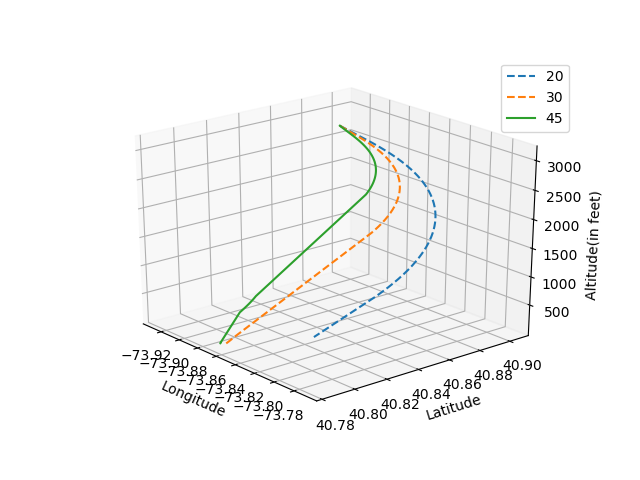}
  \caption{3D View.}
  \label{fig:sub2}
\end{subfigure}
\caption{Trajectory to LGA22 with a glide ratio of 19:1 at time t+24.}
\label{fig:22_24}
\end{figure}

\begin{figure}[!htb]
\centering
\begin{subfigure}{.5\textwidth}
  \centering
  \includegraphics[width=.8\linewidth]{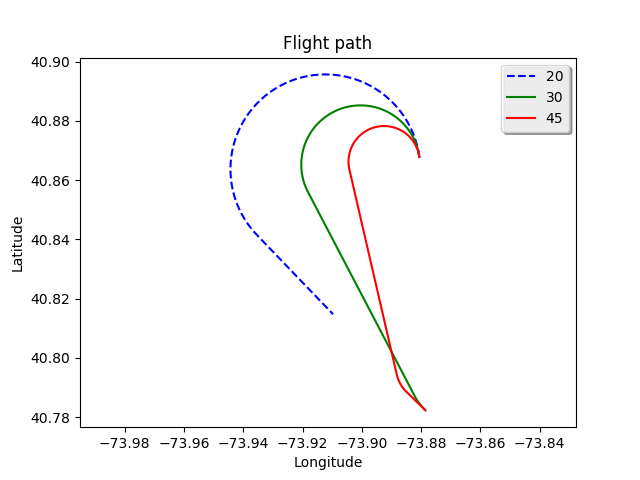}
  \caption{2D View.}
  \label{fig:sub1}
\end{subfigure}%
\begin{subfigure}{.5\textwidth}
  \centering
  \includegraphics[width=.8\linewidth]{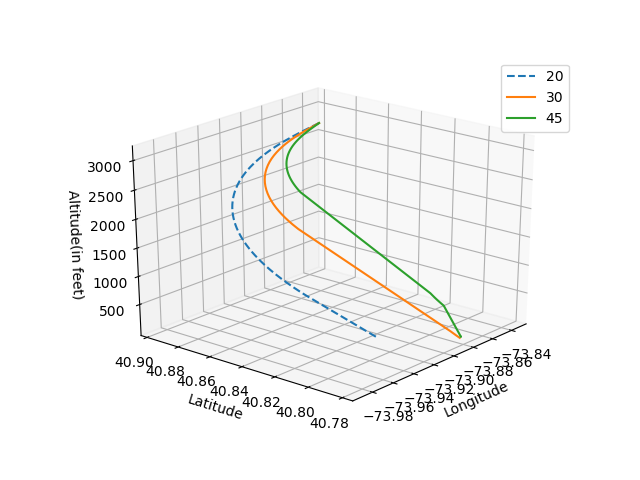}
  \caption{3D View.}
  \label{fig:sub2}
\end{subfigure}
\caption{Trajectory to LGA13 with a glide ratio of 19:1 at time t+24.}
\label{fig:13_24}
\end{figure}

\begin{figure}[!htb]
\centering
\begin{subfigure}{.5\textwidth}
  \centering
  \includegraphics[width=.8\linewidth]{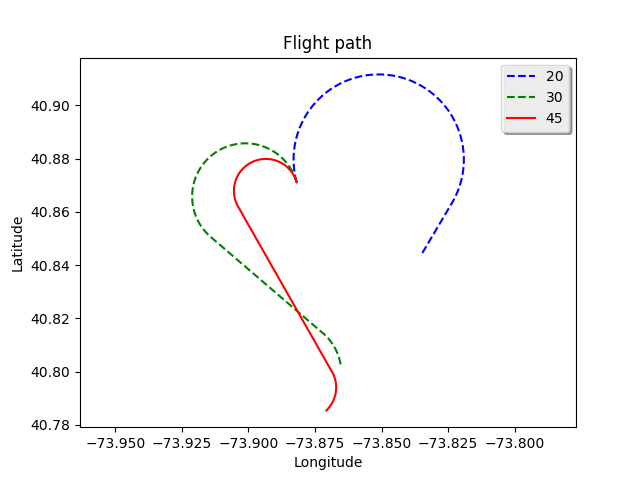}
  \caption{2D View.}
  \label{fig:sub1}
\end{subfigure}%
\begin{subfigure}{.5\textwidth}
  \centering
  \includegraphics[width=.8\linewidth]{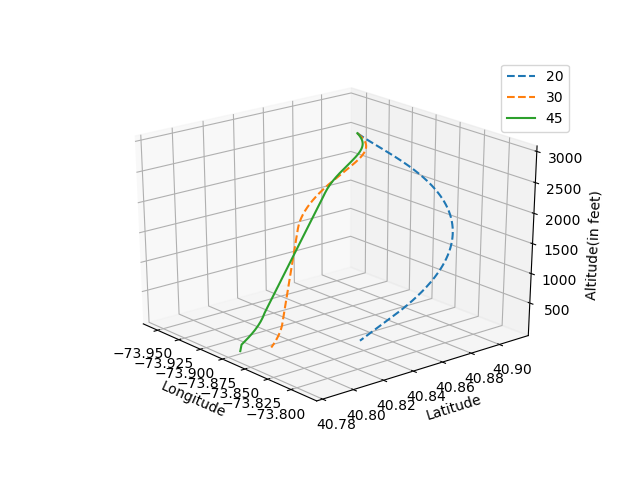}
  \caption{3D View.}
  \label{fig:sub2}
\end{subfigure}
\caption{Trajectory to LGA22 with a glide ratio of 19:1 at time t+28.}
\label{fig:22_28}
\end{figure}
\begin{figure}[!htb]
\centering
\begin{subfigure}{.5\textwidth}
  \centering
  \includegraphics[width=.8\linewidth]{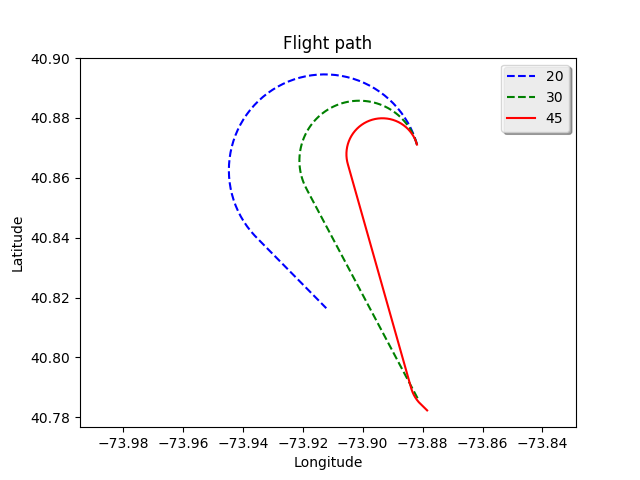}
  \caption{2D View.}
  \label{fig:sub1}
\end{subfigure}%
\begin{subfigure}{.5\textwidth}
  \centering
  \includegraphics[width=.8\linewidth]{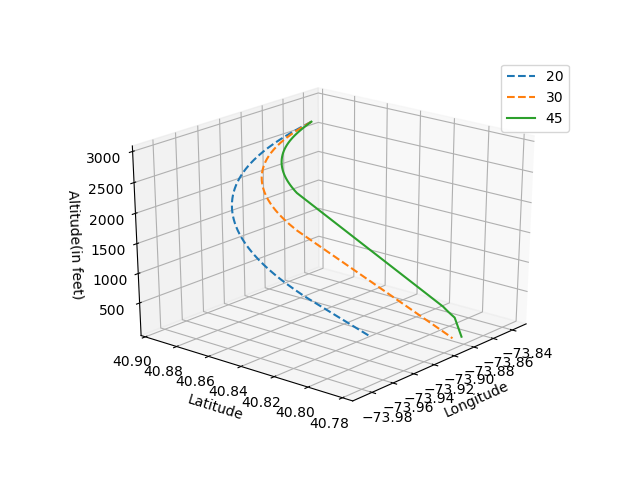}
  \caption{3D View.}
  \label{fig:sub2}
\end{subfigure}
\caption{Trajectory to LGA13 with a glide ratio of 19:1 at time t+28.}
\label{fig:13_28}
\end{figure}

\begin{figure}[!htb]
\centering
\begin{subfigure}{.5\textwidth}
  \centering
  \includegraphics[width=.8\linewidth]{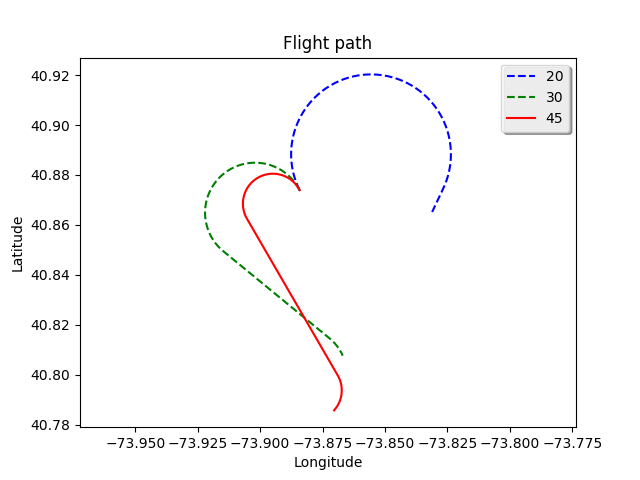}
  \caption{2D View.}
  \label{fig:sub1}
\end{subfigure}%
\begin{subfigure}{.5\textwidth}
  \centering
  \includegraphics[width=.8\linewidth]{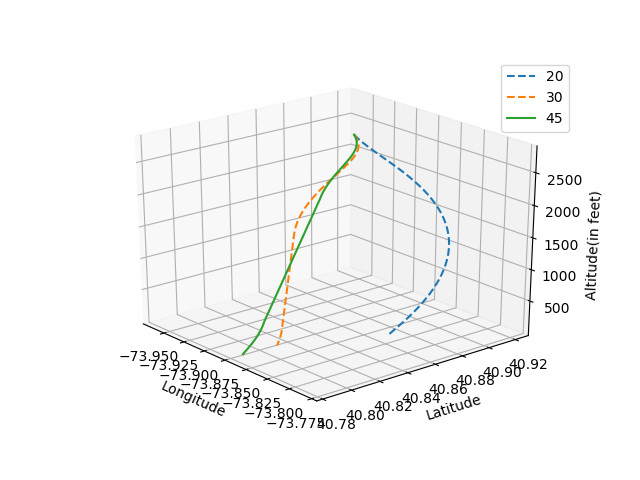}
  \caption{3D View.}
  \label{fig:sub2}
\end{subfigure}
\caption{Trajectory to LGA22 with a glide ratio of 19:1 at time t+32.}
\label{fig:22_32}
\end{figure}
\begin{figure}[!htb]
\centering
\begin{subfigure}{.5\textwidth}
  \centering
  \includegraphics[width=.8\linewidth]{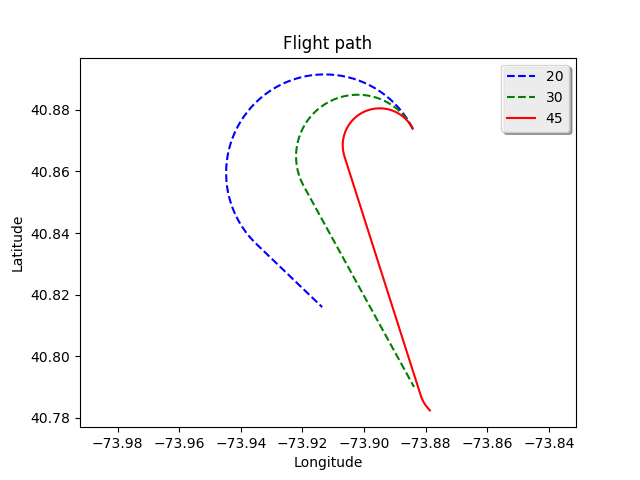}
  \caption{2D View.}
  \label{fig:sub1}
\end{subfigure}%
\begin{subfigure}{.5\textwidth}
  \centering
  \includegraphics[width=.8\linewidth]{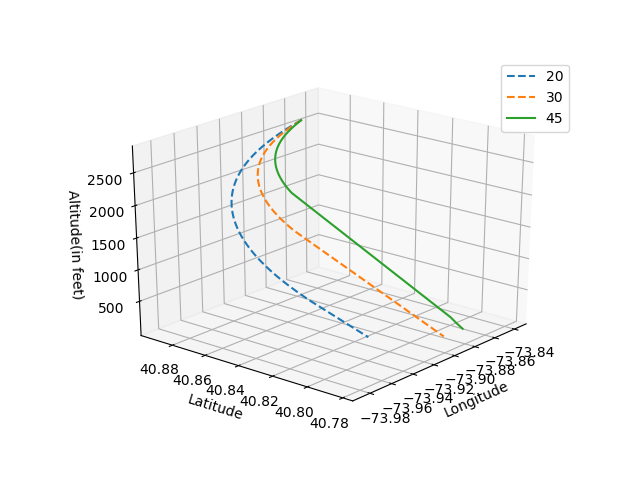}
  \caption{3D View.}
  \label{fig:sub2}
\end{subfigure}
\caption{Trajectory to LGA13 with a glide ratio of 19:1 at time t+32.}
\label{fig:13_32}
\end{figure}

\begin{figure}[!htb]
\centering
\begin{subfigure}{.5\textwidth}
  \centering
  \includegraphics[width=.8\linewidth]{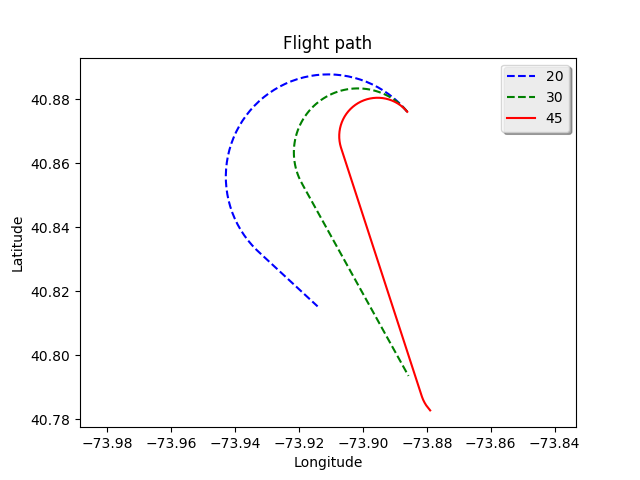}
  \caption{2D View.}
  \label{fig:sub1}
\end{subfigure}%
\begin{subfigure}{.5\textwidth}
  \centering
  \includegraphics[width=.8\linewidth]{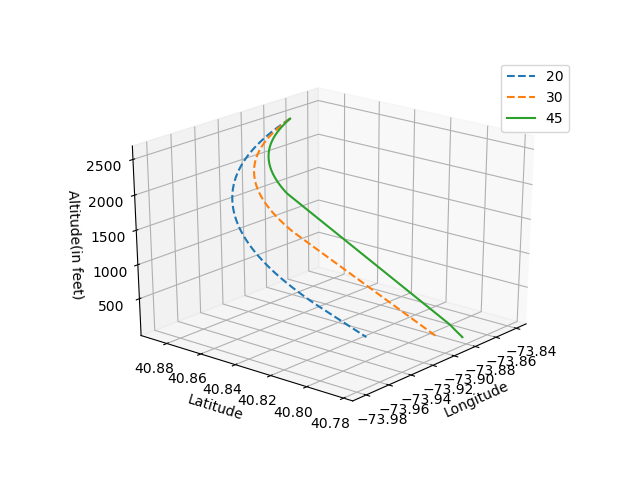}
  \caption{3D View.}
  \label{fig:sub2}
\end{subfigure}
\caption{Trajectory to LGA13 with a glide ratio of 19:1 at time t+36.}
\label{fig:13_36}
\end{figure}
We collected data from Flight Data Recorder data as published in the National Transportation Safety Board report (Table~\ref{tab:table2}) and simulated scenarios for t+4 through t+40 seconds, t being the time when the pilot said \emph{"birds"} (15:27:10 UTC) as determined from the sound recording in the Flight Data Recorder. From the data (Fig.~ \ref{1549tag}), it is clearly visible that the flight kept gaining altitude until t+16 seconds and attained a true altitude of 3352 feet before beginning to descend. We simulated two cases. For the first case, we used a glide ratio of 17.25:1 as predicted by \cite{avrenli2015green} for an Airbus A320 in no thrust conditions. For the second case, we used a glide ratio of 19:1 as calculated from the simulator data. We used a dirty configuration glide ratio of 9:1 for the extended runway segments.\\

\subsubsection{Using a glide ratio of 17.25:1}~\\
At t+4 seconds, the {latitude, longitude, true altitude, magnetic heading} configuration of the aircraft was \{40.8513, -73.8767, 3152, 0.7\} and our algorithm was able to generate trajectories to LGA22 using 30° and 45° bank angles (Fig. \ref{fig:22t4}) and LGA13 using 45° bank angle (Fig. \ref{fig:13t4}). LGA4 and LGA31 were unreachable according to our results. The trajectories were ranked according to our metrics (Table~\ref{tab:table3}).\\
At t+8 seconds, the {latitude, longitude, true altitude, magnetic heading} configuration of the aircraft was \{40.8547, -73.8781, ,3232, 0\} and LGA22 was reachable using 30° and 45° bank angles (Fig. \ref{fig:22t8}) and LGA13 was reachable using 45° bank angle (Fig. \ref{fig:13t8}). The trajectories were ranked according to our metrics (Table~\ref{tab:table4}).\\
At t+12 seconds, the {latitude, longitude, true altitude, magnetic heading} configuration of the aircraft was \{40.8581, -73.8786, 3312, 0.4\} and LGA22 was reachable using 30° and 45° bank angles (Fig. \ref{fig:22t12}) and LGA13 was reachable using 45° bank angle (Fig. \ref{fig:13t12}). The trajectories were ranked according to our metrics (Table~\ref{tab:table5}).\\ 
At t+16 seconds, the {latitude, longitude, true altitude, magnetic heading} configuration of the aircraft was \{ 40.8617, -73.8794, 3352, 358.9\} and LGA22 was reachable using 30° and 45° bank angles (Fig. \ref{fig:22t16}) and LGA13 was reachable using 45° bank angle (Fig. \ref{fig:13t16}). The trajectories were ranked according to our metrics (Table~\ref{tab:table6}).\\
At t+20 seconds, the {latitude, longitude, true altitude, magnetic heading} configuration of the aircraft was \{40.865, -73.88, 3304, 357.2\} LGA22 was reachable using 45° bank angle (Fig. \ref{fig:22t20}) and LGA13 was reachable using 45° bank angle (Fig. \ref{fig:13t20}). The trajectories were ranked according to our metrics (Table~\ref{tab:table7}).\\
At t+24 seconds, the {latitude, longitude, true altitude, magnetic heading} configuration of the aircraft was \{40.8678, -73.8806, 3180, 352.6\} and LGA22 was reachable using 45° bank angle (Fig. \ref{fig:22t24}) and LGA13 was reachable using 45° bank angle (Fig. \ref{fig:13t24}). The trajectories were ranked according to our metrics (Table~\ref{tab:table8}).\\
At t+28 seconds, the {latitude, longitude, true altitude, magnetic heading} configuration of the aircraft was \{40.8711, -73.8819, 3024, 344.5\} and LGA22 was no longer reachable while LGA13 was reachable using 45° bank angle(Fig. \ref{fig:13t28}). The trajectories were not ranked since there was only one trajectory.\\
At t+32 seconds, the {latitude, longitude, true altitude, magnetic heading} configuration of the aircraft was \{40.8789, -73.8897, 2420, 305.5\} and none of the runways at La Guardia Airport were reachable using any bank angle for turns.\\

\subsubsection{Using a glide ratio of 19:1}~\\
At t+4 seconds, the {latitude, longitude, true altitude, magnetic heading} configuration of the aircraft was \{40.8513, -73.8767, 3152, 0.7\} and our algorithm was able to generate trajectories to LGA31 using 45° bank angle (Fig. \ref{fig:31_4}), LGA22 using 30° and 45° bank angles (Fig. \ref{fig:22_4}) and LGA13 using 30° and 45° bank angles (Fig. \ref{fig:13_4}). LGA4 was unreachable according to our results. The trajectories were ranked according to our metrics (Table~\ref{tab:table9}).\\
At t+8 seconds, the {latitude, longitude, true altitude, magnetic heading} configuration of the aircraft was \{40.8547, -73.8781, 3232, 0\} and LGA31 was reachable using 45° bank angle (Fig. \ref{fig:31_8}), LGA22 was reachable using 30° and 45° bank angles (Fig. \ref{fig:22_8}) and LGA13 was reachable using 30° and 45° bank angles (Fig. \ref{fig:13_8}). The trajectories were ranked according to our metrics (Table~\ref{tab:table10}).\\
At t+12 seconds, the {latitude, longitude, true altitude, magnetic heading} configuration of the aircraft was \{40.8581, -73.8786, 3312, 0.4\} and LGA31 was reachable using 45° bank angle (Fig. \ref{fig:31_12}), LGA22 was reachable using 30° and 45° bank angles (Fig. \ref{fig:22_12}) and LGA13 was reachable using 30° and 45° bank angles (Fig. \ref{fig:13_12}). The trajectories were ranked according to our metrics (Table~\ref{tab:table11}).\\
At t+16 seconds, the {latitude, longitude, true altitude, magnetic heading} configuration of the aircraft was \{40.8617, -73.8794, 3352, 358.9\} and LGA31 was no longer reachable, LGA22 was reachable using 30° and 45° bank angles (Fig. \ref{fig:22_16}) and LGA13 was reachable using 30° and 45° bank angles (Fig. \ref{fig:13_16}). The trajectories were ranked according to our metrics (Table~\ref{tab:table12}).\\
At t+20 seconds, the {latitude, longitude, true altitude, magnetic heading} configuration of the aircraft was \{40.865, -73.88, 3304, 357.2\} LGA22 was reachable using 30° and 45° bank angles (Fig. \ref{fig:22_20}) and LGA13 was reachable using 30° and 45° bank angles (Fig. \ref{fig:13_20}). The trajectories were ranked according to our metrics (Table~\ref{tab:table13}).\\
At t+24 seconds, the {latitude, longitude, true altitude, magnetic heading} configuration of the aircraft was \{40.8678, -73.8806, 3180, 352.6\} and LGA22 was reachable using 30° and 45° (Fig. \ref{fig:22_24}) bank angles and LGA13 was reachable using 30° and 45° bank angles (Fig. \ref{fig:13_24}). The trajectories were ranked according to our metrics (Table~\ref{tab:table14}).\\
At t+28 seconds, the {latitude, longitude, true altitude, magnetic heading} configuration of the aircraft was \{40.8711, -73.8819, 3024, 344.5\} and LGA22 was reachable using 45° bank angle (Fig. \ref{fig:22_28}) and LGA13 was reachable using only 45° bank angle (Fig. \ref{fig:13_28}). The trajectories were ranked according to our metrics (Table~\ref{tab:table15}).\\
At t+32 seconds, the {latitude, longitude, true altitude, magnetic heading} configuration of the aircraft was \{40.8739, -73.8842, 2844, 333.3\} and LGA22 was reachable using 45° bank (Fig. \ref{fig:22_32}) angle and LGA13 was reachable using only 45° bank angle (Fig. \ref{fig:13_32}).
The trajectories were ranked according to our metrics (Table~\ref{tab:table16}).\\
At t+36 seconds, the {latitude, longitude, true altitude, magnetic heading} configuration of the aircraft was \{40.761, -73.8861, 2632, 320.6\} and LGA22 was no longer reachable and LGA13 was reachable using only 45° bank angle (Fig. \ref{fig:13_36}).\\
At t+40 seconds, the {latitude, longitude, true altitude, magnetic heading} configuration of the aircraft was \{40.8789, -73.8897, 2420, 305.5\} and none of the runways at La Guardia Airport were reachable using any bank angle for turns.

\begin{figure}[!htb]
\centering
\begin{subfigure}{.5\textwidth}
  \centering
  \includegraphics[width=.8\linewidth]{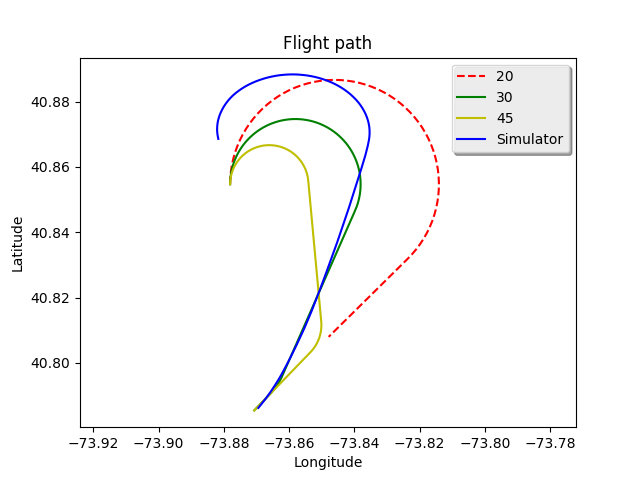}
  \caption{2D view.}
  \label{fig:sub12345}
\end{subfigure}%
\begin{subfigure}{.5\textwidth}
  \centering
  \includegraphics[width=.8\linewidth]{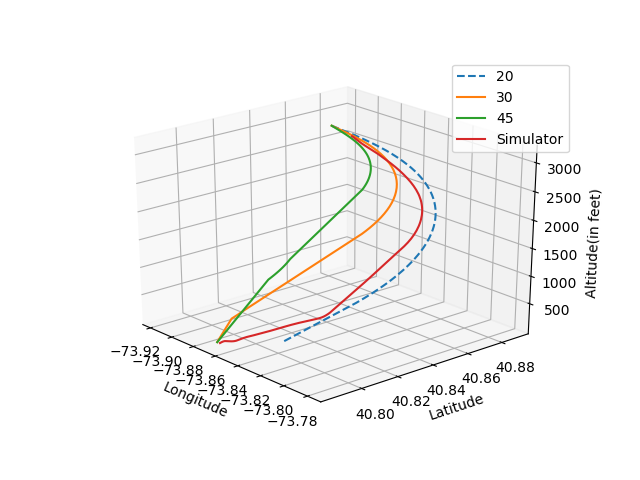}
  \caption{3D view.}
  \label{fig:sub2}
\end{subfigure}
\caption{Generated vs simulated trajectories to LGA22 at t+8.}
\label{fig:sim1}
\end{figure}

\begin{figure}[!htb]
\centering
\begin{subfigure}{.5\textwidth}
  \centering
  \includegraphics[width=.8\linewidth]{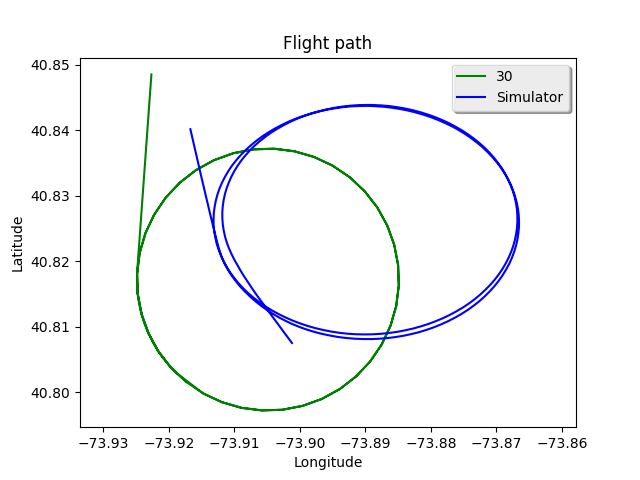}
  \caption{2D view.}
  \label{fig:sub12345}
\end{subfigure}%
\begin{subfigure}{.5\textwidth}
  \centering
  \includegraphics[width=.8\linewidth]{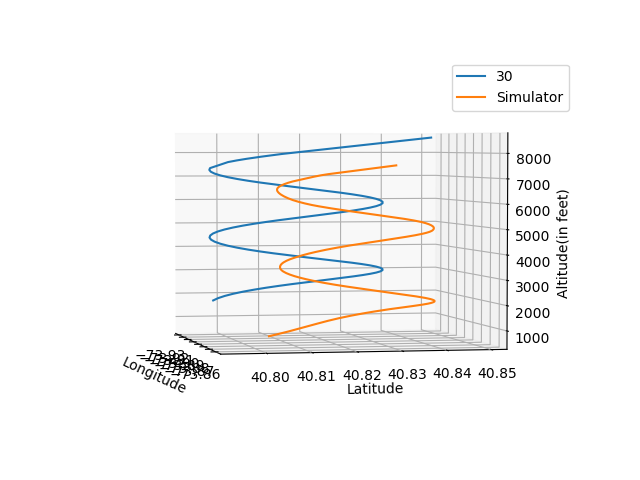}
  \caption{3D view.}
  \label{fig:sub2}
\end{subfigure}
\caption{Comparison of generated and simulated spiral segments.}
\label{fig:sim2}
\end{figure}

In order to check the viability and accuracy of the generated trajectories, we simulated the trajectories generated by our trajectory planning algorithm in a Precision Flight Controls CAT III Flight Simulator running X-Plane software with an Airbus A320 (Fig. \ref{fig:sim1}, Fig. \ref{fig:sim2}).
\section{Conclusion and Future Work}
Augmenting flight management architectures with a system to generate and rank feasible trajectories in case of LOT scenarios will help in reducing the response time of pilots for choosing a course of action to make a successful landing in case of such emergency situations. Our evaluation of US Airways flight 1549 shows that using a baseline glide ratio of 17.25:1 for straight line flight, runways were reachable up to 28 seconds after the emergency was detected and a water landing could have been avoided if this information was made available to the pilots during the actual emergency. Our algorithm was unable to generate trajectories to Teterboro airport even though we had successfully managed to land the flight at TEB24 during simulator trials. In the case of US Airways 1549, the left engine still had $30\%$ power. While it is not clear how much of this power turned into thrust for a damaged engine, a conservative baseline glide ratio of 19:1 illustrates that trajectories can be generated for up to 36 seconds after birds strike. Using a data-driven feedback loop, DDDAS based trajectory generation systems can determine the actual capabilities of the affected aircraft at the time in question and generate trajectories that are practical and also reject previously calculated trajectories that might no longer be feasible\let\thefootnote\relax\footnote{Source code available at: http://wcl.cs.rpi.edu/pilots}. 

Future directions of work includes generating terrain aware trajectory planning algorithms that take into account the features of the terrain by analyzing and processing data from terrain databases. Terrain aware algorithms may detect obstacles in the path and generate trajectories that avoid those obstacles and also detect alternative landing sites such as rivers, lakes, highways, \emph{etc}, if conventional runways are absent in the vicinity of an emergency. For example, in the case of US Airways 1549, the Hudson river was a very reasonable alternative landing zone (the pilots had concluded that no runway in the nearby airports were reachable) considering the fact that it provided the pilots with a practically unlimited runway to land the flight without having to worry about running short or overshooting the runway. In this paper, we have used Dubins airplane paths that consider only a single radius of turn for all turns in a particular trajectory. An interesting future direction of work would be to use small radii of turns (steep bank angles) in the initial curves and  larger radii of turns (shallower bank angles) in the final curves near the ground. Finally, uncertainty quantification in the trajectory planning process is another potential future direction. This would allow us to compute the probability of success of a trajectory by considering possible variations in variables such as wind, pilot error, and the actual amount of thrust being generated by the damaged engines.

\section{Acknowledgements}
This research is partially supported by the DDDAS program of the Air Force Office of Scientific Research, Grant No. FA9550-15-1-0214 and NSF Grant No. 1462342.

\bibliographystyle{unsrt}
\bibliography{references}
\end{document}